\documentclass[english,11pt,a4paper,DIV=16,numbers=noenddot]{scrartcl}
\pdfoutput=1
\usepackage[utf8]{inputenc}
\usepackage{epsf}
\usepackage{amsmath,amssymb}
\usepackage{graphicx}
\usepackage{listings}
\usepackage{array}
\usepackage{dcolumn}
\usepackage{mathtools}
\usepackage{scalefnt,ulem,soul}
\usepackage{cite}
\usepackage{color}
\usepackage[table]{xcolor}

\usepackage{etoolbox}

\makeatletter
\pretocmd\start@align
{%
  \let\everycr\CT@everycr
  \CT@start
}{}{}
\apptocmd{\endalign}{\CT@end}{}{}
\makeatother

\let\theparentequation\theequation
\patchcmd{\theparentequation}{equation}{parentequation}{}{}

\renewenvironment{subequations}[1][]{
  \refstepcounter{equation}%
  \setcounter{parentequation}{\value{equation}}
  \setcounter{equation}{0}
  \def\theequation{\theparentequation\alph{equation}}%
  \let\parentlabel\label
  \ifx\\#1\\\relax\else\label{#1}\fi
  \ignorespaces
}{%
  \setcounter{equation}{\value{parentequation}}
  \ignorespacesafterend
}

\newcommand*{\nextParentEquation}[1][]{
  \refstepcounter{parentequation}
  \setcounter{equation}{0}
  \ifx\\#1\\\relax\else\parentlabel{#1}\fi
}

\makeatletter
\newlength{\fnhskip}
\renewcommand\@makefntext[1]{
  \settowidth{\fnhskip}{\@makefnmark}
  \leftskip=\fnhskip
  \hskip-\fnhskip
  \@makefnmark#1
}
\makeatother
\usepackage{setspace}

\usepackage[numbers,sort&compress]{natbib}
\makeatletter
\def\NAT@spacechar{\,}
\makeatother

\usepackage[
colorlinks=true
,urlcolor=blue
,anchorcolor=blue
,citecolor=blue
,filecolor=blue
,linkcolor=blue
,menucolor=blue
,pagecolor=blue
,linktoc=all
,unicode
,backref=page
]{hyperref}
\usepackage{hypernat}

\usepackage[all]{hypcap}
\usepackage[format=plain, margin=0.5cm, font=footnotesize, labelfont=bf]{caption}

\usepackage{cleveref}

\usepackage{footnotebackref}

\addtokomafont{disposition}{\rmfamily}

\newrobustcmd*{\tocref}[1]{\hyperref[TOC]{\color{black}{#1}}}
\newcommand{\tocsection}[2][]{\section[\boldmath #2]{\tocref{#2#1}}}
\newcommand{\tocsubsection}[2][]{\subsection[\boldmath #2]{\tocref{#2#1}}}

\renewcommand*{\backref}[1]{}
\renewcommand*{\backrefalt}[4]{%
  \ifcase #1%
  \or [p\,#2]%
  \else [pp\,#2]%
  \fi%
}

\newif\ifbackrefshowonlyfirst
\backrefshowonlyfirsttrue
\makeatletter
\let\BR@direct@old@hyper@natlinkstart\hyper@natlinkstart
\renewcommand*{\hyper@natlinkstart}{\phantomsection\BR@direct@old@hyper@natlinkstart}
\let\BR@direct@oldBR@citex\BR@citex
\renewcommand*{\BR@citex}{\phantomsection\BR@direct@oldBR@citex}%

\long\def\hyper@page@BR@direct@ref#1#2#3{\hyperlink{#3}{#1}}

\ifx\backrefxxx\hyper@page@backref
    \let\backrefxxx\hyper@page@BR@direct@ref
    \ifbackrefshowonlyfirst
    \fi
\else
    \ifbackrefshowonlyfirst
    \fi
\fi

\patchcmd{\Hy@backout}{Doc-Start}{\@currentHref}{}{\errmessage{I can't seem to patch backref}}
\makeatother

\apptocmd{\thebibliography}{\setlength{\itemsep}{0.02cm}}{}{}
\Crefname{figure}{Fig.}{Figs.}
\usepackage{placeins}

\let\theparentequation\theequation
\patchcmd{\theparentequation}{equation}{parentequation}{}{}

\BeforeTOCHead[toc]{{\pdfbookmark[1]{\contentsname}{toc}}}
\apptocmd{\thebibliography}{\scriptsize}{}{}
\let\OLDthebibliography\thebibliography
\renewcommand\thebibliography[1]{
  \OLDthebibliography{#1}
  \setlength{\parskip}{1pt}
  \setlength{\itemsep}{1pt plus 0.3ex}
}

\newcommand{\IE}{\textit{i.\,e.}}
\newcommand{\EG}{\textit{e.\,g.}}
\newcommand{\citere}[1]{Ref.\,\cite{#1}}
\newcommand{\citeres}[1]{Refs.\,\cite{#1}}
\newcommand{\simord}{\mathord{\sim}\,}

\newcommand{\lsimord}{\mathord{\lesssim}\,}
\newcommand{\gsimord}{\mathord{\gtrsim}\,}
\newcommand{\code}[1]{\texttt{#1}}

\newcommand{\FH}{\code{FeynHiggs}}
\newcommand{\FA}{\code{FeynArts}}
\newcommand{\FC}{\code{FormCalc}}
\newcommand{\LT}{\code{LoopTools}}
\newcommand{\HB}{\code{HiggsBounds}}
\newcommand{\HS}{\code{HiggsSignals}}
\newcommand{\HBv}[1]{\code{HiggsBounds} version \code{#1}}
\newcommand{\HSv}[1]{\code{HiggsSignals} version \code{#1}}
\newcommand{\abbrev}{\scalefont{.9}}
\newcommand{\eqn}[1]{Eq.\,(\ref{#1})}
\newcommand{\eqns}[1]{Eqs.\,(\ref{#1})}
\newcommand{\fig}[1]{Fig.\,\ref{#1}}
\newcommand{\figs}[1]{Figs.\,\ref{#1}}
\newcommand{\sct}[1]{Section~\ref{#1}}
\newcommand{\appref}[1]{Appendix~\ref{#1}}

\newcommand{\lhc}{{\abbrev LHC}}
\newcommand{\lep}{{\abbrev LEP}}
\newcommand{\qcd}{{\abbrev QCD}}

\newcommand{\sm}{{\abbrev SM}}

\newcommand{\mssm}{{\abbrev MSSM}}
\newcommand{\nmssm}{{\abbrev NMSSM}}
\newcommand{\munmssm}{{\abbrev $\mu$NMSSM}}
\newcommand{\gnmssm}{{\abbrev GNMSSM}}

\newcommand{\susy}{{\abbrev SUSY}}

\newcommand{\atlas}{{\abbrev ATLAS}}
\newcommand{\cms}{{\abbrev CMS}}

\newcommand{\cp}{{\abbrev $\mathcal{CP}$}}

\newcommand{\nklo}[1]{{\abbrev N$^{#1}$LO}}
\newcommand{\drbar}{{\abbrev $\overline{\text{DR}}$}}

\newcommand{\betaf}[1]{\ensuremath{\beta{\left(#1\right)}}}

\newcommand{\MeV}{\textrm{MeV}}
\newcommand{\GeV}{\textrm{GeV}}
\newcommand{\TeV}{\textrm{TeV}}
\newcommand{\Mass}{{\mathcal{M}}}

\newcommand{\mue}{\mu_{\text{eff}}}
\newcommand{\sign}{\ensuremath{\operatorname{sign}}}

\newcounter{notecount}
\makeatletter
\DeclareRobustCommand\em{%
  \@nomath\em \ifdim \fontdimen\@ne\font >\z@\scshape
  \else \slshape \fi}
\makeatother

\renewcommand{\emph}[1]{{\em #1}}

\hypersetup{
  pdfauthor={Wolfgang G. Hollik, Stefan Liebler, Gudrid Moortgat-Pick, Sebastian Paßehr and Georg Weiglein}
  ,pdftitle={Phenomenology of the inflation-inspired NMSSM at the electroweak scale}
  ,pdfsubject={}
  ,pdfkeywords={}
}

\title{\vspace*{-4em}
  \begin{flushright}
    {\textsf{\small
        arXiv:1809.07371,
        DESY-17-075, KA-TP-26-2018, TTP18--035}\\
    }
  \end{flushright}
  \vspace*{2em}
  Phenomenology of the inflation-inspired {\Huge\nmssm{}} at the electroweak scale}

\author{Wolfgang Gregor Hollik${}^{a,b,c}$, Stefan Liebler${}^d$, Gudrid Moortgat-Pick${}^{a,e}$,\\ Sebastian Pa{\ss}ehr${}^{f}$, Georg Weiglein${}^a$\\[.8em]
  \normalsize\textit{${}^a$DESY, Notkestra{\ss}e 85, D-22607 Hamburg, Germany}\\[.4em]
  \normalsize\textit{${}^b$Institute for Nuclear Physics (IKP),}\\[-1ex]
  \normalsize\textit{Karlsruhe Institute of Technology, D-76021 Karlsruhe, Germany}\\[.4em]
  \normalsize\textit{${}^c$Institute for Theoretical Particle Physics (TTP),} \\[-1ex]
  \normalsize\textit{Karlsruhe Institute of Technology, D-76128 Karlsruhe, Germany}\\[.4em]
  \normalsize\textit{${}^d$Institute for Theoretical Physics (ITP),} \\[-1ex]
  \normalsize\textit{Karlsruhe Institute of Technology, D-76131 Karlsruhe, Germany}\\[.4em]
  \normalsize\textit{${}^e$II. Institut f\"ur Theoretische Physik, Universit\"at Hamburg,}\\[-1ex]
  \normalsize\textit{Luruper Chaussee 149, D-22761 Hamburg, Germany}\\[.4em]
  \normalsize\textit{$^f$Sorbonne Université, CNRS,}\\[-1ex]
  \normalsize\textit{Laboratoire de Physique Théorique et Hautes Énergies (LPTHE),}\\[-1ex]
  \normalsize\textit{4 Place Jussieu, F--75252 Paris CEDEX~05, France}\\[.8em]
  {\small\texttt{hollik@kit.edu}}\\[-0.3em]
  {\small\texttt{stefan.liebler@kit.edu}}\\[-0.3em]
  {\small\texttt{gudrid.moortgat-pick@desy.de}}\\[-0.3em]
  {\small\texttt{passehr@lpthe.jussieu.fr}}\\[-0.3em]
  {\small\texttt{georg.weiglein@desy.de}}
}

\date{\today}

\begin{document}
\maketitle
\thispagestyle{empty}

\begin{abstract}
\noindent
The concept of Higgs inflation can be elegantly incorporated in the
Next-to-Minimal Supersymmetric Standard Model~(\nmssm{}). A linear
combination of the two Higgs-doublet fields plays the role of the
inflaton which is non-minimally coupled to gravity. This non-minimal
coupling appears in the low-energy effective superpotential and
changes the phenomenology at the electroweak scale. While the field
content of the inflation-inspired model is the same as in the~\nmssm,
there is another contribution to the~\(\mu\)~term in addition to the
vacuum expectation value of the singlet. We explore this extended
parameter space and point out scenarios with phenomenological
differences compared to the pure~\nmssm. A special focus is set on the
electroweak vacuum stability and the parameter dependence of the Higgs
and neutralino sectors. We highlight regions which yield a \sm{}-like
$125$\,GeV Higgs boson compatible with the experimental observations
and are in accordance with the limits from searches for additional
Higgs bosons.  Finally, we study the impact of the non-minimal
coupling to gravity on the Higgs mixing and in turn on the decays of
the Higgs bosons in this model.

\end{abstract}

\newpage
\hypersetup{linkcolor=black}

\setstretch{.86}
\tableofcontents\label{TOC}

\hypersetup{linkcolor=blue}

\setstretch{.98}

\tocsection[\label{sec:introduction}]{Introduction}

In the history of our universe, there has been a period in which the
size of the universe exponentially increased. This short period is
known as inflationary epoch, and many models have been developed in
order to explain the inflation of the early universe. Unfortunately,
most of these models of inflation cannot be tested directly in the
laboratory; the observation of the universe is the only discriminator
to disfavor or support such models. Therefore, testing the
phenomenology of a particle physics model of inflation at the
electroweak scale with colliders is of interest both from the point of
view of particle physics and cosmology.

One possibility to describe inflation is the extension of a particle
physics model by additional scalar fields which drive inflation but
are removed from the theory afterwards. A more economical approach is
the idea of using the Higgs field of the Standard Model~(\sm{}) as
inflaton~\cite{Bezrukov:2007ep, Bezrukov:2008ej, Bezrukov:2009db}. The
simplest version, however, is under tension as it suffers from a
fine-tuning and becomes unnatural~\cite{Barbon:2009ya}. A less minimal
version of Higgs-portal inflation with an additional complex scalar
field can in addition solve further problems of the \sm{}, see
\citeres{Ballesteros:2016euj, Ballesteros:2016xej}. Also the concept
of critical Higgs inflation can raise the range of perturbativity to
the Planck scale and solve further problems of the SM,
see \citeres{Salvio:2015cja,Salvio:2015kka,Salvio:2017oyf}. Other
solutions are offered by scale-free extensions of the~\sm{}. A natural
way of such an implementation can be realized in canonical
superconformal supergravity~(CSS) models as proposed
by~\citeres{Ferrara:2010yw, Ferrara:2010in} based on earlier work
by~\citere{Einhorn:2009bh}.

The Higgs inflation in the supergravity framework is triggered by a
non-minimal coupling to Einstein gravity. For the supergravity
Lagrangian this can be achieved with an additional
term~\(X(\hat\Phi)\,R\) of chiral superfields~\(\hat\Phi\) and the
curvature multiplet~\(R\) (the supersymmetrized field version of the
Ricci scalar which contains the scalar curvature in the Grassmannian
coordinate~$\theta^2$), following the notation
of~\citere{Einhorn:2009bh}. The Lagrangian then reads
\begin{equation}\label{eq:nonminSUGRA}
\mathcal{L}_X = -6 \int \operatorname{d^2\theta}\mathcal{E}
\left[
R + X(\hat\Phi)\,R - \frac{1}{4} \left( {\bar{\mathcal{D}}}^2 - 8\,R \right)
\hat{\Phi}^\dag\,\hat\Phi + \mathcal{W}(\hat\Phi) \right]
+ \text{h.\,c.} + \ldots,
\end{equation}
where~\(X(\hat\Phi)\) as well as the
Superpotential~\(\mathcal{W}(\hat\Phi)\) are holomorphic functions of
the (left) chiral superfields~\(\hat\Phi\), \(\mathcal{E}\) is the
vierbein multiplet and \(\bar{\mathcal{D}}\) a covariant
derivative. The ellipses encode further gauge terms. The only possible
choice of such a non-minimal coupling suitable for inflation is given
by \cite{Einhorn:2009bh}
\begin{equation}\label{eq:Xfunction}
  X = \chi\,\hat{H}_u \cdot \hat{H}_d,
\end{equation}
where~$\chi$ is a dimensionless coupling and~$\hat{H}_{d,u}$ contain
the two~$SU(2)_{\text{L}}$~Higgs doublets of the Next-to-Minimal
Supersymmetric Standard Model~(\nmssm{}).\footnote{The field content
of the~\mssm{} alone (without the Higgs singlet) is not sufficient to
describe inflation successfully as pointed out
in \citere{Einhorn:2009bh}.}  The extension by an additional scalar
singlet like in the~\nmssm{} has been shown to be a viable model for
inflation, although this version suffers from a tachyonic
instability~\cite{Lee:2010hj}. In order to avoid this instability, a
stabilizer term has been introduced in \citeres{Lee:2010hj,
Ferrara:2010in} that is suppressed at low energies.  The stabilizer
term can be avoided in a model with minimal supergravity couplings
where the K\"ahler potential has a shift symmetry in the doublet
fields~\cite{BenDayan:2010yz}; however, cosmological phenomenology and
observations have meanwhile ruled out this
possibility~\cite{BenDayan:2010yzadd}.

The simplest implementation of a superconformal model which can
accommodate the non-minimal coupling term~\mbox{\(\chi\,\hat{H}_u \cdot
\hat{H}_d\)} is the well-known $\mathbb{Z}_3$-invariant~\nmssm{}
augmented by an additional $\mu$~term, which we call $\mu$-extended
\nmssm{}~(\munmssm{}) in the following. We neglect all additional
$\mathbb{Z}_3$-violating parameters in the superpotential at the tree
level (see the discussion below). These terms are not relevant for the
physics of inflation: the function~$X$ could potentially also contain
an~\(\hat S^2\) term, since it has the same structure as~\mbox{$\hat
H_u\cdot \hat H_d$} and is allowed by gauge symmetries. However,
inflation driven by this term does not lead to the desired properties
as pointed out in \citere{Einhorn:2009bh}. The other term, which is
not present in the~\nmssm, is a singlet tadpole proportional to~\(\hat
S\) that is not quadratic or bilinear in the chiral superfields and
thus would need a dimensionful coupling to supergravity instead of the
dimensionless~\(\chi\).

In this work, we are going to study the low-energy electroweak
phenomenology of the model outlined in
\citeres{Ferrara:2010yw,Ferrara:2010in} and \citere{Lee:2010hj}, where
previously the focus was put on the description of inflation and the
superconformal embedding of the~\nmssm{} into supergravity. We have
generated a model file for \FA~\cite{Kublbeck:1990xc, Hahn:2000kx},
where \code{SARAH}~\cite{Staub:2009bi, Staub:2010jh, Staub:2012pb,
  Staub:2013tta} has been used to generate the tree-level couplings of
the~\munmssm{}, and we have implemented the one-loop counterterms. The
loop calculations have been carried out with the help of
\FC~\cite{Hahn:1998yk} and \LT~\cite{Hahn:1998yk}. In order to predict
the Higgs-boson masses, we have performed a one-loop renormalization
of the Higgs sector of the~\munmssm{} which is compatible with the
renormalization schemes that have been employed
in~\citeres{Fritzsche:2013fta,Domingo:2017rhb} for the cases of
the~\mssm and~\nmssm, respectively. This allowed us to add the leading
MSSM-like two-loop corrections which are implemented in
\FH~\cite{Heinemeyer:1998np, Heinemeyer:1998yj, Degrassi:2002fi,
  Frank:2006yh, Hahn:2010te, Bahl:2016brp, Bahl:2017aev, Bahl:2018FH}
in order to achieve a state-of-the-art prediction for the Higgs masses
and mixing. The parameter space is checked for compatibility with the
experimental searches for additional Higgs bosons using
\HBv{5.1.0beta}~\cite{Bechtle:2008jh, Bechtle:2011sb, Bechtle:2013gu,
  Bechtle:2013wla, Bechtle:2015pma} and with the experimental
observation of the \sm-like Higgs boson via
\HSv{2.1.0beta}~\cite{Bechtle:2014ewa}. In addition, we check the
electroweak vacuum for its stability under quantum tunneling to a
non-standard global minimum and for tachyonic Higgs states in the
tree-level spectrum. Finally, we investigate some typical scenarios
and study their collider phenomenology at
the~\mbox{Large~Hadron~Collider}~(\lhc) and a future electron-positron
collider. For this purpose in some analyses we
use \code{SusHi}~\cite{Harlander:2002wh,Harlander:2003ai}
for the calculation of neutral Higgs-boson production cross-sections.
We emphasize the possibility of light~\cp-even singlets in the
spectrum with masses below~$100\,\GeV$ that could be of interest in
view of slight excesses observed in the existing data of
the~\mbox{Large~Electron--Positron~collider}~(\lep)~\cite{Schael:2006cr}
and the~\mbox{Compact~Muon~Solenoid}~(\cms)~\cite{CMS:2017yta} which
are compatible with bounds
from~\mbox{A~Toroidal~LHC~ApparatuS}~(\atlas)~\cite{ATLAS-CONF-2018-025}.
For one scenario that differs substantially from the usual~\nmssm, we
exemplarily discuss the total decay widths and branching ratios of the
three lightest Higgs bosons and their dependence on the additional
parameters of the~\munmssm.

The paper is organized as follows: we start with a description of our
model and the theoretical framework in \sct{sec:framework} by
discussing analytically the phenomenological differences of the Higgs
potential in the~\munmssm{} compared to
the~$\mathbb{Z}_3$-invariant~\nmssm{}. We study vacuum stability and
the incorporation of higher-order corrections for the Higgs boson
masses. Then, we derive the trilinear self-couplings of the Higgs
bosons and comment on the remaining sectors of the model which are
affected by the additional~$\mu$~term. In \sct{sec:analysis}, we focus
on the parameter space of interest and investigate the Higgs-boson
masses as well as the stability of the electroweak vacuum numerically
and also show the neutralino spectrum. Furthermore, we study the
effect of the additional~$\mu$~parameter on Higgs-boson production and
decays. Lastly, we conclude in \sct{sec:conclusions}. In the Appendix
we present the beta functions for the superpotential and some
soft-breaking parameters of the
general~\nmssm{}~(GNMSSM)~\cite{Ellwanger:2009dp, Ross:2011xv,
  Ross:2012nr} including all~\(\mathbb{Z}_3\)-breaking~terms.

\tocsection[\label{sec:framework}]{Theoretical framework}

In this section we introduce the model under consideration, the
\munmssm{}, which differs by an additional~$\mu$~term from the
scale-invariant~\nmssm{}. We derive the Higgs potential and
investigate vacuum stability and the prediction for the Higgs-boson
masses of the model. Furthermore, we discuss the trilinear
self-couplings of the Higgs bosons and comment on the
electroweakinos---\IE~charginos and neutralinos---as well as on the
sfermion sector. We constrain our analytical investigations in this
section mostly to tree-level relations.  Higher-order contributions,
\EG~for the Higgs-boson masses, are explained generically and are
evaluated numerically in the subsequent phenomenological section.

\tocsubsection[\label{sec:model}]{Model description}

For the Higgs sector of the~\nmssm{} the superpotential is of the
form\footnote{Compared to \citeres{Ferrara:2010yw, Ferrara:2010in}, we
  flip the sign of~\(\lambda\) to follow the conventions of
  the~\nmssm{} literature---see \EG~\citere{Ellwanger:2009dp}---and
  thus have~\(\lambda > 0\). As shown in \citere{Ferrara:2010yw}, the
  product of~$\kappa$ and~$\lambda$ needs to be positive for that
  convention.}
\begin{align}
  \mathcal{W}_{\text{Higgs}} &=
  \lambda\,\hat{S}\,\hat{H}_u\cdot \hat{H}_d
  + \tfrac{1}{3}\,\kappa\,\hat{S}^3\,.
\end{align}
where~$\hat{H}_u$ and~$\hat{H}_d$ are the well-known
$SU(2)_{\text{L}}$~doublets of the~\mssm{}, and~$\hat{S}$ is the
additional
$SU(2)_{\text{L}}$~singlet. The~$SU(2)_{\text{L}}$-invariant
product~\mbox{$\hat{H}_u\cdot\hat{H}_d$} is defined
through~\mbox{$\hat{H}_u\cdot\hat{H}_d=\sum_{a,b}\epsilon_{ab}\,\hat{H}_d^a\,\hat{H}_u^b$}
with~\mbox{$\epsilon_{21}=1$}, \mbox{$\epsilon_{12}=-1$}
and~\mbox{$\epsilon_{aa}=0$} with~\mbox{\(a,b \in \{1,2\}\)}. As
outlined in \citere{Ferrara:2010in}, a K\"ahler transformation
starting from Jordan-frame supergravity introduces a correction in the
superpotential, which is of the form
\begin{align}\label{eq:inflationmu}
 \mathcal{W}_{\text{Higgs}} &\rightarrow
 \mathcal{W}_{\text{Higgs}}
 + \tfrac{3}{2}\,m_{3/2}\,\chi\, \hat{H}_u\cdot \hat{H}_d\,.
\end{align}
The parameter~$m_{3/2}$ denotes the gravitino mass, and~$\chi$ is the
coupling of \eqn{eq:Xfunction}.  The scalar Higgs fields are denoted
by~$H_u$, $H_d$ and~$S$ in the following. During electroweak symmetry
breaking, they receive the vacuum expectation values~(vevs)~$v_u$,
$v_d$ and~$v_s$, respectively. Expanding around the vevs, we decompose
the fields as follows:
\begin{subequations}\label{eq:higgsfields}
\begin{align}
H_u &\equiv \begin{pmatrix} h_u^+ \\ h_u \end{pmatrix}
= \begin{pmatrix} \eta_u^+ \\
  v_u + \tfrac{1}{\sqrt{2}}\left(\sigma_u + i\,\phi_u\right)
\end{pmatrix}, \qquad
H_d \equiv \begin{pmatrix} h_d \\ h_d^- \end{pmatrix}
= \begin{pmatrix}
  v_d + \tfrac{1}{\sqrt{2}}\left(\sigma_d + i\,\phi_d\right)\\
  \eta_d^- \end{pmatrix},\\
S &\equiv v_s + \tfrac{1}{\sqrt{2}}\left(\sigma_s + i\,\phi_s\right) \,.
\end{align}
\end{subequations}
The additional bilinear contribution to the superpotential in
\eqn{eq:inflationmu} generates a term which is analogous to the
$\mu$~term of the~\mssm{}, but with
\begin{align}
  \label{eq:muchi}
  \mu&=\tfrac{3}{2}\,m_{3/2}\,\chi\,.
\end{align}
When the singlet~$S$ acquires its vev, an
effective~\mbox{$\mue=\lambda\,v_s$} is dynamically generated. Often,
the sum~\mbox{$\left(\mu+\mue\right)$} is the phenomenologically more
relevant parameter of the model. It takes the form
\begin{align}
  \mu + \mue &= \tfrac{3}{2}\,m_{3/2}\,\chi\, + \lambda\,v_s\,
\label{eq:effmu}
\end{align}
and corresponds to the \mssm{}-like higgsino mass term. In the
following, we consider both quantities~$\mu$ and~$\mue$ as independent
input parameters, where \(\mu\) is linearly dependent on the gravitino
mass \(m_{3/2}\). In order to be a viable dark-matter candidate, the
gravitino mass can range from a few~eV to multiple~TeV, see
\EG~\citere{Steffen:2008qp}. The value of~$\chi$ is \textit{a~priori}
not fixed; for cosmological reasons we adopt
\begin{align}\label{eq:chi}
\chi \simeq 10^5 \; \lambda
\end{align}
according to \citeres{Lee:2010hj, Ferrara:2010in}. The additional
contribution to the superpotential in the~\munmssm{} is thus mainly
steered by the gravitino mass, whereas~\(v_s\) can be traded
for~$\mue$. If we require a~$\mu$~parameter above the electroweak
scale,~\mbox{$\mu\gtrsim 1$\,TeV}, and in addition a sizable
coupling~\mbox{$\lambda\gtrsim 0.1$}, the typical gravitino mass turns
out to be much below the electroweak scale at~\mbox{$m_{3/2}\gtrsim
10$\,MeV}. However, if we allow for very small values
of~\mbox{$\lambda \ll 10^{-2}$} and very large values
of~\mbox{\(\mu \gg 1\,\TeV\)}, the gravitino mass could as well be
above the~TeV~scale. In the latter case, the phenomenology of
the~\munmssm{} is not necessarily similar to the~\mssm{}: the singlets
only decouple for~\mbox{$\lambda\to 0$}
with~\mbox{$\kappa\propto\lambda$} and
therefore~\mbox{$v_s\to\infty$}. If the
constraint~\mbox{$\kappa\propto\lambda$} is dropped, interesting
effects can occur; \EG~we will discuss a scenario with small~$\lambda$
and small~$\mue$ in our numerical studies. In contrast to the~\nmssm,
the higgsino mass can be generated by~\(\mu\) alone and thus even a
vanishing~\(v_s\) is not in conflict with experimental bounds.

In order to avoid the cosmological gravitino
problem~\cite{Moroi:1993mb}, where the light gravitino dark matter
overcloses the universe~\cite{Pagels:1981ke, Weinberg:1982zq}, one has
to control the reheating temperature in order to keep the production
rate of the light gravitinos low~\cite{Ellis:1984eq}. This potential
problem may affect the model under consideration for gravitino masses
in the range from~MeV to~GeV; it disappears for much heavier
gravitinos~(\mbox{\(\gsimord 10\,\TeV\)}). In the latter case the
inflationary~\(\mu\)~term would dominate over the~\nmssm-like~\(\mue\)
and drive the higgsino masses to very high values (unless~$\mue$ is
tuned such that the sum~\mbox{\((\mu+\mue)\)} remains small). For
gravitino masses~\mbox{\(m_{3/2} > 1 \,\GeV\)} it affects
Big~Bang~Nucleosynthesis via photo-deconstruction of light elements,
see \citere{Moroi:1993mb}. As discussed in~\citere{Ferrara:2010in}, in
the~\munmssm{} there is no strict constraint on the reheating
temperature~\(T_R\). We note that a reheating temperature
below~\mbox{$T_R \lesssim 10^8$--$10^9\,\GeV$}, as advocated
in~\citere{Khlopov:1984pf}, avoids the gravitino problem. The rough
estimate of~\mbox{\(m_{3/2} \sim 10\,\MeV\)} even
needs~\mbox{\(T_R \lesssim 10^5\,\GeV\)} in order to not overclose the
universe with thermally produced gravitinos after
inflation~\cite{Bolz:2000fu, Pradler:2006qh, Pradler:2006hh,
Hook:2018sai}. Interestingly, such low reheating temperatures preserve
high-scale global minima after inflation, see \citere{Falk:1996zt},
and disfavor the preparation of the universe in a meta-stable state
after the end of inflation~\cite{Savoy:2007jb}. In any case, the
reheating temperature at the end of inflation is very model dependent
and rather concerns the inflationary physics. A study to estimate the
reheating temperature~\(T_R\) is given
in \citere{Kofman:1994rk}. Therein, a relation is drawn between the
decay width of the inflaton and~$T_R$. Interestingly, if we na\"ively
assume that this width at the end of inflation is equal to
the~\sm-like Higgs width~\mbox{\(\Gamma_h \approx 4 \times
10^{-3}\,\GeV\)}, we can estimate a rather low reheating
temperature~\mbox{\(T_R \sim \sqrt{\Gamma_h M_\text{Pl}} \approx
10^7\,\GeV\)} with the Planck~mass~\mbox{$M_{\text{Pl}}\approx
2.4 \times 10^{18}\,\GeV$}. For our studies below we assume that a
reheating temperature as low as~\mbox{\(T_R\lesssim 10^9\,\GeV\)} can
be achieved even with large couplings.

Since the bilinear~$\mu$~term breaks the~$\mathbb{Z}_3$~symmetry,
additional parameters are allowed compared to the~\nmssm. In the
general~\nmssm~(\gnmssm)---including the bilinear singlet mass
parameter~$\nu$ and the singlet tadpole coefficient~$\xi$---the Higgs
sector of the superpotential is given by
\begin{align}\label{eq:GNMSSM}
  \mathcal{W}_{\text{Higgs}}
  &= \lambda\,\hat{S}\,\hat{H}_u\cdot \hat{H}_d
  + \tfrac{1}{3}\,\kappa\,\hat{S}^3 + \mu\,\hat{H}_u\cdot \hat{H}_d
  + \tfrac{1}{2}\,\nu\,\hat{S}^2 + \xi\,\hat{S}\,.
\end{align}
However, we assume that the non-minimal coupling of the Higgs doublets
to supergravity is the only source of superconformal and
thus~\(\mathbb{Z}_3\)~symmetry breaking---as outlined in
Section~\href{https://arxiv.org/pdf/1008.2942.pdf#section.5}{5}
of \citere{Ferrara:2010in}. In this case, all other superpotential
parameters that are forbidden by~\(\mathbb{Z}_3\)~symmetry remain
exactly zero at all scales: the beta~functions for the parameters of
the superpotential are proportional to the respective parameter itself
and thus they cannot be generated radiatively.

Because the~$\mathbb{Z}_3$~symmetry is broken (which avoids the
typical domain-wall problem of the \nmssm~\cite{Vilenkin:1984ib}),
another symmetry at the high scale is required in order to solve the
tadpole problem~\cite{Ferrara:1982ke, Polchinski:1982an,
Nilles:1982mp, Lahanas:1982bk, Hall:1983iz, AlvarezGaume:1983gj}:
without such a symmetry, Planck-scale corrections could possibly
induce large contributions to the tadpole term~\cite{Abel:1996cr}. The
superconformal embedding of the~\munmssm, where the~\(\mu\) term is
generated from the K\"ahler potential, serves as this symmetry. As
pointed out in \citere{Abel:1996cr}, other possibilities consist of
discrete or continuous non-gauge symmetries,
so-called~\(R\)~symmetries. Imposing
discrete~$\mathbb{Z}_4$~or~$\mathbb{Z}_8$~$R$~symmetries as proposed
in \citeres{Lee:2010gv,Lee:2011dya,Ross:2011xv} provide a viable
solution, since dimensionful linear and bilinear terms are forbidden
as long as the symmetry is not broken.\footnote{There is an interplay
between discrete~\(R\)~symmetries, SUSY~breaking and hence the
gravitino mass in supergravity, which favors
the~\(\mathbb{Z}_4\)~\(R\)~symmetry~\cite{Kumekawa:1994gx}. Note,
however, that our model at hand is fundamentally different from
\citere{Kumekawa:1994gx} as the inflaton is related to the Higgs
fields of the~\nmssm{}.}

Furthermore, each parameter in the superpotential induces a
corresponding soft-breaking term; additional mass terms are allowed:
\begin{align}\label{eq:break}
  \begin{split}
  -\mathcal{L}_{\text{soft}} &= \left[A_\lambda\,\lambda\,S\,H_u\cdot H_d + \tfrac{1}{3}\,A_\kappa\,\kappa\,S^3 + B_\mu\,\mu\,H_u\cdot H_d + \tfrac{1}{2}\,B_\nu\,\nu\,S^2 + C_\xi\,\xi\,S + \text{h.\,c.}\right]\\
  &\quad + m_{H_d}^2\,\lvert H_d\rvert^2 + m_{H_u}^2\,\lvert H_u\rvert^2 + m_s^2\,\lvert S\rvert^2\,.
\end{split}
\end{align}
It should be noted that the beta~functions for soft-breaking
parameters are not only proportional to themselves, but also receive
contributions from the other soft-breaking parameters. Thus, in
contrast to the terms in the superpotential, finite contributions may
emerge even if a soft-breaking parameter is set to zero at the tree
level. The beta~functions for the parameters of the superpotential in
\eqn{eq:GNMSSM} and its corresponding soft-breaking parameters in
\eqn{eq:break} can be found in~\citeres{King:1995vk, Masip:1998jc,
  Ellwanger:2009dp}; however, since we employ different conventions we
list them in~\appref{sec:betaf}.

Contrary to studies in the~\gnmssm{} (see \citeres{Ellwanger:2009dp,
  Ross:2011xv, Ross:2012nr, Kaminska:2013mya}), where
the~\mssm-like~\(\mu\)~term can be easily shifted away and absorbed in
a redefinition of the other parameters---especially the tadpole
contribution---we cannot do so in the
inflation-inspired~\munmssm. First of all, the~$\mu$~term is
introduced via the~\(R\)~symmetry-breaking non-minimal coupling to
supergravity only. The other parameters in the singlet sector are not
supposed to be generated by this breaking. Secondly, by redefining the
parameters, we would introduce a tadpole term and shift the effect
simply there. Note that the authors of \citere{Ross:2011xv} perform
this shift in order to eliminate the linear (\IE~tadpole) term in the
superpotential and keep~\(\mu\), while others
(\EG~\citere{Badziak:2016tzl}) shift the~\(\mu\)~term to zero and keep
the tadpole and bilinear terms for the singlet in the
superpotential. As discussed above, in the~\munmssm{} considered in
this paper due to the superconformal symmetry breaking at the Planck
scale solely the~\(\mathbb{Z}_3\)-breaking~\(\mu\)~term is present.

\tocsubsection[\label{sec:potential}]{Higgs potential}

With the superpotential of \eqn{eq:GNMSSM} and the soft-breaking
Lagrangian of \eqn{eq:break}, we derive the following Higgs potential,
where we stick to real parameters:
\begin{align}\label{eq:higgspotential}
\begin{split}
  V &= \left[m_{H_d}^2 +
    \left(\mu + \lambda\,S\right)^2\right] \lvert H_d\rvert^2
  + \left[m_{H_u}^2 + \left(\mu
      + \lambda\,S\right)^2\right] \lvert H_u\rvert^2
  + \left(m_{S}^2 + B_\nu\,\nu\right) S^2 \\
  &\quad + 2\,C_\xi\,\xi\,S + \tfrac{2}{3}\,\kappa\,A_\kappa\,S^3
    + \left[\xi + \nu\,S + \kappa\,S^2 + \lambda\,H_u\cdot H_d\right]^2
    + 2\left(B_\mu\,\mu + \lambda\,A_\lambda\,S\right) H_u\cdot H_d\\
  &\quad + \tfrac{1}{8}\left(g_1^2 + g_2^2\right)
  \left(\lvert H_d\rvert^2 - \lvert H_u\rvert^2\right)^2
  + \tfrac{1}{2}\,g_2^2\,\lvert H_d^\dagger\,H_u\rvert^2\,.
\end{split}
\end{align}
This potential can be expanded in the components of the Higgs fields
in \eqn{eq:higgsfields}. Defining the vectors in field
space~\mbox{$\mathcal{S}^{\text{T}} =
  \left(\sigma_d,\sigma_u,\sigma_s\right)$},
\mbox{$\mathcal{P}^{\text{T}} = \left(\phi_d,\phi_u,\phi_s\right)$}
and~\mbox{$\mathcal{C}^{\text{T}} = \left(\phi_d^-,\phi_u^-\right) =
  \left(\eta_d^+,\eta_u^+\right)^*$}, it reads
\begin{align}\label{eq:pot_exp}
  \begin{split}
    V &= \text{const} - \mathcal{T}_S^{\text{T}}\,\mathcal{S}
    - \mathcal{T}_P^{\text{T}}\,\mathcal{P}
    + \tfrac{1}{2}\,\mathcal{S}^{\text{T}}\,\Mass_S^2\,\mathcal{S}
    + \tfrac{1}{2}\,\mathcal{P}^{\text{T}}\,\Mass_P^2\,\mathcal{P}
    + \mathcal{C}^{\text{T}}\,\Mass_C^2\,\mathcal{C}^{*}\\
    &\quad +
    \sum\limits_{ijk\,=\,1}^{6}
    \tfrac{1}{\sqrt{2}}\,\lambda_{ijk}^\prime
    \left(\mathcal{S},\mathcal{P}\right)_i
    \left(\mathcal{S},\mathcal{P}\right)_j
    \left(\mathcal{S},\mathcal{P}\right)_k
    +
    \sum\limits_{i\,=\,1}^{6}
    \sum_{jk\,=\,1}^{2}\tfrac{1}{\sqrt{2}}\,\tilde{\lambda}_{ijk}^\prime
    \left(\mathcal{S},\mathcal{P}\right)_i
    \left(\mathcal{C}\right)_j\left(\mathcal{C}^{*}\right)_k
    + \cdots\,,
  \end{split}
\end{align}
where the~\cp-even and~\cp-odd tadpole coefficients~$\mathcal{T}_S$
and~$\mathcal{T}_P$, the~\cp-even, \cp-odd and charged squared mass
matrices~\(\mathcal{M}_S^2\), \(\mathcal{M}_P^2\)
and~\(\mathcal{M}_C^2\) are given below, and the trilinear
couplings~\(\lambda_{ijk}^\prime\)
and~\(\tilde{\lambda}_{ijk}^\prime\) are specified in
Section~\ref{sec:selfcoupling}, though in a basis where the Goldstone
mode corresponds to a mass eigenstate and does not mix with the other
states at lowest order. The ellipses denote quadrilinear terms which
are immaterial for the following.

We substitute the electroweak vevs~$v_u$ and~$v_d$ by their
ratio~\mbox{\(\tan\beta = v_u/v_d\)} and the sum of their
squares~\mbox{\(v^2 \equiv v_u^2 + v_d^2 =(174\,\GeV)^2\)}. The
symbols~$t_\beta$, $c_\beta$ and~$s_\beta$ denote~$\tan\beta$,
$\cos\beta$ and~$\sin\beta$, respectively. Furthermore,~$g_1$
and~$g_2$ are substituted by the~$W$ and $Z$~gauge-boson masses,
\begin{align}\label{eq:gaugemasses}
  m_W^2 &= \tfrac{1}{2}\,g_2^2\,v^2\,, &
  m_Z^2 &= \tfrac{1}{2}\left(g_1^2 + g_2^2\right) v^2\,.
\end{align}
Using the abbreviations
\begin{subequations}\label{eq:abbrev}
\begin{align}
  a_1 &= B_\mu\,\mu + \xi\,\lambda + \mue\left(\nu + \frac{\kappa}{\lambda}\,\mue + A_\lambda\right),\label{eq:abbrev_a1}\\
  a_2 &= 2\,v\,\lambda\left(\mu + \mue\right),\label{eq:abbrev_a2}\\
  a_3 &= v\,\lambda\left(\nu + 2\,\frac{\kappa}{\lambda}\,\mue + A_\lambda\right),\\
  a_4 &= \frac{1}{\mue}\left[v^2\,\lambda^2\,c_\beta\,s_\beta\left(\nu + \frac{\kappa}{\lambda}\,\mue + A_\lambda\right) - v^2\,\lambda^2\,\mu - \xi\,\lambda\left(\nu + C_\xi\right)\right],\\
  a_5 &= 4\left(\frac{\kappa}{\lambda}\right)^2\mue^2 + \frac{\kappa}{\lambda}\left[\mue\left(A_\kappa + 3\,\nu\right) - v^2\,\lambda^2\,c_\beta\,s_\beta\right],\label{eq:abbrev_a5}\\
  a_6 &= v\,\lambda\left(\nu + 2\,\frac{\kappa}{\lambda}\,\mue - A_\lambda\right),\\
  a_7 &= -6\left(\frac{\kappa}{\lambda}\right)^2\mue^2 + 2\,\frac{\kappa}{\lambda}\left(\xi\,\lambda - 4\,\nu^2\right) + B_\nu\,\nu\,,\label{eq:abbrev_a7}
\end{align}
\end{subequations}
we can write the explicit expressions for the tadpole
coefficients~$\mathcal{T}_{S,P}$ as
\begin{align}
  \label{eq:tadpoles}
  \mathcal{T}_S &= \begin{pmatrix}
    \sqrt{2}\,v\left\{
    s_\beta\,a_1 -
    c_\beta\left[m_{H_d}^2 + \left(\mu + \mue\right)^2 +
      v^2\,\lambda^2\,s_\beta^2 + \tfrac{1}{2}\,m_Z^2\,c_{2\beta}\right]
    \right\}\\[1.5ex]
    \sqrt{2}\,v\left\{
    c_\beta\,a_1 -
    s_\beta\left[m_{H_u}^2 + \left(\mu + \mue\right)^2 +
    v^2\,\lambda^2\,c_\beta^2 - \tfrac{1}{2}\,m_Z^2\,c_{2\beta}\right]
    \right\}\\[1.5ex]
    \sqrt{2}\,\frac{\mue}{\lambda}\left[a_4 - m_S^2 - a_5 - a_7 - v^2\,\lambda^2
      - \left(\nu + 2\,\mue\,\frac{\kappa}{\lambda}\right)^2\right]
  \end{pmatrix}, &
  \mathcal{T}_P &= \begin{pmatrix} 0\\ 0\\ 0 \end{pmatrix} \equiv \mathbf{0}\,.
\end{align}
The minimization of the Higgs potential requires all tadpole
coefficients in \eqn{eq:tadpoles} to be equal to zero. With the
conditions~\mbox{$\mathcal{T}_S=\mathbf{0}$} we choose to
eliminate~$m_{H_d}^2$, $m_{H_u}^2$ and~$m_S^2$ according to
\begin{subequations}
\label{eq:softHiggsmasses}
\begin{align}
  m_{H_d}^2 &=
   -\left(\mu + \mue\right)^2 - v^2\,\lambda^2\,s_\beta^2 - \tfrac{1}{2}\,m_Z^2\,c_{2\beta} + a_1\,t_\beta\,,\\
  m_{H_u}^2 &=
   -\left(\mu + \mue\right)^2 - v^2\,\lambda^2\,c_\beta^2 + \tfrac{1}{2}\,m_Z^2\,c_{2\beta} + \frac{a_1}{t_\beta}\,,\\
  m_S^2 &= a_4 - a_5 - a_7 - v^2\,\lambda^2 - \left(\nu + 2\,\frac{\kappa}{\lambda}\,\mue\right)^2\,.
\end{align}
\end{subequations}
Substituting these expressions in the symmetric mass
matrices~$\Mass_{S,P,C}$ we find
\begin{subequations}\label{eq:Higgsmasses}
\begin{align}
  \mathcal{M}_S^2 &=
  \begin{pmatrix}
    m_Z^2\,c_\beta^2 + a_1\,t_\beta & \left(2\,v^2\,\lambda^2 - m_Z^2\right) c_\beta\,s_\beta - a_1 & a_2\,c_\beta - a_3\,s_\beta\\
    \cdot & m_Z^2\,s_\beta^2 + a_1/t_\beta & a_2\,s_\beta - a_3\,c_\beta\\
    \cdot & \cdot & a_4 + a_5
  \end{pmatrix},
  \label{eq:MassS}\\
  \mathcal{M}_P^2 &=
  \begin{pmatrix}
    a_1\,t_\beta & a_1 & -a_6\,s_\beta\\
    \cdot & a_1/t_\beta & -a_6\,c_\beta\\
    \cdot & \cdot & a_4 - 3\,a_5 - 2\,a_7
  \end{pmatrix},
  \label{eq:MassP}\\
  \mathcal{M}_C^2 &=
  \left[\left(m_W^2 - v^2\,\lambda^2\right)\,c_\beta\,s_\beta + a_1\right]
  \begin{pmatrix}
    t_\beta & 1\\
    \cdot & 1/t_\beta
  \end{pmatrix}.
  \label{eq:MassC}
\end{align}
\end{subequations}
Diagonalizing \eqn{eq:MassC} yields zero for the massless charged
Goldstone boson, and the charged Higgs-boson mass~$m_{H^\pm}$ at the
tree level is given by
\begin{align}\label{eq:chargedHi}
m_{H^\pm}^2 &= m_W^2 - v^2\,\lambda^2 + \frac{a_1}{c_\beta\,s_\beta}\,,
\end{align}
which we employ as an input parameter. Inserting \eqn{eq:abbrev_a1} we
can then eliminate~$A_\lambda$ via
\begin{align}\label{eq:Alambda}
A_\lambda &= \frac{c_\beta\,s_\beta}{\mue}\left(m_{H^\pm}^2 - m_W^2 + v^2\,\lambda^2\right) - \frac{1}{\mue}\left(B_\mu\,\mu + \xi\,\lambda\right) - \left(\nu + \frac{\kappa}{\lambda}\,\mue\right).
\end{align}
Substituting~$A_\lambda$ in the abbreviations of \eqn{eq:abbrev}
yields~($a_2$, $a_5$ and $a_7$ are not changed)
\begin{subequations}\label{eq:abbrevprime}
\begin{align}
  a_1^\prime &= c_\beta\,s_\beta\left(m_{H^\pm}^2 - m_W^2 + v^2\,\lambda^2\right),\label{eq:a1prime}\\
  a_3^\prime &= v\,\lambda\left[\frac{\kappa}{\lambda}\,\mue + \frac{1}{\mue}\left(a_1^\prime - B_\mu\,\mu - \xi\,\lambda\right)\right],\label{eq:a3prime}\\
  a_4^\prime &= c_\beta\,s_\beta\left(\frac{v\,\lambda}{\mue}\right)^2\left(a_1^\prime - B_\mu\,\mu - \xi\,\lambda\right) - \frac{1}{\mue}\left[\mu\,v^2\,\lambda^2 + \xi\,\lambda\left(\nu + C_\xi\right)\right],\label{eq:a4prime}\\
  a_6^\prime &= v\,\lambda\left[3\,\frac{\kappa}{\lambda}\,\mue + 2\,\nu - \frac{1}{\mue}\left(a_1^\prime - B_\mu\,\mu - \xi\,\lambda\right)\right] = -a_3^\prime + 2\,v\,\lambda\left(2\,\frac{\kappa}{\lambda}\,\mue + \nu\right)\label{eq:a6prime}.
\end{align}
\end{subequations}
The tree-level masses of the three neutral~\cp-even Higgs
bosons~$m_{h_{1,2,3}}^2$ are determined by diagonalizing
\eqn{eq:MassS}. Analogously, diagonalizing \eqn{eq:MassP} yields the
masses~$m_{a_{1,2}}^2$ of the~\cp-odd Higgs bosons at the tree level;
the third eigenvalue is equal to zero and belongs to the neutral
Goldstone boson.

\begin{description}

\item[Higgs doublets:] The mass-matrix elements of the doublet fields
  in the upper-left~\mbox{$\left(2\times2\right)$}~block matrices
  of \eqns{eq:MassS}--\eqref{eq:MassP} contain the
  abbreviation~$a_1^\prime$. From \eqn{eq:a1prime} it is apparent that
  they are determined by \sm{}~parameters and~$m_{H^\pm}$, $\lambda$
  and~$t_\beta$ like in the~\nmssm. Neglecting the mixing between the
  doublet and singlet sector, the mass of the light~\cp-even doublet
  state has an upper bound
  of~\mbox{$m_Z^2\,c^2_{2\beta}+\lambda^2\,v^2\,s^2_{2\beta}$}. In the
  limit~\mbox{$m_{H^\pm}\gg m_Z$}, the other two doublet fields
  decouple and obtain a mass close to~$m_{H^\pm}$. Smaller values
  of~$m_{H^\pm}$ increase the mixing of both~\cp-even doublet
  fields. Also~$t_\beta$ needs to be close to one for large doublet
  mixing.

\item[Higgs singlets:] The~$\left(3,3\right)$~elements
  of~$\mathcal{M}_S$ and~$\mathcal{M}_P$ in \eqns{eq:MassS}
  and~\eqref{eq:MassP} set the mass scale of the Higgs singlets. They
  contain the terms~$a_4^\prime$ from \eqn{eq:a4prime}, $a_5$ from
  \eqn{eq:abbrev_a5}, and~$a_7$ from
  \eqn{eq:abbrev_a7}. All~$\mathbb{Z}_3$-violating parameters
  besides~$\mu$ and~$B_\mu$ appear in these terms; in our later
  analysis we set these parameters besides~$\mu$ and~$B_\mu$ to zero,
  but for completeness we mention them in the following discussion of
  this section.

The parameter~$A_\kappa$ appears only in the term~$a_5$,
whereas~$B_\nu$ only appears in~$a_7$. Thus it is obvious that the
diagonal mass-matrix elements for the singlet fields---and therefore
their masses---can be controlled by these two quantities, without
changing any other matrix element. If all~$\mathbb{Z}_3$-violating
parameters except~$\mu$ and~$B_\mu$ were set to zero, we would
rediscover the~\nmssm-specific feature that~$A_\kappa$ is bound from
below and above to avoid tachyonic singlet states at the tree level.

The ratio~$\kappa/\lambda$ which appears in both terms,~$a_5$
and~$a_7$, has sizable impact on the mass scale of the
singlets. If~\mbox{$\kappa\ll\lambda$} the~\cp-even singlet entry is
purely controlled by~$a_4^\prime$, which in turn is proportional
to~$1/\mue$; in the same limit, the~\cp-odd singlet entry is
controlled by~$a_4^\prime$ and the remainder of~$a_7$ which
is~$B_\nu\,\nu$. Also note that~$a_4^\prime$ contains a term which is
linear in~$\mu$. In the opposite case~\mbox{$\kappa\gtrsim\lambda$},
the term~$a_5$ is likely to dominate the~$\left(3,3\right)$~matrix
element for the~\cp{}-even singlet due to the suppression
of~$a_4^\prime$ by~$\mue$ if it is of the order of a
few~$100$\,GeV. The term~$a_5$ is proportional
to~$(\kappa/\lambda)^2\,\mue^2$, such that the~\cp{}-even singlet
exhibits a strong dependence on~$\mue$. On the other hand
for~\mbox{\(\mu\gtrsim\mue\)}, the term~\(a_4'\) can balance the
large~\(\kappa\)-enhanced contribution in~$a_5$; thus, possible upper
bounds on~$\kappa$ as derived in \citere{Ellwanger:1996gw} might be
evaded.

For the case of the~\cp-odd singlet, the terms in~$a_5$ and~$a_7$ that
are quadratic in~$\mue$ cancel each other. Then the size of the other
parameters (especially~$A_\kappa$, $\mu$ and~$\mue$) determines which
contribution is dominant. For moderate values of~\mbox{$\kappa\approx
  \lambda \gtrsim 0.1$} together with small~$A_\kappa$ the~\cp{}-odd
singlet develops a dependence on~$\mu/\mue$, as we will discuss
later. Lastly, we note that in the case of~\mbox{$\kappa\gg\lambda$}
and~\mbox{$A_\kappa\neq 0$\,\GeV} the~\cp{}-even and~\cp{}-odd singlet
masses are controlled through~$(\kappa/\lambda)^2\,\mue^2$
and~$(\kappa/\lambda)\,\mue\,A_\kappa$, respectively. Later, this will
allow us to present a rescaling procedure that keeps both singlet
masses constant over a large parameter range.

\item[Doublet--singlet mixing:] The masses of the doublet-like and the
  singlet-like Higgs states can be significantly shifted by mixing
  between both sectors. The relevant matrix elements are the ones in
  the third columns of \eqns{eq:MassS} and~\eqref{eq:MassP}. They
  contain the abbreviations~$a_2$, $a_3^\prime$ and~$a_6^\prime$, see
  \eqns{eq:abbrev_a2}, \eqref{eq:a3prime} and~\eqref{eq:a6prime},
  respectively. The mixing vanishes in the limit~\mbox{$\lambda\to 0$}
  with constant~$\kappa/\lambda$, and it is enhanced for larger values
  of~$\lambda$. For fixed~$\lambda$ it is also strongly enhanced in
  the limit~\mbox{$\mue\to 0$\,\GeV}.

In the~\cp-even sector, two terms contribute to the doublet--singlet
mixing:~$a_2$ which depends on the sum~\mbox{$(\mu+\mue)$},
and~$a_3^\prime$ which does not directly depend on~$\mu$, but only on
the soft-breaking term~$B_\mu\,\mu$. In the case of large~$\mu$
and~$\mue$ of the same sign,~$a_2$~often dominates the mixing with the
lighter doublet, eventually yielding a tachyonic singlet or doublet
Higgs; this behavior can be avoided by choosing a proper value
for~$B_\mu$ (or~$\xi$) to cancel the large effect in~$a_2$
by~$a_3^\prime$. In the case of similar~$\mu$ and~$\mue$ of opposite
signs,~$a_3^\prime$~will always dominate the mixing. Again, the mixing
strength can be adjusted by setting~$B_\mu$ (or~$\xi$).

The doublet--singlet mixing in the~\cp-odd sector contains only one
term~$a_6^\prime$ which is similar to~$a_3^\prime$ with opposite
sign. Furthermore, the~\cp-odd mixing elements can be modified by
non-zero~$\xi$ and~$\nu$. As indicated above, due to the dependences
of~$a_3^\prime$ and~$a_6^\prime$ on~$1/\mue$, a small~\mbox{$\mue\ll
100$\,GeV} yields a strong mixing between singlets and doublets.

\end{description}

We subsequently discuss vacuum structure and vacuum stability bounds
in the~\munmssm{} around the electroweak scale. We do not discuss
tachyonic instabilities during inflation or the stabilization of the
inflationary direction, since they are not of relevance for our study
(see \EG~\citeres{Ferrara:2010in, Lee:2010hj}).

\tocsubsection[\label{sec:vacuum}]{Vacuum structure and vacuum
  stability bounds}

The space of model parameters can be constrained using experimental
exclusion limits and theoretical bounds. Those constraints can be
applied to rule out certain parts of the parameter space. In this
context, constraints from the stability of the electroweak vacuum
appear to be very robust and theoretically well motivated. It has
already been noticed in the early times of supersymmetry that
constraints from the electroweak vacuum stability on the trilinear
soft SUSY-breaking parameters can be important~\cite{Frere:1983ag,
  Claudson:1983et, Kounnas:1983td, Drees:1985ie, Gunion:1987qv,
  Komatsu:1988mt, Casas:1995pd, Baer:1996jn, LeMouel:1997tk}. Recently
they have been rediscussed in light of the Higgs
discovery~\cite{Altmannshofer:2012ks, Carena:2012mw, Blinov:2013fta,
  Chowdhury:2013dka, Camargo-Molina:2013sta}. These constraints are
usually associated with non-vanishing vacuum expectation values of
sfermion fields (\EG~staus or stops) and thus known under the phrase
``charge- and color-breaking minima''. Such minima can invalidate the
electroweak vacuum and therefore lead to unphysical parameter
configurations (see below).

However, the existence of charge- and color-breaking minima is only a
necessary condition for the destabilization of the electroweak
vacuum. Clearly one has to compare the value of the potential at this
new minimum with the desired electroweak one, and only if the
non-standard vacuum is deeper the corresponding scenario is
potentially excluded. In fact, some of the points with a deeper
non-standard vacuum may be valid when accepting meta-stable vacua under
the condition that the transition time from the local electroweak
vacuum to the global true vacuum appears to be longer than the age of
the universe~\cite{Kusenko:1996jn}. However, the possibility of the
existence of meta-stable vacua is of limited practical relevance for
our analysis: typically only parameter points in close neighborhood to
the stable region are affected by such considerations; well-beyond the
boundary region, the false vacua become rather short-lived and thus
are strictly excluded. In addition, there are thermal corrections in
the early universe which give a sizable and positive contribution to
the effective potential as the one-loop corrections are proportional
to~\(m^2(\phi)\,T^2\) for the field-dependent masses~\(m(\phi)\). For
finite temperature, they shift the ground state to the symmetric phase
around~\mbox{\(\phi = 0\,\GeV\)}~\cite{Riotto:1995am,
  Kusenko:1996xt}. We presume, however, that our inflationary scenario
preselects a vacuum at field values different from zero and, thanks to
the relatively low reheating temperatures in our scenario, gets caught
in it, see \citere{Falk:1996zt}. Following the inflationary scenario
of \citere{Ferrara:2010in}, the trajectory in field space lies
at~\mbox{\(\beta = \pi/4\)} with~\mbox{\(h_u^2 = h_d^2 = h^2\)}
and~\mbox{\(s = 0\)\,\GeV}; the presence of the singlet field~\(S\) is
needed for the stabilization of the inflationary trajectory in order
to not fall into the tachyonic direction as pointed out by
\citeres{Lee:2010hj, Ferrara:2010in}. Inflation ends at field
values~\mbox{\(h = \mathcal{O}(0.01)\)} in units of the Planck
mass. For small~\mbox{\(\lambda\sim 10^{-2}\)}, the~\(D\)-flat
trajectory remains stable after inflation ends according to
\citere{Ferrara:2010in}, and will change to~\mbox{\(\beta\neq\pi/4\)}
and~\mbox{$s \neq 0$\,\GeV} when the~SUSY-breaking terms become
important. \nmssm{}-specific effects like the relevance of singlet
Higgs bosons and the additional contribution to the~$125\,\GeV$~Higgs
boson are usually connected to a large value of~$\lambda$. This is not
necessarily the case in the~\munmssm{}, where striking differences
also appear for small values of~$\mue$. Moreover, we will take it as a
working assumption that after inflation ends, even for larger values
of~$\lambda$ the universe will remain in the state with the
inflationary field direction until it settles down in a minimum
closest to this direction. If it is the global minimum of the
zero-temperature potential, reheating may not be sufficient to
overcome the barrier and to select a false (and maybe meta-stable)
vacuum. The thermal history of the universe plays then no role for the
choice of the vacuum, and in this case the universe would remain in
the global minimum. Accordingly, we adopt the prescription to exclude
all points with a global minimum that does not coincide with the
electroweak vacuum. This means that we do not consider meta-stable
electroweak vacua as they are excluded by the selection rule. A
similar discussion and argument has been given in
\citere{Strumia:1996pr}, where a selection of the vacuum with the
largest expectation values was promoted, irrespective whether or not
it is the global minimum of the theory.

We will see that actually in most cases scenarios are excluded because
of a tachyonic Higgs mass. Tachyonic masses are related to the fact
that the electroweak point---around which the potential is
expanded---is not a local minimum in the scalar potential, but rather
resembles a saddle point or even local maximum, and the true vacuum
lies at a deeper point along this tachyonic direction. Thus, the true
vacuum has vevs different from the input values, and the electroweak
breaking condition~\mbox{\(\mathcal{T}_S = \mathbf{0}\)} in
\eqn{eq:tadpoles} does not select a minimum.

We briefly sketch how to get constraints on the relevant model
parameters in the (neutral) Higgs sector of the~\munmssm. Similar
observations for the~\nmssm{} have been intensively discussed in the
literature~\cite{Kanehata:2011ei, Beuria:2016cdk}. Already the
presence of an additional Higgs singlet~(see
\EG~\citeres{Branco:2000dq, Hugonie:2003yu, Wittbrodt:2016fuk})
invalidates the well-known results that no charge-breaking Higgs vevs
exist at lowest order in the~\mssm{}~(see \EG~\citeres{Romao:1986jy,
  Casas:1995pd}) and in two-Higgs-doublet models~(see
\EG~\citeres{Ferreira:2004yd, Barroso:2005sm}). On the other hand, in
the~\nmssm{} the inclusion of such charge-breaking minima has rather
little impact on the overall vacuum stability and gives no further
information, see~\citere{Krauss:2017nlh}. In a similar manner, we
neglect non-vanishing squark vevs (see discussion below) and therefore
we only have to deal with the following~potential:
\begin{align}\label{eq:triHiggs}
\begin{split}
V &= \kappa^2\,s^4
+ \tfrac{1}{8}\left(g_1^2 + g_2^2\right) \left(h_u^2 - h_d^2 \right)^2
+ \left(\lambda^2\,s^2 + 2\,\lambda\,\mu\,s\right) \left(h_u^2 + h_d^2 \right)
 - 2\,\lambda\left(\kappa\,s^2 + A_\lambda\,s\right)h_u\,h_d\\
&\quad + \lambda^2\,h_u^2\,h_d^2
+ \tfrac{2}{3}\,\kappa\,A_\kappa\,s^3
+ \left(m_{H_u}^2 + \mu^2\right) h_u^2
+ \left(m_{H_d}^2 + \mu^2\right) h_d^2 + m_S^2\,s^2 - 2\,B_\mu\,\mu\,h_u\,h_d\,,
\end{split}
\end{align}
where we just presented the real fields as we do not consider
spontaneous~\cp~violation.\footnote{We treat the fields as ``classical
  field values'' in the sense of vacuum-expectation values. To avoid
  confusion with the true and desired electroweak vevs, we always keep
  the fields as commuting variables~\(h_u\), \(h_d\) and~\(s\) and
  interpret them as vacuum-expectation values only at the minima.}
Notice also that we do not consider the shifted theory with all
fields~\mbox{\(\phi \to \phi - v_\phi\)} expanded around the
electroweak point,~\mbox{\(h_u = v_u, h_d = v_d, s =
  \mue/\lambda\)}. In our case for the stability analysis, the
potential vanishes at the origin, and the electroweak minimum is one
of the minima not located at the origin. It is not necessarily the
global minimum. Furthermore, compared to \eqn{eq:higgspotential}, we
neglect all additional~\(\mathbb{Z}_3\)-breaking terms besides the
contributions of~\(\mu\) and~\(B_\mu\,\mu\) of the~\munmssm\ (see the
discussion above).

The ``desired'' electroweak vacuum can be constructed by fulfilling
the minimization conditions at the tree level,~\mbox{\(\mathcal{T}_S =
  \mathbf{0}\)}, with~\(\mathcal{T}_S\) given by
\eqn{eq:tadpoles}. The vevs of the doublet fields are taken as fixed
input parameters, whereas the value of~\(\mue\) is treated as variable
similar to~\(\mu\). These equations can be solved for the
soft-breaking masses~\(m_{H_u}^2\), \(m_{H_d}^2\) and~\(m_S^2\)
according to \eqns{eq:softHiggsmasses}.

The masses of the Higgs sector are determined in such a way that the
desired vacuum with~\mbox{\(\langle h_u \rangle = v_u\)},
\mbox{\(\langle h_d \rangle = v_d\)} and~\mbox{\(\langle s \rangle =
  \mue / \lambda\)} is a viable vacuum of the potential~$V$ in
\eqn{eq:triHiggs}. However, one has to ensure that there is no deeper
minimum of~$V$. This can only be achieved reasonably-well through a
numerical evaluation. For that purpose, we determine the stationary
points of the potential~$V$ and then compare the corresponding values
of~$V$ at these points with the desired minimum given by
\begin{align}\label{eq:desmin}
\begin{split}
V_\text{min}^\text{des} &=
- \frac{1}{8}\left(g_1^2 + g_2^2\right) v^4\,c^2_{2\beta}
- \frac{1}{4}\,\lambda^2\,v^4\,s^2_{2\beta}
- v^2\,\mue^2 \left[ 1 - \frac{\kappa^2}{\lambda^2}\,s_{2\beta} \right]\\
&\quad - \frac{\kappa^2}{\lambda^4}\,\mue^4
- v^2\,\mu\,\mue
- \frac{1}{3} \frac{\kappa\,A_\kappa}{\lambda^3}\,\mue^3
+ \frac{1}{2}\,v^2\,A_\lambda\,\mue\,s_{2\beta}
- B_\mu\,\mu\,v^2\,s_{2\beta}\,.
\end{split}
\end{align}
From the expression in \eqn{eq:desmin}, one can derive a few general
results: (a)~for small values of~\(\lambda\) the desired minimum gets
deeper and---as the singlet contribution decouples from the rest of
the potential---it becomes more difficult for a non-standard vacuum to
appear and to be deeper than the desired minimum;
(b)~the~(\(\mu\))\nmssm{} potential at the desired minimum is usually
deeper than in the case of the~\mssm{}\footnote{Compare
  \eqn{eq:desmin} with the desired minimum of the~\mssm{} in
  \eqn{eq:mssmmin} which is solely determined by the~\(D\) term
  and~\(M_A^2\).} and is mainly driven by~\(\mue\); (c)~the
contribution of~\(A_\lambda\) plays a subdominant role compared
to~\(A_\kappa\) whose impact is strongly influenced by~\(\mue\)
and~\(\lambda\); (d)~parameter points
with~\mbox{\(V_\text{min}^\text{des} > 0\)} have to be excluded
because the trivial minimum at~\mbox{\(\langle h_u \rangle = \langle
  h_d \rangle = \langle s \rangle = 0\)\,\GeV} is obviously deeper.

In our analysis, we focus for clarity on constraints from the
tree-level potential, considering the appearance of global
non-standard minima and, as discussed above, disregarding the
possibility of meta-stable false vacua. Employing higher-order
(\IE~one-loop) corrections does not necessarily give more accurate
predictions of vacuum stability, see \citere{Hollik:201Xxxx}. An
approach to include one-loop effects using a certain numerical
procedure has been implemented in the public code collection of
\code{Vevacious}, see \citere{Camargo-Molina:2013qva}, including a
tunneling calculation also at finite temperature using
\code{CosmoTransitions}~\cite{Wainwright:2011kj}. The tree-level
evaluation is much faster and numerically more stable; moreover, it
has been argued that the one-loop effective potential is problematic
for tunneling rate calculations \cite{Andreassen:2016cvx}.

\paragraph{Constraints on the \nmssm{} parameters:}

There are two main constraints known for the trilinear soft
\susy{}-breaking parameters~\(A_\kappa\) and~\(A_\lambda\). The first
constraint relies on the existence of a non-vanishing singlet vev to
generate~\mbox{\(\mue \neq 0\)\,\GeV}. This can be easily derived from
the Higgs potential with only~\mbox{\(s\neq 0\)\,\GeV} and is given by
the requirement~\cite{Ellwanger:1996gw}
\begin{equation}\label{eq:Akappaconstr}
A_\kappa^2 > 9\,m_S^2\,.
\end{equation}
This lower bound on~\(A_\kappa\) is inappropriate for the~\munmssm, as
there always exists a non-vanishing higgsino mass term
from~\mbox{\(\mu = \tfrac{3}{2}\,m_{3/2}\,\chi\)}. As shown in
\sct{sec:analysis}, this constraint has hardly any impact on our
analyses. We simply keep it for illustrative reasons.

The second constraint, on~\(A_\lambda\), follows from a non-tachyonic
charged Higgs mass, since a tachyonic mass~(\mbox{\(m^2 <
0\,\GeV^2\)\,}) means that the potential has negative curvature at
this stationary point derived by the minimization conditions. Thus,
the true vacuum would have some non-zero vev for a charged Higgs
component. Configurations like this are possible in the~\nmssm,
whereas they do not exist as global or local minima in
the~\mssm~\cite{Casas:1995pd}. From the (tree-level) charged Higgs
mass in \eqn{eq:chargedHi}, we get an indirect bound
on~\(A_\lambda\). Taking~\(m_{H^\pm}\) as input value, we can
eliminate~\(A_\lambda\) as free parameter, see
\eqn{eq:Alambda}. Hence, we can ensure that~\(m_{H^\pm}^2\) is always
positive. Still, it is worth noticing that by this
procedure~\(A_\lambda\) gets strongly enhanced for small~\(\mue\)
(compared to~\(m_{H^\pm}\)) and thus drives tachyonic neutral Higgs
bosons.

\paragraph{Charge and color breaking:}

There exist quite strong constraints in the~\mssm{} from the formation
of non-standard minima which break the electric and color charges,
known as charge- and color-breaking~(CCB) minima. The famous
``\(A\)-parameter bounds'' read traditionally~\cite{Frere:1983ag,
  Gunion:1987qv, Casas:1995pd, Casas:1996de}
\begin{subequations}\label{eq:tradbounds}
\begin{align}
A_t^2 &< 3\, \big( m_{H_u}^2 + \mu^2 + m_{\tilde Q}^2 +
  m_{\tilde t}^2 \big)\,,\label{eq:tradboundsAt} \\
A_b^2 &< 3\, \big( m_{H_d}^2 + \mu^2 + m_{\tilde Q}^2 +
  m_{\tilde b}^2 \big)\,,
\end{align}
\end{subequations}
where~\(m_{\tilde Q}^2\) and~\(m_{\tilde t,\tilde b}^2\) are the soft
\susy{}-breaking masses for the superpartners of the
left-handed~$SU(2)_{\text{L}}$ quark doublet,~\(\tilde Q\), and of the
right-handed quark singlets,~$\tilde t$ and~$\tilde b$. Several
modifications and improvements of Ineqs.~\eqref{eq:tradbounds} are
present in the literature, see \EG~\citeres{Casas:1995pd,
LeMouel:1997tk,Kusenko:1996jn}. These constraints follow from the
``\(D\)-flat'' directions in the scalar potential of the~\mssm,
\IE~\mbox{\(h_u = \tilde t_L = \tilde t_R\)} and~\mbox{\(h_d
= \tilde b_L = \tilde b_R\)}, respectively. Thus the quartic terms
associated with squared gauge couplings vanish. In addition, one has
to be reminded that Ineqs.~\eqref{eq:tradbounds} are only necessary
conditions for the formation of a non-trivial minimum with
non-vanishing squark vevs in that specific direction. In the case of a
violation of Ineqs.~\eqref{eq:tradbounds}, one has to check that the
generated~CCB~vacuum is actually deeper than the electroweak
minimum. In the~\mssm{} the desired minimum takes on a comparably
small numerical value, only depending on~\(c_{2\beta}\) (and
the~\(B_\mu\)~term which can be replaced by the~\cp-odd Higgs
mass~\(M_A\)):
\begin{align} \label{eq:mssmmin}
 V_\text{min}^\text{\mssm} &=
 -\tfrac{1}{8}\left(g_1^2 + g_2^2\right) v^4\,c^2_{2\beta}
 -\tfrac{1}{2}\,M_A^2\,v^2\,s^2_{2\beta} \,.
\end{align}
In principle, the~\(A\)-parameter bounds \eqref{eq:tradbounds} can be
simply transferred to the~\munmssm, where~\(\mu\) has to be replaced
by~\mbox{\((\mu + \mue)\)}, as they can be transferred to the
NMSSM~\cite{Ellwanger:1999bv}. The net effect is roughly the same in
the~\mssm, \nmssm{} and~\munmssm; if~\(A_t\) fulfills
Ineq.~\eqref{eq:tradboundsAt}, no~CCB will appear. Constraints
on~\(\mue\) alone may get weakened, because the desired minimum also
gets deeper for larger~\(\mue\). Moreover, the additional singlet
direction stabilizes the potential with respect to~CCB~minima since
the~\(\mue\)~term originates from a quadrilinear scalar coupling, and
the vacuum with non-vanishing~\(\mue\) or~\(v_s\) is typically deeper
than a~CCB~vacuum. Generically, constraints from the coupling to the
wrong Higgs doublet relating down-type sfermion vevs to the up-type
Higgs and vice versa, see~\citeres{Bobrowski:2014dla, Hollik:2015pra},
are expected to be valid for~\mbox{\((\mu + \mue)\)} and not weakened
if the singlet is fixed at its vev. Similarly, there are bounds
on~\(A_{t,b}\) not related to~\(D\)-flat directions as discussed in
\citere{Hollik:2016dcm}. These can be reasonably-well determined only
numerically. Generically speaking, for the~\munmssm{} the risk of
generating a~CCB~vacuum is reduced because (a)~the dependence of the
desired minimum on~\(\mue\) drives the electroweak vevs to be more
stable, and (b)~not as large values of~\(A_t\) are needed to raise
the~\sm{}-like Higgs mass because of the additional \nmssm-specific
tree-level contribution.

Constraints from~CCB~minima as given in Ineqs.~\eqref{eq:tradbounds},
are less important in comparison to the~\mssm{} for both, the~\nmssm{}
and the~\munmssm, even if large stop corrections are needed to shift
the~\sm-like Higgs mass (as in the case for small~\(\lambda\)). If the
singlet-field direction were neglected and the stop~\(D\)-flat
direction~\mbox{\(\tilde t_R = \tilde t_L = \tilde t\)} defined, one
could directly apply Ineqs.~\eqref{eq:tradbounds} for the~\munmssm,
keeping~\mbox{\(v_s \neq 0\)\,\GeV} and replacing~\mbox{\(\mu \to \mu
  + \mue\)}. However, with the singlet as dynamical degree of freedom,
the stability of the electroweak vacuum is improved as the only
singlet--stop contribution is actually a quadrilinear
term~\(\lambda\,h_d\,s\,{\tilde t}^2\) and the occurrence of a true
vacuum with~\mbox{\(\langle h_{u,d} \rangle \neq v_{u,d}\)},
\mbox{\(\langle s \rangle \neq v_s\)} and~\mbox{\(\langle \tilde t
  \rangle \neq 0\)\,\GeV} is disfavored.

\paragraph{Meta-stability and tunneling rates:}

Lastly, we comment on vacuum-to-vacuum transitions in case of a local
electroweak vacuum. It is in general of interest to see how long such
a meta-stable state could survive compared with the life-time of the
universe. We have outlined some arguments why---in view of the
inflationary history of the universe---we disregard meta-stable
long-lived vacua. We will see in \sct{sec:numerics} that totally
stable points survive in a wide range of the parameter space.

For an estimate of the bounce action of the unstable
configuration~\cite{Coleman:1977py}, we define an effectively
single-field scalar potential linearly interpolating between the
electroweak local minimum and the true vacuum found by the numerical
minimization of the scalar potential at different field values and
apply an exact solution of the quartic potential given by
\citere{Adams:1993zs}. See also \citere{Hollik:2018nhq} for the
application of this method to the \munmssm.

\tocsubsection[\label{sec:masscorrections}]{Higher-order corrections
  to Higgs-boson masses and mixing}

It is well-known that perturbative corrections beyond the tree level
alter the Higgs masses and mixing significantly in supersymmetric
models. For instance, in the~\mssm{} such large corrections are needed
to lift the lightest~\cp{}-even Higgs mass beyond the~$Z$-boson
mass. On the other hand, in the~\nmssm{} and similarly the~\munmssm{}
there are scenarios where an additional tree-level term lowers the
tension between the tree-level SM-like Higgs mass and the measured
value of the~\sm{}-like Higgs boson at~$125$\,GeV. Still, since loop
corrections to the Higgs spectrum have a large impact, in our
phenomenological analysis we take into account contributions of higher
order as described in the following.

The masses of the Higgs bosons are obtained from the complex poles of
the full propagator matrix. The inverse propagator matrix is
a~\mbox{$(6 \times 6)$}~matrix that reads
\begin{align}\label{eq:ellmasslag}
  \mathbf{\hat{\Delta}}^{-1}{\left(k^2\right)} = i \left[k^2\mathbf{1}
  - \begin{pmatrix} \mathcal{M}_S^2 & 0\\ 0
  & \mathcal{M}_P^2 \end{pmatrix}
  + \begin{pmatrix} \mathbf{\hat{\Sigma}}_S{\left(k^2\right)} & 0\\ 0
  & \mathbf{\hat{\Sigma}}_P{\left(k^2\right)} \end{pmatrix} \right].
\end{align}
\noindent
Here~$\mathbf{\hat{\Sigma}}_S$ and~$\mathbf{\hat{\Sigma}}_P$ denote
the matrices of the renormalized self-energy corrections to the
neutral~\cp-even and~\cp-odd Higgs fields. In the~\cp-conserving limit
there are no transition elements between~\cp-even and~\cp-odd degrees
of freedom, which is why \eqn{eq:ellmasslag} is block diagonal.

In principle, contributions from mixing with the
longitudinal~$Z$~boson have to be considered as well. However, these
contributions as well as those from mixing with the Goldstone mode
enter the mass predictions only at subleading two-loop
level~\cite{Baro:2008bg, Williams:2011bu}. Since these contributions
are numerically small~\cite{Hollik:2002mv} we neglect them in the
following and use a~\mbox{$(5\times 5)$} propagator
matrix. The~\mbox{$(5\times 5)$}~matrices are denoted by the
symbols~$\mathbf{\hat{\Delta}}_{hh}$ for the propagators
and~$\mathbf{\hat{\Sigma}}_{hh}$ for the renormalized self-energies in
the following. The complex poles of the propagator are given by the
values of the squared external momentum~$k^2$ for which the
determinant of~$\mathbf{\hat{\Delta}}_{hh}^{-1}$ vanishes,
\begin{align}
  \det{\big[
    \mathbf{\hat{\Delta}}^{-1}_{hh}{\left(k^2\right)}
    \big]_{k^2\;=\;M_{h_i}^2\;+\;i\,\Gamma_{h_i}\,M_{h_i}}} &\overset{!}{=} 0\,.
\end{align}
The real part,~$M_{h_i}^2$, of each pole yields the loop-corrected
mass of the corresponding Higgs boson~$h_i$.

In this work, a model file for~\FA~\cite{Kublbeck:1990xc, Hahn:2000kx}
of the~\gnmssm{} at the tree level has been generated with the help
of~\code{SARAH}~\cite{Staub:2009bi, Staub:2010jh, Staub:2012pb,
  Staub:2013tta}. In addition, the one-loop counterterms for all
vertices and propagators have been implemented, and a renormalization
scheme which is consistent with
\citeres{Fritzsche:2013fta,Domingo:2017rhb} for the cases of
the~\mssm{} and~\nmssm{} has been set up. All~$\mathbb{Z}_3$-violating
parameters are renormalized in the~\drbar{}~scheme, see
\appref{sec:betaf} for a list of the respective beta~functions. The
numerical input values of all~\drbar-renormalized parameters are
understood to be given at a renormalization scale which equals the
top-quark pole mass. The renormalized self-energies of the Higgs
bosons~$\mathbf{\hat{\Sigma}}_{hh}$ are evaluated with the help
of~\FC~\cite{Hahn:1998yk} and~\LT~\cite{Hahn:1998yk} by taking into
account the full contributions from the~\gnmssm{} at the one-loop
order. For other variations of the~\nmssm{}, similar calculations of
Higgs-mass contributions up to the two-loop order have been performed
in \citeres{Staub:2010ty, Ender:2011qh, Graf:2012hh, Baglio:2013iia,
  Muhlleitner:2014vsa, Goodsell:2014pla, Drechsel:2016jdg,
  Goodsell:2016udb, Biekotter:2017xmf}. A comparison of results from
public codes using different renormalization schemes can be found in
\citeres{Staub:2015aea, Drechsel:2016htw}.

As an approximation, we have added the leading two-loop contributions
in the~\mssm{}
of~$\mathcal{O}{\left(\alpha_t\alpha_s\right)}$~\cite{Heinemeyer:2007aq}
and~$\mathcal{O}{\left(\alpha_{t}^2\right)}$~\cite{Hollik:2014wea,
Hollik:2014bua} at vanishing external momentum to their~\mssm-like
counterparts in the~\munmssm{} (for a discussion of this approximation
in the~\nmssm{} see Ref.~\cite{Drechsel:2016jdg}). They are taken from
their current implementation in~\FH~\cite{Heinemeyer:1998np,
Heinemeyer:1998yj, Degrassi:2002fi, Frank:2006yh, Hahn:2010te,
Bahl:2016brp, Bahl:2017aev, Bahl:2018FH}.\footnote{Additional
contributions from the~\mssm{} at the two-loop order or
beyond---\EG~further fixed-order results~\cite{Passehr:2017ufr,
Borowka:2018anu} or resummation of large logarithms for heavy
sfermions~\cite{Hahn:2013ria, Bahl:2016brp, Bahl:2017aev}---are
available. However, we will confine our discussion in this paper to
the leading two-loop contributions.} We thus have
\begin{align}
  \label{eq:SEapprox}
  \mathbf{\hat{\Sigma}}_{hh}{\left(k^2\right)}
  \approx
  \left.
  \mathbf{\hat{\Sigma}}^{(\text{1L})}_{hh}{\left(k^2\right)}
  \right|^{\text{\gnmssm{}}} +
  \left.
  \mathbf{\hat{\Sigma}}^{(\text{2L})}_{hh}{\left(k^2\right)}
  \right|_{k^2=0}^{\text{\mssm{}, leading}}.
\end{align}
We note that the two-loop contributions
of~$\mathcal{O}{\left(\alpha_b\alpha_s\right)}$ to the~\mssm{}-like
Higgs self-energies are not included in our calculation. However, in
the definition of the bottom-Yukawa coupling we employ a
running~\drbar~bottom mass at the scale~$m_t$~\cite{Williams:2011bu}
which
enters~$\mathbf{\hat{\Sigma}}^{(\text{1L})}_{hh}{\left(k^2\right)}\big|^{\text{\gnmssm{}}}$,
and we take into account large~$t_\beta$-enhanced contributions to the
bottom mass as discussed in~\citeres{Banks:1987iu, Hall:1993gn,
  Hempfling:1993kv, Carena:1994bv, Carena:1999py, Eberl:1999he,
  Williams:2011bu}. We expect that the missing two-loop piece
of~$\mathcal{O}{\left(\alpha_b\alpha_s\right)}$ is numerically
subleading (for a discussion in the~\mssm{}
see~\cite{Dedes:2003km,Heinemeyer:2004xw}).

Higher-order propagator-type corrections are not only needed for
predicting the Higgs-boson masses, but also for the correct
normalization of~$S$-matrix elements involving Higgs bosons as
external particles. The wave-function normalization factors
incorporating the effects of the mixing between the different Higgs
bosons can be written as a non-unitary matrix~$Z_{ij}^{\mbox{\tiny
mix}}$. It is constructed from the Higgs self-energies and their
derivatives with respect to~$k^2$, evaluated at the various physical
poles; for details we refer the reader to~\citeres{Chankowski:1992er,
Heinemeyer:2001iy, Fuchs:2016swt, Fuchs:2017wkq,Domingo:2017rhb}. A
recent application in the framework of the~\nmssm{} can be found in
\citere{Domingo:2018uim}. Here, we follow the setup outlined in
Section~\href{https://arxiv.org/pdf/1706.00437.pdf#subsection.2.6}{$2.6$}
of~\citere{Domingo:2017rhb} and determine the matrix elements
of~$Z_{ij}^{\text{\tiny mix}}$ from the eigenvalue equation
\begin{align}
  \Big[\text{diag}{\left(m_{h_1}^2,m_{h_2}^2,m_{h_3}^2,m_{a_1}^2,m_{a_2}^2\right)}
  -\mathbf{\hat{\Sigma}}_{hh}\big|_{k^2\;=\;M_{h_i}^2\;+\;i\,\Gamma_{h_i}\,M_{h_i}}\Big]_{kl}
  \,Z^{\mbox{\tiny mix}}_{il} =
  \left(M_{h_i}^2+i\,\Gamma_{h_i}\,M_{h_i}\right)Z^{\mbox{\tiny mix}}_{ik}\,.\\[-.2ex]
\intertext{The normalization of each eigenvector is fixed by}
  \Bigg[
    \frac{\mathrm{d}\mathbf{\hat{\Delta}}^{-1}_{hh}}{\mathrm{d}k^2}\bigg|_{k^2\;=\;M_{h_i}^2\;+\;i\,\Gamma_{h_i}\,M_{h_i}}
    \Bigg]_{kl}\,Z^{\mbox{\tiny mix}}_{ik}\,Z^{\mbox{\tiny mix}}_{il}
  = \Bigg[\mathbf{1}
  + \frac{\mathrm{d}\mathbf{\hat{\Sigma}}_{hh}}{\mathrm{d}k^2}\bigg|_{k^2\;=\;M_{h_i}^2\;+\;i\,\Gamma_{h_i}\,M_{h_i}}\Bigg]_{kl}
  \,Z^{\mbox{\tiny mix}}_{ik}\,Z^{\mbox{\tiny mix}}_{il}
  = 1\,.
\end{align}
In our numerical analysis we denote the three~\cp-even mass
eigenstates~$h_i$ as~$h^0$, $H^0$ and~$s^0$, and the two~\cp-odd mass
eigenstates~$a_i$ as~$A^0$ and~$a_s$. These assignments become
ambiguous as soon as loop corrections are included. In our analysis we
use the largest admixture to a loop-corrected mass state in order to
define the assignment. For this purpose we employ the previously
discussed loop-corrected mixing matrix~$Z^{\text{\tiny mix}}_{ij}$. In
this way~\(s^0\) denotes the dominantly singlet-like state. The light
doublet-like state is named~$h^0$ and the heavy doublet-like state
is~$H^0$. The~\cp-odd Higgs bosons are the predominantly singlet-like
state~$a_s$ and the doublet-like state~$A^0$.

\tocsubsection[\label{sec:selfcoupling}]{Trilinear Higgs-boson self-couplings}

In order to discuss possible distinctions between the~\nmssm{} and
the~\munmssm{}, the Higgs-boson self-couplings are particularly
relevant. Experimentally these self-couplings can be probed through
Higgs pair production or through decays of a heavier Higgs boson to
two lighter ones. Through electroweak symmetry breaking there is also
a strong correlation with Higgs-boson decays into Higgs bosons and
gauge bosons, \EG~\mbox{$A^0\to Zh^0$} or~\mbox{$H^0\to Za_s$}. For
both, the Higgs mixing between singlets and doublets is essential. We
take both types of decays into account when checking against
experimental limits from Higgs boson searches, but only exemplify the
parameter dependence for the decays involving only Higgs bosons in our
numerical analysis below.

The Higgs self-couplings are introduced in \eqn{eq:pot_exp}. In order
to simplify their presentation in the neutral sector we
define~$\phi_i$ to be the~$i$-th~component
of~\mbox{$\Phi=\left(\sigma_d,\sigma_u,\sigma_s,A,\phi_s\right)$},
where in the~\cp{}-odd sector the Goldstone boson is in a basis where
it does not mix with the other Higgs bosons at lowest order (see
discussion in Section~\ref{sec:potential}).\footnote{The state~$A$
  differs from the mass eigenstate~$A^0$ that we defined in the
  previous section.} We denote the couplings as~$\lambda_{ijk}$ for
the interactions among three Higgs bosons~$\phi_i\phi_j\phi_k$ in the
basis~$\Phi$. For the couplings among the~\cp{}-even
components---expressed in gauge couplings~(see \eqn{eq:gaugemasses}
for the relation to the gauge-boson masses)---we obtain at the tree
level
\begin{subequations}
\begin{align}
\lambda_{111}&=-\tfrac{3}{2}\left(g_1^2+g_2^2\right)c_{\beta}\,v\,, &
\lambda_{112}&=\tfrac{1}{2}\left(g_1^2+g_2^2-4\,\lambda^2\right)s_{\beta}\,v\,,\\
\lambda_{113}&=-2\,\lambda\left(\mu+\mue\right)\,, &
\lambda_{122}&=\tfrac{1}{2}\left(g_1^2+g_2^2-4\,\lambda^2\right)c_{\beta}\,v\,,\\
\lambda_{123}&=\lambda\left(\nu+A_\lambda + 2\,\frac{\kappa}{\lambda}\,\mue\right), &
\lambda_{133}&= 2\,\lambda\left(\kappa\,s_{\beta}\,v-\lambda\,c_{\beta}\,v\right),\\
\lambda_{222}&=-\tfrac{3}{2}\left(g_1^2+g_2^2\right)s_{\beta}\,v\,, &
\lambda_{223}&=-2\,\lambda\left(\mu+\mue\right),\\
\lambda_{233}&= 2\,\lambda\left(\kappa\,c_{\beta}\,v-\lambda\,s_{\beta}\,v\right), &
\lambda_{333}&=-2\,\kappa\left(A_\kappa + 3\,\nu\right)
              - 12\,\frac{\kappa}{\lambda}\,\mue.\\
\intertext{The couplings of~\cp{}-even components to~\cp{}-odd components
  are given by}
\lambda_{144}&=-\tfrac{1}{2}\left(g_1^2+g_2^2\right)c_{\beta}\,v\,, &
\lambda_{244}&=\tfrac{1}{2}\left(g_1^2+g_2^2-4\,\lambda^2\right)s_{\beta}\,v\,,\\
\lambda_{344}&=-2\,\lambda\left(\mu+\mue\right)\,, &
\lambda_{345}&=-\lambda\left(\nu+A_{\lambda}+2\,\frac{\kappa}{\lambda}\,\mue\right),\\
\lambda_{155}&=\tfrac{1}{2}\left(g_1^2+g_2^2-4\,\lambda^2\right)c_{\beta}\,v\,, &
\lambda_{255}&=-\tfrac{1}{2}\left(g_1^2+g_2^2\right)s_{\beta}\,v,\\
\lambda_{355}&=-2\,\lambda\left(\mu+\mue\right)\,. \label{eq:lambda355}
\end{align}
\end{subequations}
Similarly we can write down the couplings~$\tilde{\lambda}_i$ for the
interaction~$\phi_iH^+H^-$ of the neutral Higgs bosons in the
basis~$\Phi$ to the physical charged Higgs bosons (the Goldstone
bosons are again in a basis where they do not mix) as follows:
\begin{subequations}
\begin{align}
\tilde{\lambda}_1&= \lambda^2\,s_{\beta}\,s_{2\beta}\,v
                   + \tfrac{1}{2}\left(g_1^2+g_2^2\right)c_{\beta}\,c_{2\beta}\,v
                   - g_2^2\,c_{\beta}\,v,\\
\tilde{\lambda}_2&= \lambda^2\,c_{\beta}\,s_{2\beta}\,v
                   - \tfrac{1}{2}\left(g_1^2+g_2^2\right)s_{\beta}\,c_{2\beta}\,v
                   - g_2^2\,s_{\beta}\,v,\\
\tilde{\lambda}_3&=-\lambda\left[2\left(\mu+\mue\right)
                     + \left(\nu + 2\,\frac{\kappa}{\lambda}\,\mue
                       + A_\lambda\right)s_{2\beta}\right].
\end{align}
\end{subequations}
The remaining couplings which are not present above are equal to
zero. Again~$s_x$ and~$c_x$ are defined as~\mbox{$s_x=\sin(x)$}
and~\mbox{$c_x=\cos(x)$}. In most of the cases when~$\mu$ or~$\mue$
appear, the coupling depends on the
sum~\mbox{$\left(\mu+\mue\right)$}. For the interactions of the
neutral Higgs bosons, only a few couplings carry an (additional)
proportionality to~$\mue$ itself, see~\(\lambda_{123}\),
\(\lambda_{345}\) and \(\lambda_{333}\) which all involve the singlet
state. This dependence manifests itself for the former two couplings
in the Higgs-to-Higgs decays~\mbox{$s^0\to h^0\,h^0$}, \mbox{$H^0\to
  s^0\,h^0$} and~\mbox{$A^0\to s^0\,a_s$}. In the charged Higgs
sector, the decay~\mbox{$s^0\to H^+\,H^-$} has a direct dependence
on~$\mue$ at the tree level in addition to~\mbox{$(\mu+\mue)$} for a
dominantly singlet-like state~\(s^0\), as can be seen in~\(\tilde
\lambda_3\). For both cases a very pronounced mixing of the singlet
states with the Higgs doublets, and an individual dependence on~$\mue$
and on the sum~\mbox{$(\mu+\mue)$} can also occur in other
Higgs-to-Higgs decays. We will emphasize later that Higgs mixing is
crucial for the observed dependences on~$\mue$ and~$\mu$. We consider
the decays at the tree level, however, including the external
corrections to Higgs-boson masses and mixing as discussed in
\sct{sec:masscorrections}. Though, we emphasize that higher-order
contributions to Higgs-boson self-couplings and Higgs-boson decays can
be large, see \citeres{Nhung:2013lpa, Baglio:2015noa,
  Muhlleitner:2015dua, Domingo:2018uim} for corresponding calculations
in the~\nmssm{}.

\tocsubsection[\label{sec:ewinos}]{Neutralino and chargino masses}

We write the neutralino and chargino sector in the gauge-eigenstate
bases
\begin{align}
\big(\psi^0\big)^{\text{T}}&=\big(\tilde B^0, \tilde W_3^0, \tilde h_d^0, \tilde h_u^0, \tilde s\big)\,,&
\big(\psi^+\big)^{\text{T}}&=\big(\tilde W^+,\tilde h_u^+\big) &\text{and}&&
\big(\psi^-\big)^{\text{T}}&=\big(\tilde W^-,\tilde h_d^-\big)\,,
\end{align}
which includes the bino~component~$\tilde B^0$, the neutral and
charged wino~components~$\tilde W_3^0$ and~$\tilde W^\pm$, the neutral
and charged higgsino~components~$\tilde h_{u,d}^0$ and~$\tilde
h_{u,d}^\pm$, and the singlino~component~$\tilde s^0$ in the form of
Weyl spinors. Their mass terms in the Lagrangian can be written in the
form
\begin{align}
 -\mathcal{L}_{\chi\text{-masses}}&=\tfrac{1}{2}\big(\psi^0\big)^{\text{T}} \Mass_{\chi}\,\psi^0
 +\tfrac{1}{2}\big[\big(\psi^-\big)^{\text{T}}\Mass_{\chi^\pm}\,\psi^++
 \big(\psi^+\big)^{\text{T}}\Mass_{\chi^\pm}^{\text{T}}\,\psi^-\big]+\text{h.\,c.}\,.
\end{align}
The symmetric mass matrix of the neutralinos and the mass matrix of
the charginos are given by
\begin{subequations}\label{eq:mass_chi}
\begin{align}\label{eq:mass_chi0}
 \Mass_{\chi} &= \begin{pmatrix}
   M_1 & 0 & -m_Z\,s_{\text{w}}\,c_\beta & m_Z\,s_{\text{w}}\,s_\beta & 0\\
   \cdot & M_2 & m_Z\,c_{\text{w}}\,c_\beta & -m_Z\,c_{\text{w}}\,s_\beta & 0\\
   \cdot & \cdot & 0 & -\left(\mu + \mue\right) & -\lambda\,v\,s_\beta\\
   \cdot & \cdot & \cdot & 0 & -\lambda\,v\,c_\beta\\
   \cdot & \cdot & \cdot & \cdot & 2\,\tfrac{\kappa}{\lambda}\,\mue + \nu
       \end{pmatrix},\\
 \Mass_{\chi^\pm} &= \begin{pmatrix}
   M_2 & \sqrt{2}\,m_W\,s_\beta\\
   \sqrt{2}\,m_W\,c_\beta & \mu+\mue
 \end{pmatrix}.
\end{align}
\end{subequations}
The abbreviations~\mbox{$s_{\text{w}}=g_2/\sqrt{g_1^2+g_2^2}$}
and~\mbox{$c_{\text{w}}=g_1/\sqrt{g_1^2+g_2^2}$} denote the sine and
cosine of the weak-mixing angle, respectively.  We see that the mass
scale of the~\mssm-like higgsinos is given by the
sum~\mbox{$(\mu+\mue)$}, and the mass scale of the singlino is
controlled by~\mbox{$(2\,\kappa/\lambda\,\mue+\nu)$}. If only the
electroweakinos were taken into account at the tree level, it is
apparent that the~\munmssm{} would be indistinguishable from
the~\nmssm{}, since any shift in masses and mixing induced
through~$\mu$ could be compensated through shifts in~$\mue$. However,
such shifts will induce differences in the Higgs sector.

Including the singlino elements (with~\mbox{\(\nu=0\)\,\GeV} as
discussed in Section~\ref{sec:model}), an~\nmssm-like neutralino
spectrum can be generated, where~\mbox{$(\mu + \mue)$} serves as
the~\nmssm-like~\(\mue\)~term and~\(\kappa\) is rescaled as
\begin{align}\label{eq:Lieblerrescaling}
  \kappa &\to \tilde{\kappa}=\kappa\,\frac{\mu+\mue}{\mue} \,.
\end{align}
This rescaling on the other hand affects the Higgs spectrum, thus
giving a possible handle to distinguish the~\munmssm{} from
the~\nmssm.

\begin{figure}[t]
  \centering
  \includegraphics[width=.5\textwidth]{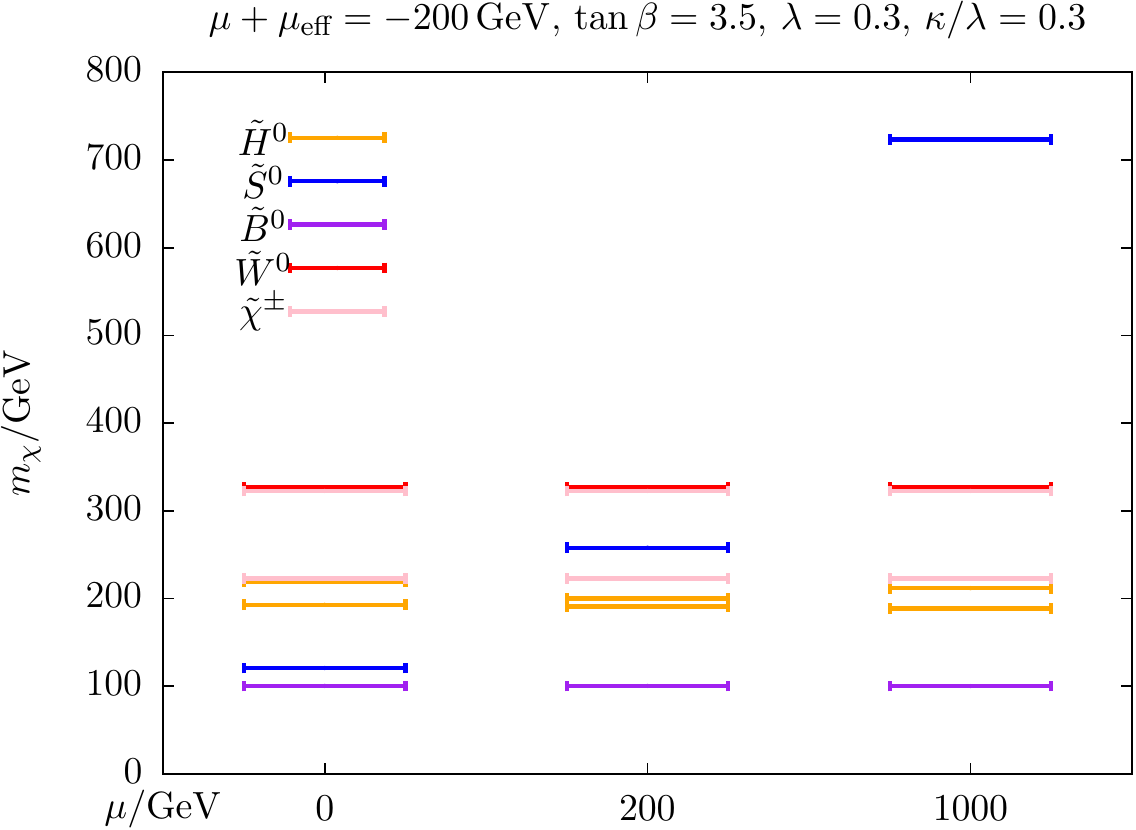}
  \caption{\label{fig:inospec}
  The masses of the neutralinos and charginos are shown for different
  values of~$\mu$. The effective higgsino mass parameter is fixed
  at~\mbox{\(\mu + \mue = -200\,\GeV\)}, and the mass parameters for
  the gauginos are set to~\mbox{\(M_1 = 100\,\GeV\)} and~\mbox{\(M_2 =
  300\,\GeV\)}. The other relevant parameters are given in the legend
  of the figure. The mostly bino- and wino-like
  states~$\tilde{B}^0$~(purple) and~$\tilde{W}^0$~(red) as well as the
  charginos~$\tilde{\chi}^{\pm}$~(rose) have (nearly) constant
  masses. The masses of the two mostly higgsino-like
  states~$\tilde{H}^0$~(orange) and the mostly singlino-like
  state~$\tilde{S}^0$~(blue) vary visibly.}
  \vspace{-1ex}
\end{figure}

For the case where~$\kappa$ and~$\lambda$ are kept fixed, an
interesting behavior can be observed for light higgsinos. For
small~\mbox{$(\mu + \mue)$} huge cancellations may occur between the
two contributions with large~\mbox{\(\mu > 0\)\,\GeV} and~\(\mue\) of
the same size but opposite sign. As a consequence, the singlino state
becomes much heavier compared to the case of the~\nmssm{} (of the
order of~\(\mue\)). Such a scenario is displayed in \fig{fig:inospec}
where the neutralino--chargino spectrum is shown for the
cases~\mbox{$\mu\in\{0,200,1000\}$}\,GeV ($\nu$ is set equal to
zero). The left column with~\mbox{$\mu=0$\,\GeV} corresponds to the
case of the~\nmssm. The masses are obtained by diagonalizing the
tree-level mass matrices in \eqn{eq:mass_chi}. With respect to
the~\(\mathbb{Z}_3\)-invariant~\nmssm{}, the most significant
alteration is visible in the singlino component~(blue): the mass shows
an about-linear increase with~$\mu$ since the sum~\mbox{$(\mu+\mue)$}
is kept fixed. Due to the varying mixing, some influence on the masses
of the other two neutral higgsino states~(orange) can be seen despite
a constant higgsino mass parameter~\mbox{$(\mu+\mue)$}; the impact on
the gaugino states~(red and purple) remains negligible. The chargino
masses~(rose) are not influenced by the different choices.

In a scenario as discussed above, with light higgsinos as well as
large~$\mu$ and~$\mue$ of opposite signs, the lightest neutralino is
typically not the singlino state as the singlino mass is pushed up,
see \fig{fig:inospec}. The lightest supersymmetric particle~(LSP),
however, tends to be the gravitino, which is at risk to overclose the
universe as dark matter candidate. In this case, the inflationary
scenario has to be such that the reheating temperature stays below a
certain value and gravitinos are not overproduced in the early
universe, see our discussion in \sct{sec:model}.

\tocsubsection[\label{sec:sfermions}]{Sfermion masses}

The mass term for each charged sfermion---for which we distinguish the
superpartners of the left- and right-handed components by the
notation~$\tilde f_{\text{L}}$ and~$\tilde f_{\text{R}}$,
respectively---takes the following form in the Lagrangian
\begin{align}
 -\mathcal{L}_{\tilde f\text{-masses}}=\left(\tilde f_{\text{L}}^\dagger, \tilde f_{\text{R}}^\dagger\right)\Mass^2_{\tilde f}\begin{pmatrix}\tilde f_{\text{L}}\\\tilde f_{\text{R}}\end{pmatrix},
\end{align}
where the squared mass matrix reads
\begin{subequations}
\begin{align}
  \Mass^2_{\tilde{f}} &= \begin{pmatrix}
    m_f^2 + m_{\tilde{f}_{\text{L}}}^2 + m_Z^2\,c_{2\beta} \left(T^{(3)}_f - Q_f\,s_{\text{w}}^2\right) & m_f \left[A_f - \theta_f \left(\mu + \mue\right)\right]\\
    m_f \left[A_f - \theta_f \left(\mu + \mue\right)\right] & m_f^2 + m_{\tilde{f}_{\text{R}}}^2 + m_Z^2\,c_{2\beta}\,Q_f\,s_{\text{w}}^2
 \end{pmatrix},\\
  \theta_f &= \left\{\begin{matrix*}[l]
  t_\beta\,, & f\in\{e,\mu,\tau,d,s,b\}\,,\\
  \frac{1}{t_\beta}\,, & f\in\{u,c,t\}\,.
  \end{matrix*}\right.
\end{align}
\end{subequations}
Therein we denote the fermion mass by~$m_f$, the bilinear
soft-breaking parameters by~$m_{\tilde{f}_{\text{L,R}}}$, the
trilinear soft-breaking parameter by~$A_f$, and the electric and weak
charges by~$Q_f$ and~\makebox(19,6.5){$T^{(3)}_f$}.

In this sector we encounter the sum~\mbox{$(\mu+\mue)$} in the
off-diagonal elements of the sfermion mass matrices as the only
difference compared to the~\nmssm{} or~\mssm. If this sum becomes
large,~$A_f/\theta_f$ needs to be adjusted in order to avoid tachyonic
sfermions in particular for the third generation squarks. In that
case, bounds from vacuum stability (see
\EG~Ineqs.~\eqref{eq:tradbounds}) can also constrain the viable size
of~\mbox{$\left(\mu+\mue\right)$}.

\tocsection[\label{sec:analysis}]{Phenomenological analysis}

In this section we investigate various scenarios of the~\munmssm{}
with a particular focus on the $\mu$~parameter. We will point out
differences between the~\munmssm{} and the ordinary
$\mathbb{Z}_3$-preserving~\nmssm, where the latter corresponds to the
limit~\mbox{$\mu = 0$\,\GeV} of the~\munmssm. At first we
qualitatively define the investigated scenarios, before we numerically
analyze them.

\tocsubsection[\label{sec:parameters}]{Viable parameter space compatible
  with theoretical and experimental bounds}

In the previous sections we have analytically discussed the relevant
sectors of the~\munmssm{} with respect to effects of the
inflation-inspired~$\mu$ parameter. Before we provide a
phenomenological analysis---including the higher-order effects
specified in \sct{sec:masscorrections}---we discuss the viability of
various parameter regions. As discussed in \sct{sec:model} we focus on
scenarios with non-zero~$\mu$ and~$B_\mu$, but set all
other~$\mathbb{Z}_3$-violating parameters in the
superpotential~\eqref{eq:GNMSSM} and soft-breaking
Lagrangian~\eqref{eq:break}, \IE~$\xi$, $C_\xi$, $\nu$ and~$B_\nu$,
equal to zero.

The~$\mu$ parameter of the model is positive by construction in the
inflation-inspired model, see Eqs.~\eqref{eq:muchi}
and~\eqref{eq:chi}. Furthermore, we only investigate scenarios
with~\mbox{$\mu\lesssim 2$\,TeV} to stay in the phenomenologically
interesting region for the collider studies. Still, we point out that
also much larger scales are viable from the inflationary point of
view. As discussed in \sct{sec:model},~\mbox{$\mu\simeq\frac{3}{2} \,
  m_{3/2} \, 10^5 \, \lambda$} implies that much larger values
of~$\mu$ are possible depending on~$\lambda$ and~$m_{3/2}$. However,
large values of~$\mu$ can cause tachyonic states as discussed in
\sct{sec:potential}.

We characterize the scenarios in the following parameter regions:
small values of~\mbox{$\mu\simeq 1\,\GeV$},\footnote{We do not
set~$\mu$ exactly to zero for purely technical reasons:
the \mssm{}-like two-loop contributions to the Higgs masses are taken
from~\FH{} where the~$\mu$~parameter of the~\munmssm{} is identified
with the~$\mu$~parameter of the~\mssm{}. In the limit~\mbox{$\mu\to
0$\,\GeV} numerical instabilities appear.} large values
of~\mbox{$\mu\gtrsim 1\,\TeV$} with~\mbox{$\mue\simeq-\mu$}, and
values of~\mbox{$\mu$ $\gsimord 100\,\GeV$} with moderate or
small~\mbox{$\lvert\mue\rvert$ $\lsimord 100\,\GeV$}.

\begin{description}
\item[small \boldmath{$\mu\simeq 1\,\GeV$}:] in the case of
  small~$\mu$ also the soft-breaking term~$B_\mu\,\mu$ becomes
  small. Since in addition we set all other~$\mathbb{Z}_3$-violating
  parameters to zero, we recover the standard~\nmssm{} in this limit
  (see the discussion in \fig{fig:inospec}). Thus, differences between
  the~\nmssm{} and the~\munmssm{} can directly be deduced by comparing
  scenarios with zero and non-zero~$\mu$ parameter.
\item[large \boldmath{$\mu\pmb{\gtrsim} 1\,\TeV$} with
  \boldmath{$\mue\simeq-\mu$}:] as discussed in \sct{sec:ewinos}, the
  higgsino masses depend only on the
  sum~\mbox{$\left(\mu+\mue\right)$} at the tree level. The same
  combination contributes to the sfermion mixing in combination with
  the trilinear soft SUSY-breaking terms. In order to keep these
  quantities small at a large value of~$\mu$, one can assign the same
  value with opposite sign to~$\mue$; note, however, that the
  region~\mbox{$\lvert\mu+\mue\rvert\lesssim100\,\GeV$} is
  experimentally excluded by direct searches for
  charginos~\cite{LEPelectroweakino,Patrignani:2016xqp}. An immediate
  consequence of large, opposite sign~$\mue$ and~$\mu$ is that the
  singlino and the singlet-like Higgs states receive large masses of
  the order of~$\lvert\mue\rvert$ [see the~$(5,5)$~entry in
    Eq.~\eqref{eq:mass_chi0} and the~$(3,3)$~elements in
    Eqs.~\eqref{eq:MassS} and~\eqref{eq:MassP}], which provides a
  potential distinction from the standard~\nmssm{}. Similar to the
  increase of the singlino mass, fixing~\mbox{$\left(\mu+\mue\right)$}
  together with an increase in~$\mu$---and thus an increase in the
  absolute value of~$\mue$---lifts the masses of the singlet states
  also in the Higgs sector. In the neutralino sector these
  contributions can be absorbed by a rescaling of~$\kappa$, see
  \sct{sec:ewinos}; however, in the Higgs sector~$\mue$ also appears
  in other combinations, thus leaving traces which can potentially
  distinguish the~\munmssm{} from the~\nmssm.
\item[\boldmath{$\mu\,\pmb{\gsimord} 100\,\GeV$} with
  \boldmath{$\pmb{\lvert}\mue\pmb{\rvert}\,\pmb{\lsimord}100\,\GeV$}:]
  if we allow for a large~$\mu$~parameter without constraining the
  sum~\mbox{$\left(\mu+\mue\right)$}, the spectra of higgsinos,
  sfermions and Higgs bosons are changed at the same time. A large
  sum~\mbox{$\left(\mu+\mue\right)$} causes very large mixing between
  the singlet and doublet sectors (see discussion in
  \sct{sec:potential}), eventually driving one Higgs state
  tachyonic. In some part of the parameter space this can be avoided
  by tuning~$B_\mu$ accordingly. Another constraint arises from the
  sfermion sector, most notably the sbottoms and staus: a
  large~\mbox{$\left(\mu+\mue\right)$} induces large terms in the
  off-diagonal elements of the sfermion mass matrices (enhanced
  by~$\tan\beta$ for the case of down-type sfermions) which can
  potentially cause tachyons, also depending on the values of the
  trilinear soft-breaking parameters~$A_f$. As discussed in
  \sct{sec:vacuum}, constraints from charge- and color-breaking minima
  induced by too large soft-breaking trilinear parameters (see
  Ineqs.~\eqref{eq:tradbounds} with~$\mu$ promoted
  to~\mbox{$\left(\mu+\mue\right)$}), have a much smaller impact in
  the~\munmssm{} as compared to the~\mssm~\cite{Ellwanger:1999bv}.

  A special case of this scenario is the possibility of having~$\mu$
  at the electroweak scale in combination with an almost
  vanishing~\mbox{$\lvert\mue\rvert\ll\mu$}. This implies
  that~\mbox{$\left(\mu+\mue\right)$} remains at the electroweak
  scale. In contrast to the standard~\nmssm{} this scenario allows the
  occurrence of both,~\mbox{$\kappa\gg\lambda$} and a light singlet
  sector. As discussed in \sct{sec:potential}, the mixing between
  singlets and doublets is in this case dominated by terms
  proportional to~$\mue^{-1}$. We will explicitly discuss such a
  scenario in \sct{sec:decays}.
  
\end{description}

\begin{table}[b!]
\centering
\caption{\label{tab:param}The input parameters which are fixed
  throughout our numerical analysis (interpreted as on-shell
  parameters if not specified otherwise). The gaugino mass parameters
  are denoted as~$M_i$ with \(i=1,2,3\). The trilinear soft-breaking
  terms for the sfermions~$A_{f_g}$ carry the generation
  index~\mbox{$g=1,2,3$}. The charged Higgs mass~$m_{H^\pm}$ is fixed
  to the shown value, if not mentioned otherwise.}
\par\rule{0.94\textwidth}{0.4pt}\\[-2ex]
\setlength{\abovedisplayskip}{0pt} \setlength{\belowdisplayskip}{0pt}
\setlength{\abovedisplayshortskip}{0pt}
\setlength{\belowdisplayshortskip}{0pt}
\begin{align*}
  m_{H^\pm} &=800\,\GeV, & m_t &= 173.2\,\GeV, & \alpha_s (m_Z) &= 0.118,\\
  G_F&=1.16637\cdot 10^{-5}\,\GeV^{-2}, &  m_Z &= 91.1876\,\GeV, & m_W &= 80.385\,\GeV,\\
  M_3 &= 2.5\,\TeV, & M_2 &= 0.5\,\TeV, & M_1 &= \frac{5}{3}\frac{g_1^2}{g_2^2}M_2,\\
  m_{\tilde{f}_{\text{L}}} &= 2\,\TeV, & m_{\tilde{f}_{\text{R}}} &= 2\,\TeV, & A_{f_3} &= 4\,\TeV, \quad A_{f_{1,2}} = 0\,\TeV.
\end{align*}
\par\rule{0.94\textwidth}{0.4pt}
\end{table}

There are more parameters that are relevant for the following
phenomenological studies. We keep those fixed which behave similarly
as in the~\mssm{} and~\nmssm. The choice of our constant input values
is given in Tab.~\ref{tab:param}. Furthermore, we specify the values
of~$t_\beta$, $\kappa$, $\lambda$, and~\(A_\kappa\) directly at the
respective places. Besides the analyses where we explicitly study the
dependence on~\(B_\mu\), we use~\mbox{\(B_\mu = 0\,\GeV\)} as default
value.

As our analysis is focused on the impact of the inflation model, we
are not going to discuss the influence of the sfermion parameters. If
not mentioned otherwise, we use~\mbox{\(m_{\tilde f} \equiv m_{\tilde
    f_L} = m_{\tilde f_R}\)} and~\mbox{\(A_{f_3} / m_{\tilde f} =
  2\)}, which maximizes the prediction for the~\sm{}-like Higgs-boson
mass at~\mbox{$\mu+\mue=0\,\GeV$}. The gluino mass parameter~$M_3$ is
set well above the squark masses of the third generation which is in
accordance with the existing~\lhc~bounds. For completeness, we also
give the parameters of the~\sm{} which are most relevant for our
numerical study in Tab.~\ref{tab:param}.

The gaugino-mass parameters~$M_1$ and~$M_2$ do not play a big role in
the following analysis, but are necessary input parameters for the
mass matrices of the charginos and neutralinos in
Eqs.~\eqref{eq:mass_chi}. We set~\mbox{$M_2 = 500\,\GeV$} and
fix~\(M_1\) via the usual~GUT~relation, see Tab.~\ref{tab:param}. Our
phenomenological analysis is most sensitive to the neutralino and
chargino spectrum if a Higgs boson can decay into them. This is in
particular the case if the particle spectrum contains light higgsinos,
whose masses are controlled
through~\mbox{$\left(\mu+\mue\right)$}. For a scenario with light
higgsinos and a light singlino we will later also discuss the
electroweakino phenomenology at a linear collider, see
\sct{sec:production}.

As we use~\mbox{$\mu \simeq \frac{3}{2}\,m_{3/2}\,10^5\,\lambda$} and
focus on~\mbox{$\mu \lesssim 2\,\TeV$}, we are considering scenarios
where the gravitino typically is the~LSP. We do not specify the
mediator mechanism of SUSY~breaking; however, we assume that such a
light gravitino is always possible. Although the gravitino is the
Dark~Matter candidate, traditional collider searches for a
neutralino~LSP do apply in our case: for instance, if
the~next-to~LSP~(NLSP) is gaugino-like, it can decay into a photon and
the gravitino, where the~NLSP lifetime is typically so large that it
can escape the detector~\cite{CahillRowley:2012cb}. We roughly
estimate the~NLSP phenomenology via the approximate partial decay
width of the neutralino~NLSP into a photon or~$Z$~boson and
gravitino~\(\psi_{3/2}\) according to \citeres{Feng:2004zu,
  Covi:2009bk, Hasenkamp:2009zz}
\begin{gather}\label{eq:NLSPgam}
  \Gamma_{\tilde \chi^0_1 \to \gamma \psi_{3/2}} \simeq \frac{\left|N_{11} \,
      c_{\text{w}} + N_{12} \, s_\text{w}\right|^2}{48\,\pi\,M_{\text{Pl}}^2}
  \frac{m_{\tilde \chi_1^0}^5}{m_{3/2}^2}\,,
\quad
  \Gamma_{\tilde \chi^0_1 \to Z \psi_{3/2}} \simeq \frac{\left| -N_{11} \,
        s_{\text{w}} + N_{12} \, c_\text{w}\right|^2}{48\,\pi\,M_{\text{Pl}}^2}
  \frac{m_{\tilde \chi_1^0}^5}{m_{3/2}^2}\left(1-\frac{m_Z^2}{m_{\tilde\chi_1^0}^2}\right)^4,
\end{gather}
where we expanded in a small gravitino mass~$m_{3/2}$ and
use~$s_{\text{w}}$ and~$c_{\text{w}}$ for the sine and cosine of the
weak mixing angle, respectively. The neutralino mixing matrix
elements~$N_{ij}$ follow from the diagonalization of
\eqn{eq:mass_chi0}. As an example for the decay of the~NLSP
with~\mbox{\(m_{\tilde\chi_1^0} \simeq 100 \, \GeV\)}
and~\mbox{\(m_{3/2} \simeq 10\,\MeV\)}, we find a lifetime
of~\mbox{\(\tau \equiv 1/\Gamma = \mathcal{O}(1 \,
  \mathrm{s})\)}. Thus, the~NLSP decays outside of the detector and is
counted as missing energy. Nevertheless, such decays might be of
certain interest with respect to future experimental searches for
long-lived particles like
the~MATHUSLA~experiment~\cite{Curtin:2018mvb}. Note that for a
higgsino-like~NLSP the decay into a~$Z$~boson and the gravitino is
obtained by replacing the mixing factor in \eqn{eq:NLSPgam}
by~\mbox{$\lvert{-}N_{13}c_\beta+N_{14}s_\beta\rvert^2$}. If
kinematically open, also the decay into a (singlet-like)~\cp-even
or~\cp-odd Higgs boson and the gravitino can occur (see
\citere{Covi:2009bk}), but this decay mode does not change the
qualitative features described above.

We have chosen~$m_{H^\pm}$ as an input parameter and
adjust~$A_\lambda$ according to Eq.~\eqref{eq:Alambda}. If not denoted
otherwise, we set~\mbox{$m_{H^\pm} = 800\,\GeV$}. We
use~\HBv{5.1.0beta}~\cite{Bechtle:2008jh, Bechtle:2011sb,
  Bechtle:2013gu, Bechtle:2013wla, Bechtle:2015pma} in order to
implement the constraints on the parameter space of each of our
scenarios resulting from the search limits for additional Higgs
bosons. In this context, the exclusion limits from~\mbox{$H,A \to
  \tau\tau$} decays are particularly important. For relatively low
values of~$\tan\beta$ the choice of~\mbox{$m_{H^\pm}=800\,\GeV$} is
well compatible with these bounds. The code~\HB{} determines for each
parameter point the most sensitive channel and evaluates whether the
parameter point is excluded at the~$95\%$~confidence level~(C.L.). We
use those exclusion bounds as a hard cut in the parameter spaces of
our analyses.

We also indicate the regions of the parameter space which provide a
Higgs boson that is compatible with the observed state at~$125$\,GeV.
These regions are obtained with the help
of~\HSv{2.1.0beta}~\cite{Bechtle:2014ewa}. The code~\HS{} evaluates a
total~$\chi^2$~value, obtained as a sum of the~$\chi^2$~values for
each of the~$85$~implemented observables. Four~more observables are
added, which test the compatibility of the predicted Higgs-boson mass
with the observed value of~$125$\,GeV. This latter test includes a
theoretical uncertainty on the predicted Higgs-boson mass of
about~$3$\,GeV, such that a certain deviation from the~four~measured
mass values (from the two channels with either a~$\gamma\gamma$ or
a~$ZZ^{(*)}$ final state from both experiments~\atlas{} and~\cms) is
acceptable. Thus, in total~\HS{} tests~$89$~observables.

Since all our two-dimensional figures include a region with
a~\sm{}-like Higgs boson,\footnote{The minimal~$\chi^2$~value obtained
  in our numerical analysis is~\mbox{$\chi_m^2=74.6$}. All
  subsequently discussed benchmark planes include a parameter region
  with~\mbox{$\chi_m^2<80$}. Further details are provided below.} we
classify the compatibility with the observed state as follows: we
determine the minimal value of~$\chi^2$, denoted by~$\chi_m^2$, in the
two-dimensional plane and then calculate the deviation~\mbox{$\Delta
  \chi^2=\chi^2-\chi_m^2$} from the minimal value in each parameter
point. We allow for a maximal deviation of~\mbox{$\Delta \chi^2 <
  5.99$}, which corresponds to the~$95\%$~C.L.~region in the
Gaussian~limit. All parameter points that fall in this
region~\mbox{$\Delta \chi^2<5.99$} are considered to successfully
describe the observed~\sm{}-like Higgs boson.

Lastly, we note that~\HB{} and~\HS{} are operated through an
effective-coupling input. We will comment on the results of the two
codes where appropriate.

For our implementation of the constraints from the electroweak vacuum
stability we refer to \sct{sec:vacuum}. For informative reasons, we
distinguish long-lived vacua from short-lived ones in the numerical
analysis. We do not explicitly enforce a perturbativity bound
on~$\kappa$ and~$\lambda$, but discuss this issue below.

\tocsubsection[\label{sec:higgspheno}]{Higgs-boson and neutralino mass
  spectra}

In this section, we point out the differences of the Higgs-boson and
neutralino mass spectra in the~\munmssm{} with respect to
the~\nmssm{}. Similar to the case of the~\mssm{}, the charged and
the~\cp-even heavy doublet as well as the~\mssm-like~\cp-odd Higgs
bosons are (for sufficiently large~\mbox{$m_{H^\pm}\gg M_Z$})
quasi-degenerate.

\begin{figure}[t!]
  \centering
  \includegraphics[width=.66\textwidth]{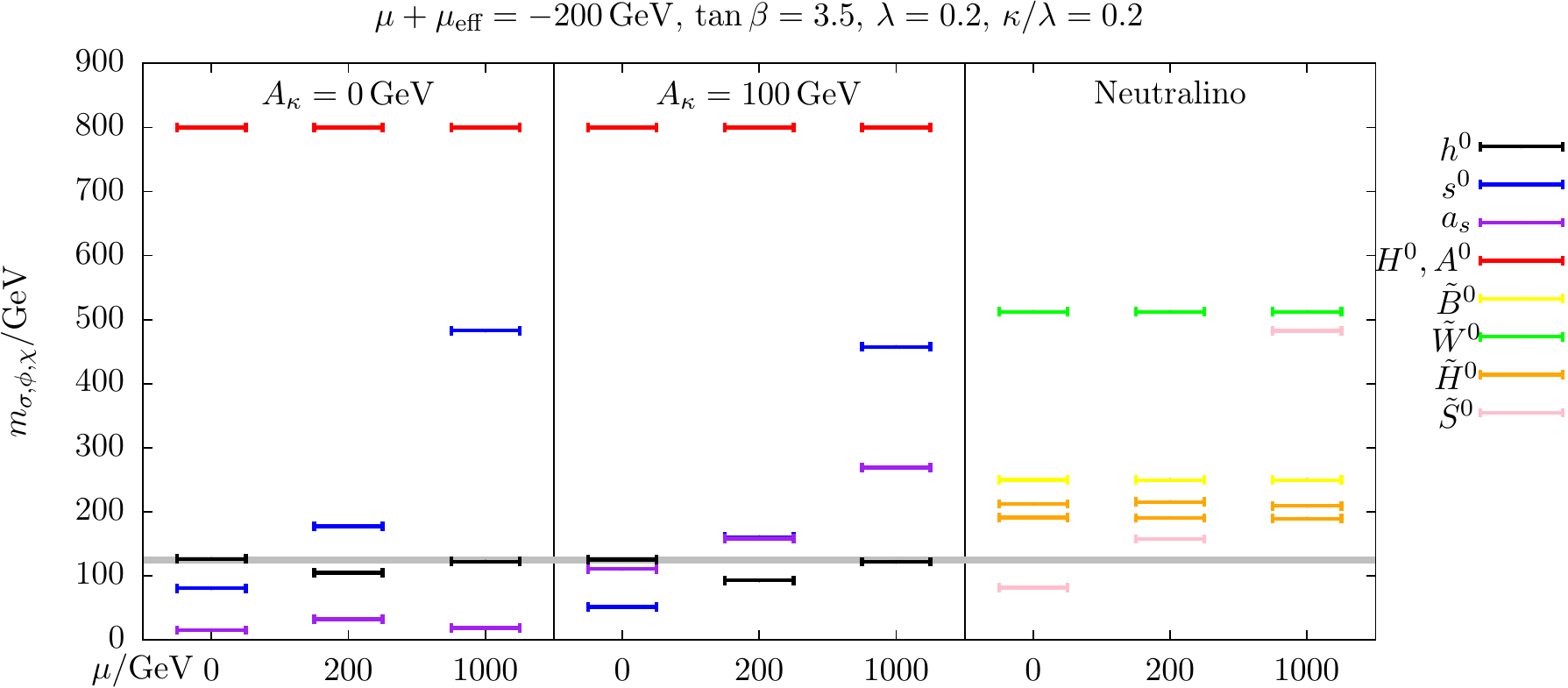} 

  \caption{\label{fig:Higgsspec} The loop-corrected Higgs-boson
    spectrum and the tree-level neutralino spectrum are shown in
    the~\munmssm{} for scenarios
    with~\mbox{$\mu\in\{0,200,1000\}\,\GeV$}
    and~\mbox{$\mu+\mue=-200\,\GeV$} fixed. The parameters are chosen
    such that the state~$h^0$~(black) that is mostly~\sm-like has a
    mass around~\(125\,\GeV\); the gray band shows
    a~$3\,\GeV$~interval around the experimentally measured Higgs
    mass. Furthermore, the masses of the~\cp-even singlet-like
    state~$s^0$~(blue), the~\cp-odd singlet-like state~$a_s$~(purple),
    and the heavy~\cp-even Higgs doublet and~\mssm-like~\cp-odd
    components~$H^0,A^0$ with values close to the
    input~\mbox{$m_{H^\pm}\sim 800\,\GeV$}~(red) are shown, where the
    assignments are made according to the loop-corrected mixing
    matrix~$Z^{\text{\tiny mix}}_{ij}$ for the Higgs sector, see
    \sct{sec:masscorrections}. For the neutralino sector on the right,
    yellow lines show the dominantly bino-like state~$\tilde{B}^0$,
    and green lines the wino-like state~$\tilde{W}^0$. The
    singlino~$\tilde{S}^0$ is shown in rose and the two (doublet)
    higgsinos~$\tilde{H}^0$ appear in orange. The assignments are
    determined by the tree-level mixing matrix. The parameter values
    are given in the plot and in Tab.~\ref{tab:param}.}

  \vspace{2ex} \hrule \vspace{2ex} \capstart

  \includegraphics[width=.5\textwidth]{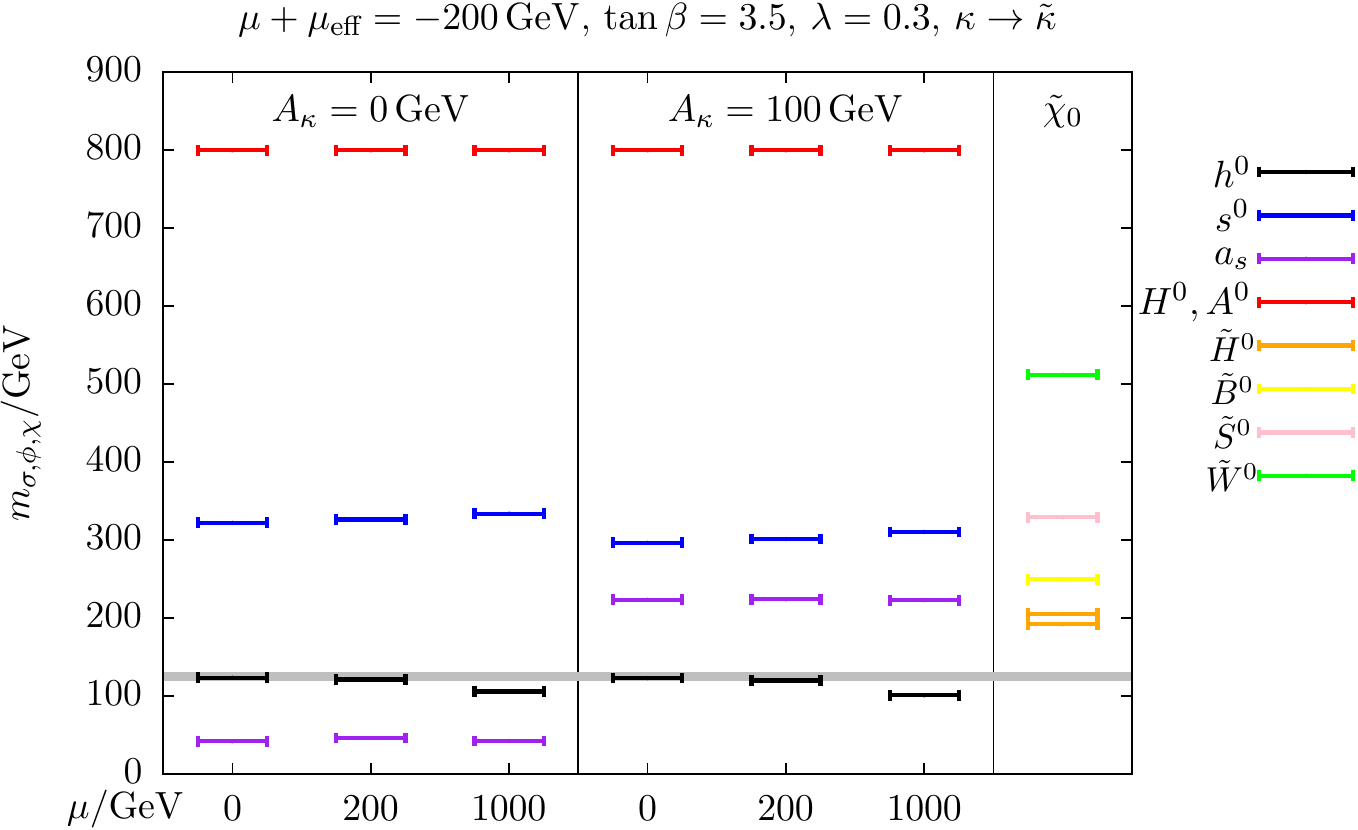} 

  \caption{\label{fig:Higgsspecrescale} In a similar manner as in
    \fig{fig:Higgsspec}, the spectra of Higgs bosons and neutralinos
    are shown in the~\munmssm. The neutralino masses are invariant
    under changes in~$\mu$ by identifying the
    sum~\mbox{$\left(\mu+\mue\right)$} of the~\munmssm{} with
    the~$\mue$~term of the~\nmssm, and by rescaling~$\kappa$ according
    to Eq.~\eqref{eq:Lieblerrescaling}. We
    set~\mbox{$\kappa=0.8\,\lambda$}, and for~\mbox{$\mu=0$\,\GeV} we
    assign~\mbox{$\mue=-200\,\GeV$}. The Higgs mass spectra are
    slightly affected by the rescaling.}
  \vspace{-1.3ex}
\end{figure}

In \fig{fig:Higgsspec}, we show the masses of the Higgs bosons for
vanishing~$A_\kappa$ in the left, \mbox{\(A_\kappa = 100\,\GeV\)} in
the middle frame, and the masses of the neutralinos in the right
frame. Each frame contains three different scenarios which are
characterized by the three values~\mbox{$\mu \in \{0, 200,
  1000\}\,\GeV$} while keeping all other parameters fixed: \mbox{$\mu
  + \mue = -200\,\GeV$}, \mbox{$t_\beta=3.5$}, \mbox{$\lambda=0.2$},
\mbox{$\kappa=0.2\,\lambda$}, and the other parameters as given in
Tab.~\ref{tab:param}. The additional~\(\mu\)~term has the biggest
influence on the singlet-like states~$s^0$ and~$a_s$, as well as the
singlino-like state~$\tilde{S}^0$. In analogy to the discussion in
\fig{fig:inospec}, the reason for this behavior is the fixed
sum~\mbox{$\left(\mu+\mue\right)$}: an increase in~$\mu$ causes a
larger negative~$\mue$ which primarily drives the singlet-mass terms
in the~$(3,3)$~elements of Eqs.~\eqref{eq:MassS} and~\eqref{eq:MassP},
and the singlino-mass term in the~$(5,5)$~element of
Eq.~\eqref{eq:mass_chi0} to large values. In the investigated
parameter region, the mass of the~\cp-odd singlet is also very
sensitive to~\(A_\kappa\): in order to avoid a tachyonic state~$a_s$
over a large fraction of the parameter space, it is essential to
keep~$A_\kappa$ sufficiently large. However, in the left frame a
scenario is shown where even a vanishing~\(A_\kappa\) is possible. It
generates a rather light~\cp-odd singlet-like state, whereas a
sizable~\mbox{$A_\kappa = 100\,\GeV$}~(middle) lifts this mass
up. There is thus the potential for a distinction between
the~\nmssm-limit for~\mbox{$\mu = 0\,\GeV$} and the~\munmssm{} with a
large~\mbox{$\mu = 1\,\TeV$}. Note that in the middle frame
for~\mbox{\(\mu = 200\,\GeV\)}, the purple and blue lines are on top
of each other.

The masses of the neutralino sector do not depend on~$A_\kappa$ at the
tree level. Concerning the Higgs sector, only the two cases in
\fig{fig:Higgsspec} with~\mbox{$\mu=0$\,GeV}
and~\mbox{$A_\kappa\in\{0, 100\}\,\GeV$} yield a~\sm{}-like Higgs
boson that is compatible with the experimental data with~$\chi^2$
values of maximal~$77$. These two cases are also compatible with
searches for additional Higgs bosons probed by~\HB. The two cases
with~\mbox{$\mu=1$\,TeV} and~\mbox{$A_\kappa\in\{0, 100\}\,\GeV$}
yield minimal~$\chi^2$ values of~$82.6$ and~$84.0$, respectively. The
larger values of~$\chi^2$ mainly arise because the~\sm{}-like
Higgs-boson mass is slightly below~$122$\,\GeV. The large variation
with~$\mu$ for the mass prediction of the mostly~\sm-like Higgs boson
is mainly induced by a large mixing with the~\cp-even singlet. The
mixing for~\mbox{$\mu=200\,\GeV$} in this scenario becomes very large
for both values of~$A_\kappa$ such that these cases are outside the
parameter region that is compatible with the constraints by~\HS. Note
that the apparent preference for~\mbox{$\mu=0\,\GeV$}
over~\mbox{$\mu\in\{200, 1000\}\,\GeV$} in this scenario is purely
accidental and could be reversed by a slight shift in the input
parameters, see the discussion below.

As already mentioned in \sct{sec:ewinos}, the electroweakino sector
alone, at least at the tree level, does not allow one to distinguish
the~\munmssm{} from the~\nmssm{}: one can keep the
neutralino--chargino spectrum at the tree level invariant by
identifying the sum~\mbox{$\left(\mu+\mue\right)$} with
the~$\mue$~term of the~\nmssm{}, and rescaling~$\kappa$ according to
Eq.~\eqref{eq:Lieblerrescaling}. However, as pointed out above, the
rescaling does have an impact on the Higgs spectrum. We show in
\fig{fig:Higgsspecrescale} spectra for~\mbox{$\mu\in\{0, 200,
  1000\}\,\GeV$} and~\mbox{$A_\kappa\in\{0, 100\}\,\GeV$} with
fixed~\mbox{$\mu+\mue=-200\,\GeV$}. The neutralino spectrum is shown
in only one column in the very right frame. In analogy to
\fig{fig:Higgsspec}, the left and middle frames show the Higgs-boson
masses for the two values of~\(A_\kappa\) where one still can see the
effect of a varying~\(\mu\)~term. While contributions to the mass
matrices in Eqs.~\eqref{eq:Higgsmasses} which are proportional
to~\mbox{$\left(\mu+\mue\right)$} or~$\kappa\,\mue$ are kept constant,
other terms~${\propto}\,\mue^{-1},\mue^{-2}$ induce
variations. Accordingly, the singlet-like Higgs masses in
\fig{fig:Higgsspecrescale} are only slightly sensitive to~\(\mu\),
much less than the changes observed in \fig{fig:Higgsspec}. A
rising~\(\mu\) slightly increases the mass splitting between the
singlet-like and the~\sm{}-like Higgs state.

Still, while the Higgs masses remain almost constant for not too
small~$\mue$, the doublet--singlet mixing can be strongly affected by
varying~$\mu$ and~$\mue$ (but keeping their sum constant), in
particular if the doublet--singlet mixing almost vanishes at a certain
choice of~$\mu$ and~$\mue$. In general, the mixing between the singlet
and doublet states is affected by a large~$\lvert\mue\rvert$. However,
by rescaling~$\kappa$ according to Eq.~\eqref{eq:Lieblerrescaling} all
contributions linear in~$\mue$ are absorbed, while the
contributions~${\propto}\,\mue^{-1}$ depend on the values
of~$t_\beta$,~$M_{H^\pm}$ and~$B_\mu$, see \eqns{eq:a1prime} and
\eqref{eq:a3prime}.\footnote{In the~\gnmssm{}, there are further
  possibilities of absorbing shifts in~$\mue$ through a redefinition
  of other $\mathbb{Z}_3$-violating parameters.} In \sct{sec:decays}
we will further investigate scenarios with very small~$\mue$ and
enhanced Higgs-boson mixing.

In \fig{fig:Higgsspecrescale} only the case~\mbox{$A_\kappa=100$\,GeV}
in combination with~\mbox{$\mu=0$\,GeV} is allowed by~\HB{}
and~\HS~(\mbox{$\chi^2=80.1$}), since the other scenarios are either
ruled out by the decay of the~\sm-like Higgs into a pair of
light~\cp-odd singlets or by a too large deviation of the~\sm-like
Higgs-boson mass from~$125$\,GeV. In addition to our discussion above,
we emphasize that in particular the latter exclusion can be easily
avoided through a slight adjustment of the input parameters.

\tocsubsection[\label{sec:numerics}]{Parameter scan}

We have discussed above the dependence of the Higgs masses and of the
condition for the stability of the electroweak vacuum on the model
parameters. Apart from the fixed parameters in Tab.~\ref{tab:param},
we choose seven ``free'' parameters that we vary in the following
regimes for our analyses:%
\begin{gather}\label{eq:ranges}
\begin{aligned}
  \mue &\in [ -2, 2 ]\,\TeV\,, &
  \mu &\in [ 0, 2 ]\,\TeV\,, &
  B_\mu &\in [ -3, 3 ]\,\TeV\,, \\[-1ex]
  \lambda &\in [ 10^{-4}, 1 ]\,, &
  \kappa &\in [ 10^{-4}, 1 ]\,, &
  A_\kappa &\in \{0, 100\}\,\GeV\,, &
  \tan\beta &\in [ 1.5, 3.5 ]\,,
\end{aligned}
\end{gather}
where the largest values of~\(\lambda\) and~\(\kappa\) in the
specified range of \eqref{eq:ranges} violate the approximate
perturbativity bound~\mbox{\(\lambda^2 + \kappa^2 \lesssim
  0.5\)}.\footnote{This perturbativity bound was explicitly derived
  for the~\nmssm{} in~\citere{Miller:2003ay}. According to the beta
  functions for~$\lambda$ and~$\kappa$ (see appendix~\ref{sec:betaf})
  no additional scale-dependent contribution is introduced by
  the~\munmssm{} at the one-loop order.} For the results presented in
the following, this bound is always fulfilled and lies outside the
plot ranges. Values of~\mbox{\(\tan\beta \gtrsim 4\)} push the model
into the~\mssm-like regime and are of less interest for studying
the~\munmssm{} effects.

We have performed a scan over the parameter space defined in
\eqref{eq:ranges} and identified regions which are compatible with
current observations concerning the properties of the~\sm-like Higgs
boson at~$125\,\GeV$ and the limits from searches for additional Higgs
bosons with~\HB{} and~\HS{} as described above. In the following, we
present a selection of results from this scan; different regions of
vacuum stability are illustrated, and the experimental constraints
from Higgs physics are indicated. While we display some typical
examples, it should be noted that similar observations hold for other
regions in the parameter space as well.

\begin{figure}[tp!]
\centering
\includegraphics[width=.49\textwidth]{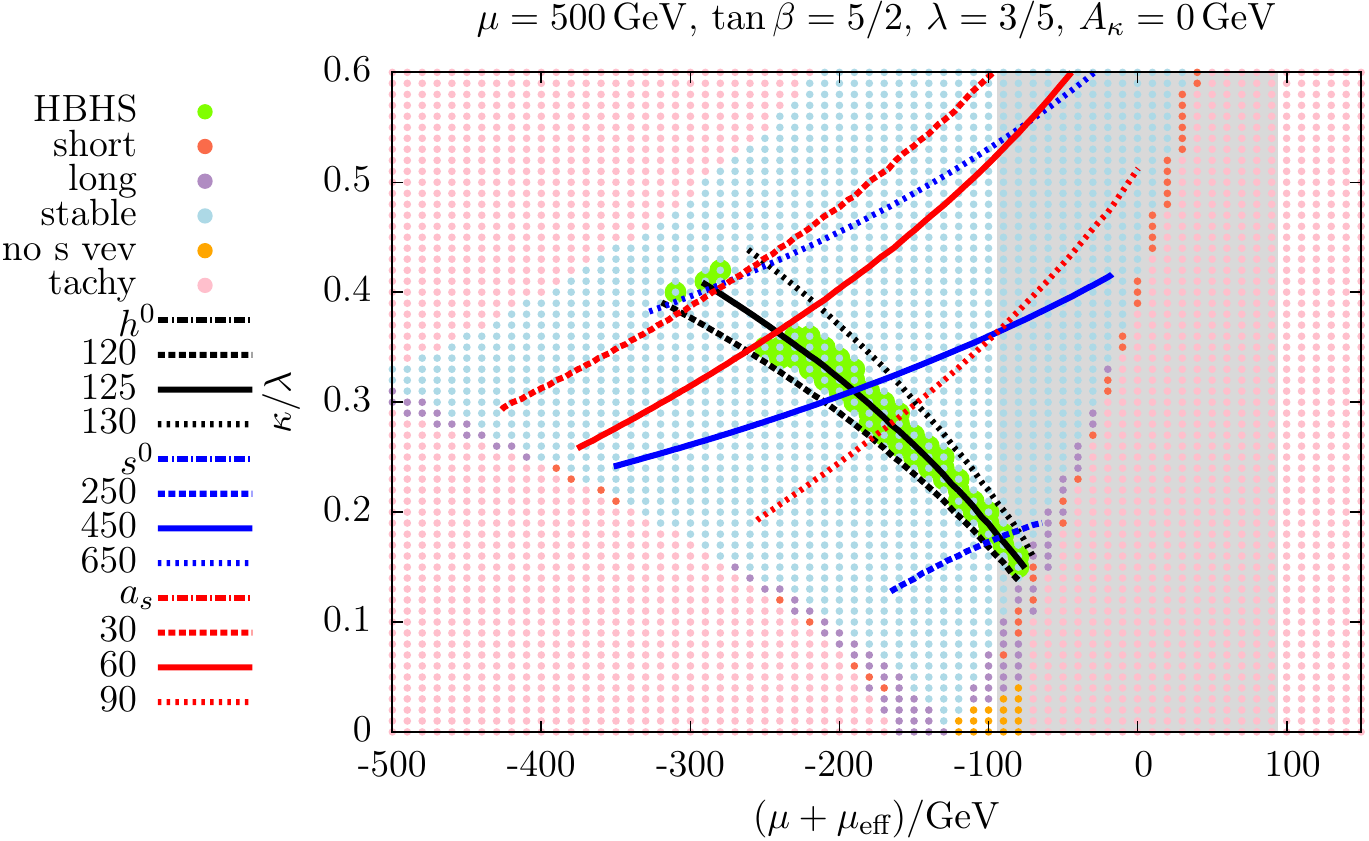} 
\hfill
\includegraphics[width=.49\textwidth]{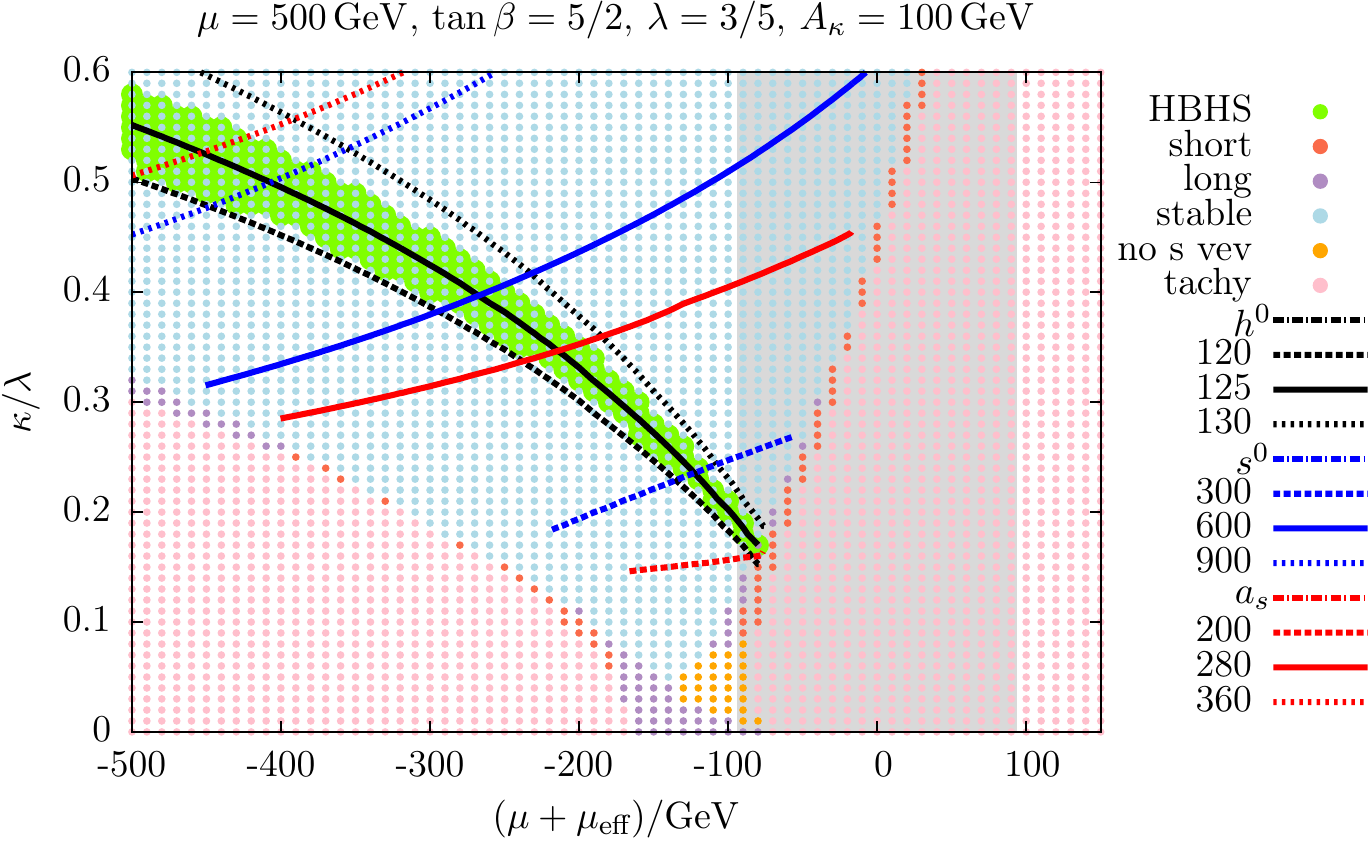}
\caption{\label{fig:example1}Contours for the~\sm-like Higgs mass
  (black) and the masses of the two singlet-like states (\cp-even in
  blue and~\cp-odd in red) in the plane~\(\kappa/\lambda\)
  versus~\mbox{\((\mu + \mue)\)}, where~\mbox{\(\lambda = 0.6\)}
  and~\mbox{\(\mu = 500\,\GeV\)} are kept fixed and~\(\kappa\)
  and~\(\mue\) vary. In the left plot~\mbox{\(A_\kappa = 0\)\,\GeV} is
  used; in the right one~\mbox{\(A_\kappa =
  100\,\GeV\)}. Furthermore,~\mbox{\(\tan\beta = 2.5\)} is set in both
  plots. The other relevant parameters are listed in
  Tab.~\ref{tab:param}. The few red and purple points have a short-
  and long-lived meta-stable electroweak vacuum, respectively, whereas
  blue points have a stable electroweak vacuum. Rose points are
  excluded because of tachyonic tree-level masses. The orange points
  cannot reproduce a non-vanishing~\(\mue\) at the electroweak vacuum
  via the constraint of Ineq.~\eqref{eq:Akappaconstr}. With the gray
  vertical band we mark a na\"ive direct experimental exclusion bound
  from the chargino mass~\mbox{\(m_{\chi^\pm} > 94\,\GeV\)}. Green
  areas are allowed by~\HB{} and~\HS{} (indicated as~``HBHS'' in the
  legend).}

\vspace{3ex} \hrule \vspace{3ex} \capstart

\includegraphics[width=.49\textwidth]{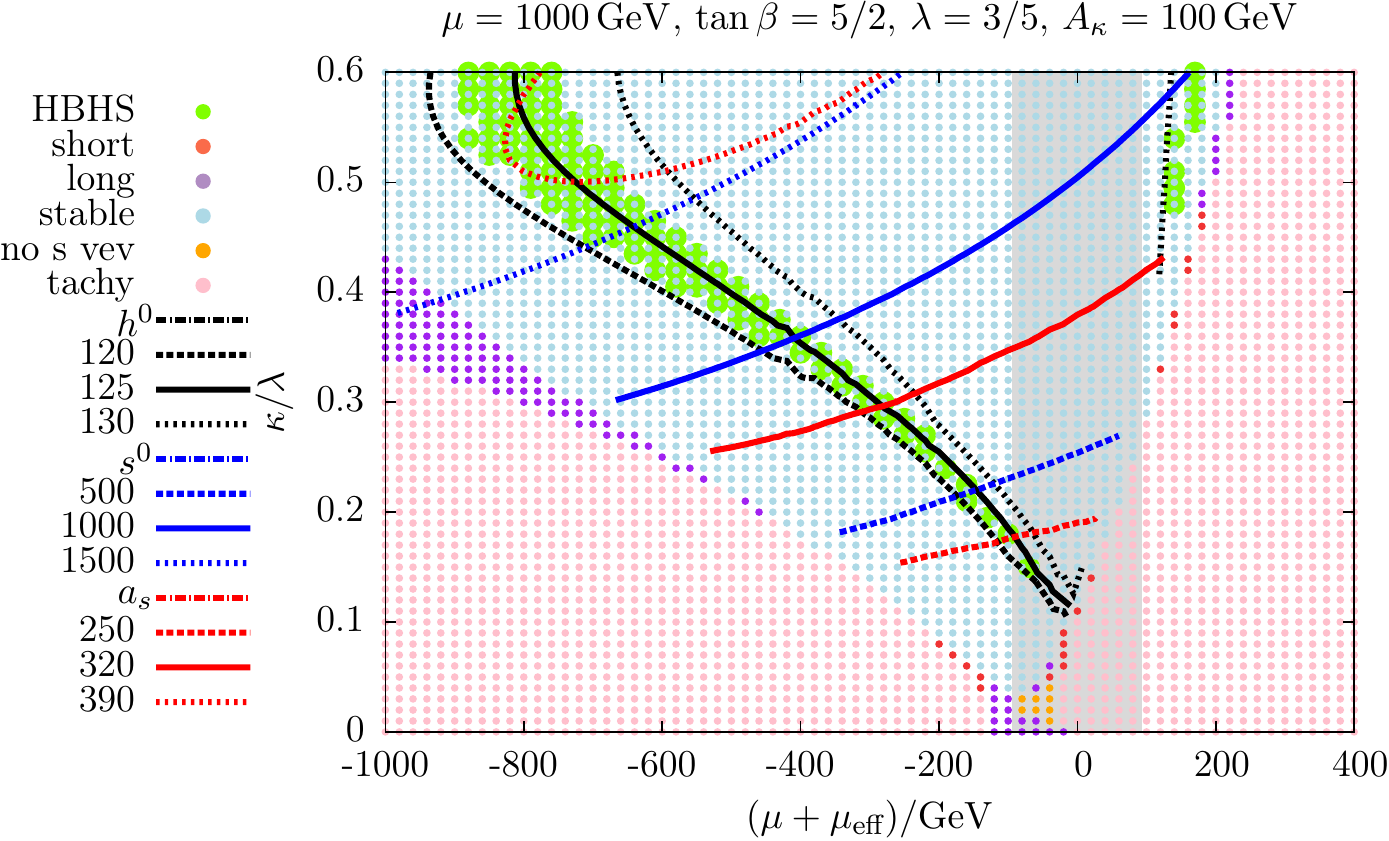} 
\hfill
\includegraphics[width=.49\textwidth]{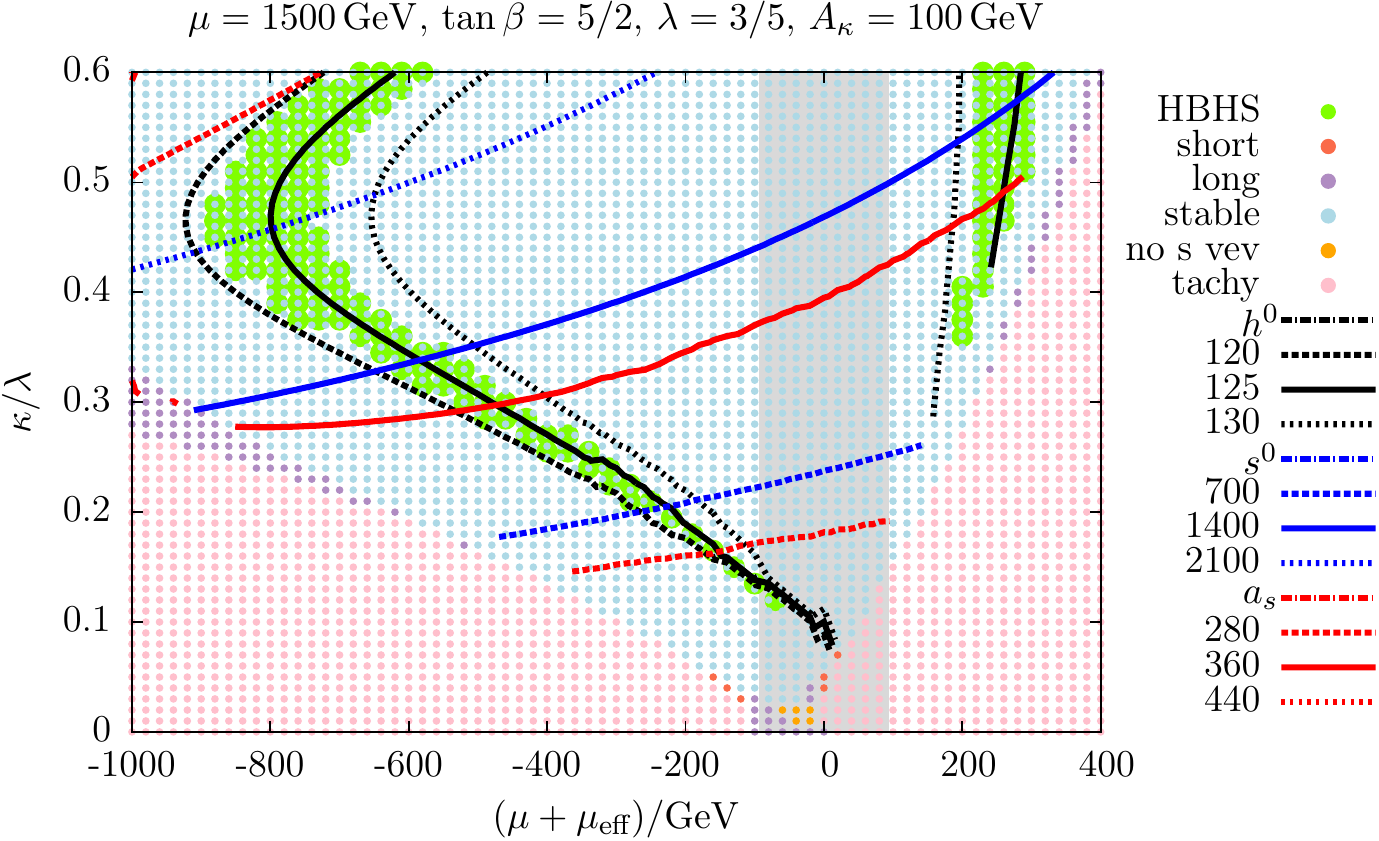}
\caption{\label{fig:example2}The same as \fig{fig:example1}, except
  that~\mbox{\(A_\kappa = 100\,\GeV\)} is used in both plots, and the
  parameter~$\mu$ is set to~\mbox{\(\mu = 1000\,\GeV\)} (left)
  and~\mbox{\(1500\,\GeV\)} (right).}

\vspace{3ex} \hrule \vspace{3ex} \capstart

\includegraphics[width=.49\textwidth]{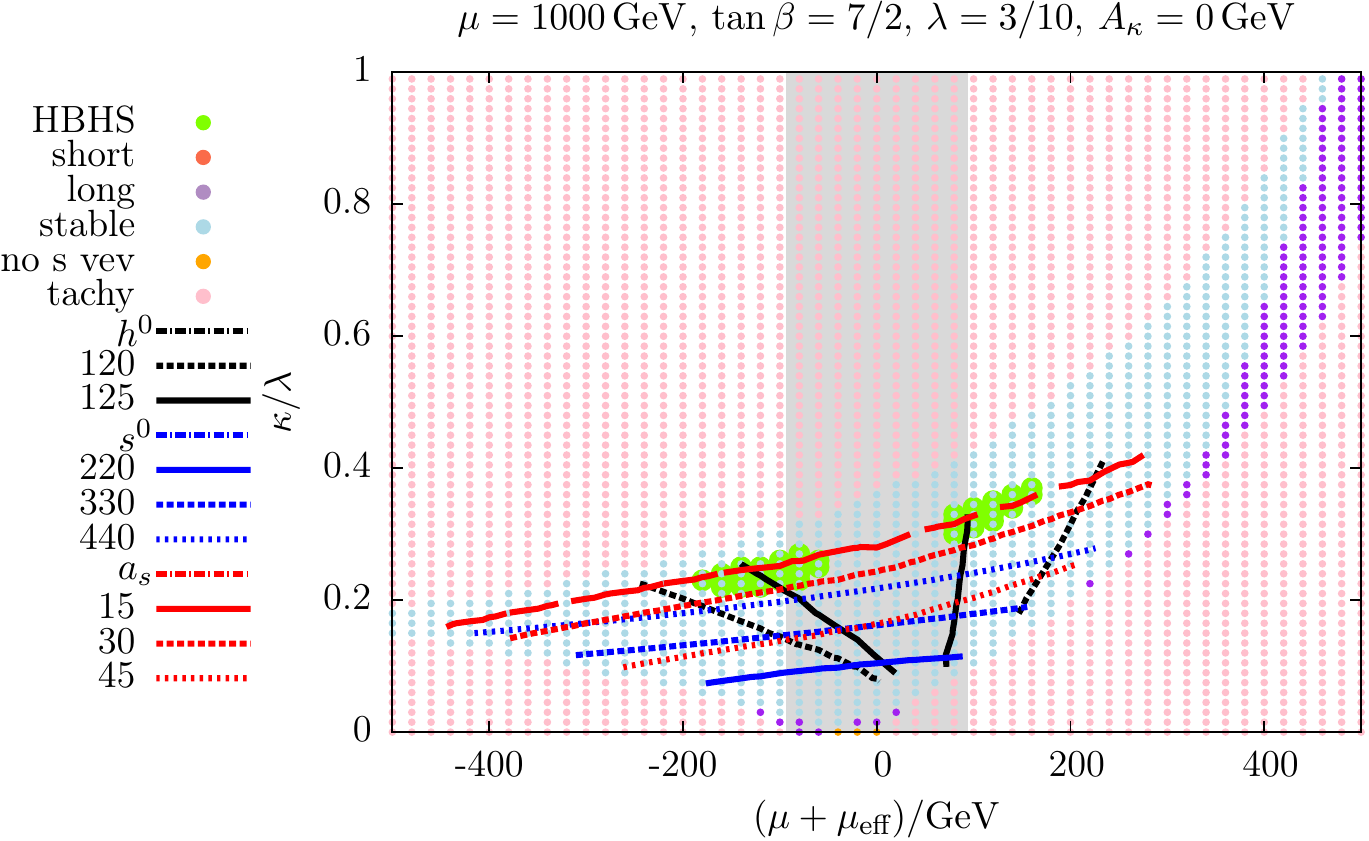} 
\hfill
\includegraphics[width=.49\textwidth]{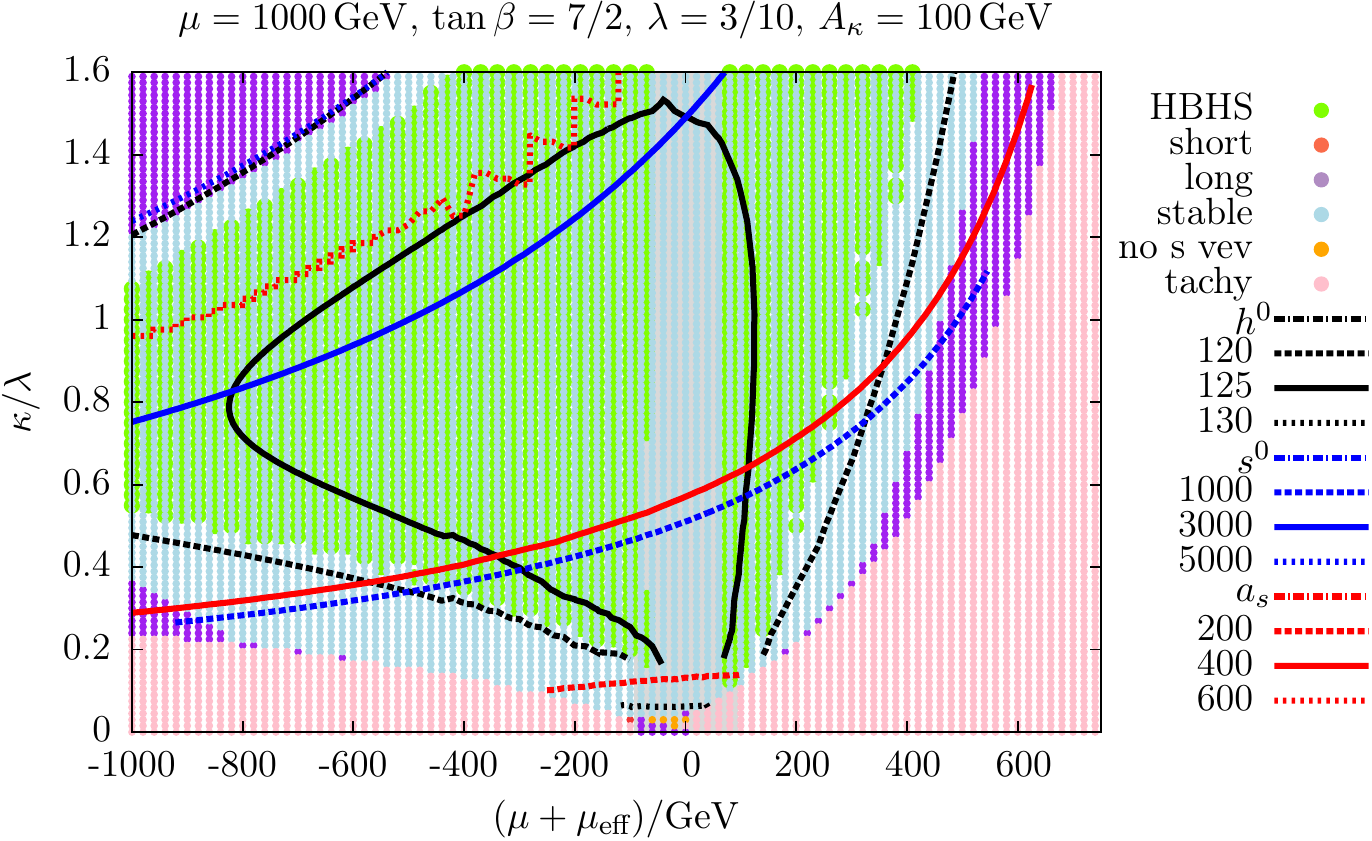}
\caption{\label{fig:example3}The same as \fig{fig:example1} but
  for~\mbox{\(\mu = 1000\,\GeV\)}, \mbox{\(\tan\beta = 3.5\)}
  and~\mbox{\(\lambda = 0.3\)}.}
\end{figure}

\begin{figure}[b!] 
\centering
\includegraphics[width=.48\textwidth]{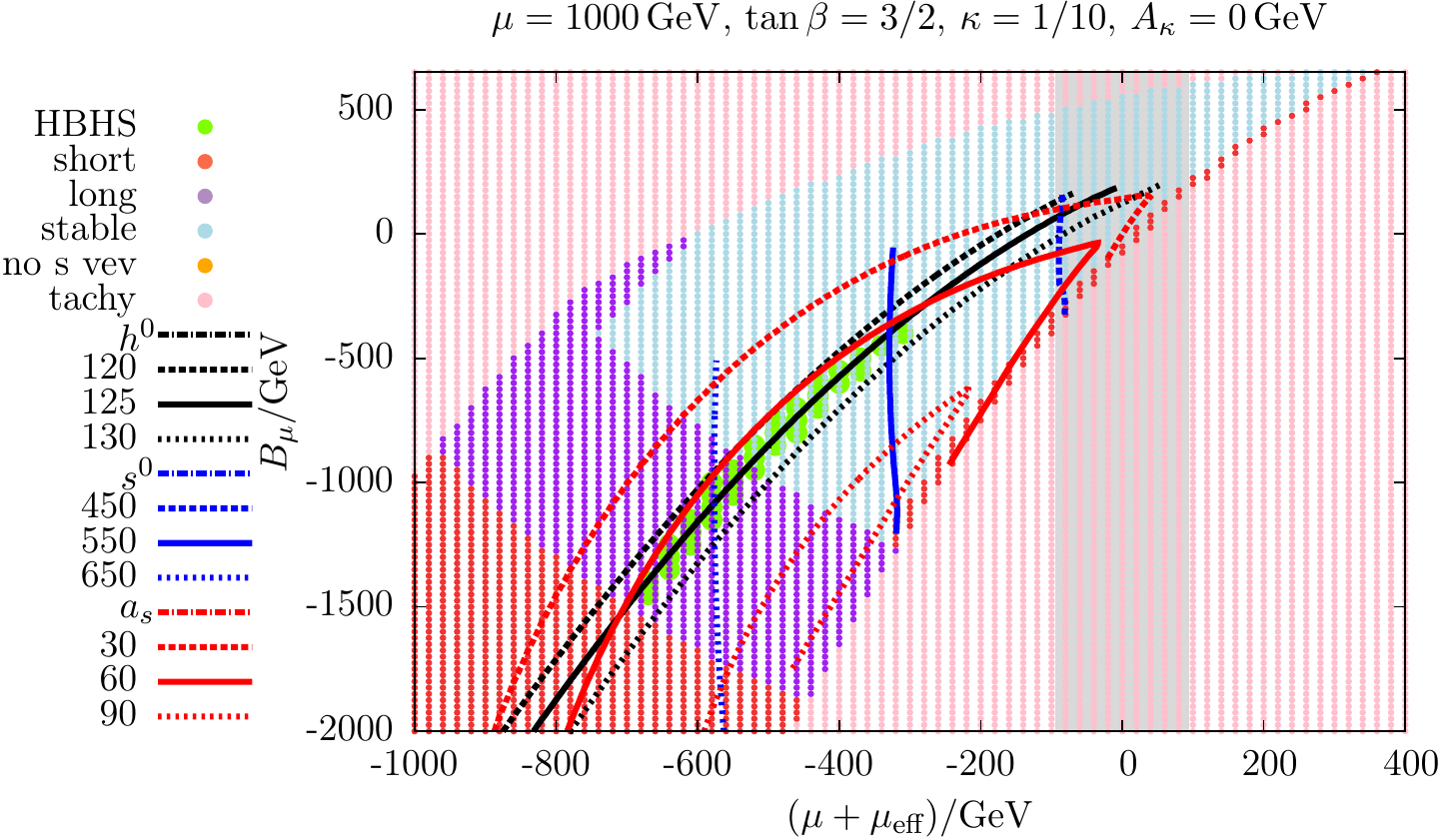}
\hfill
\includegraphics[width=.48\textwidth]{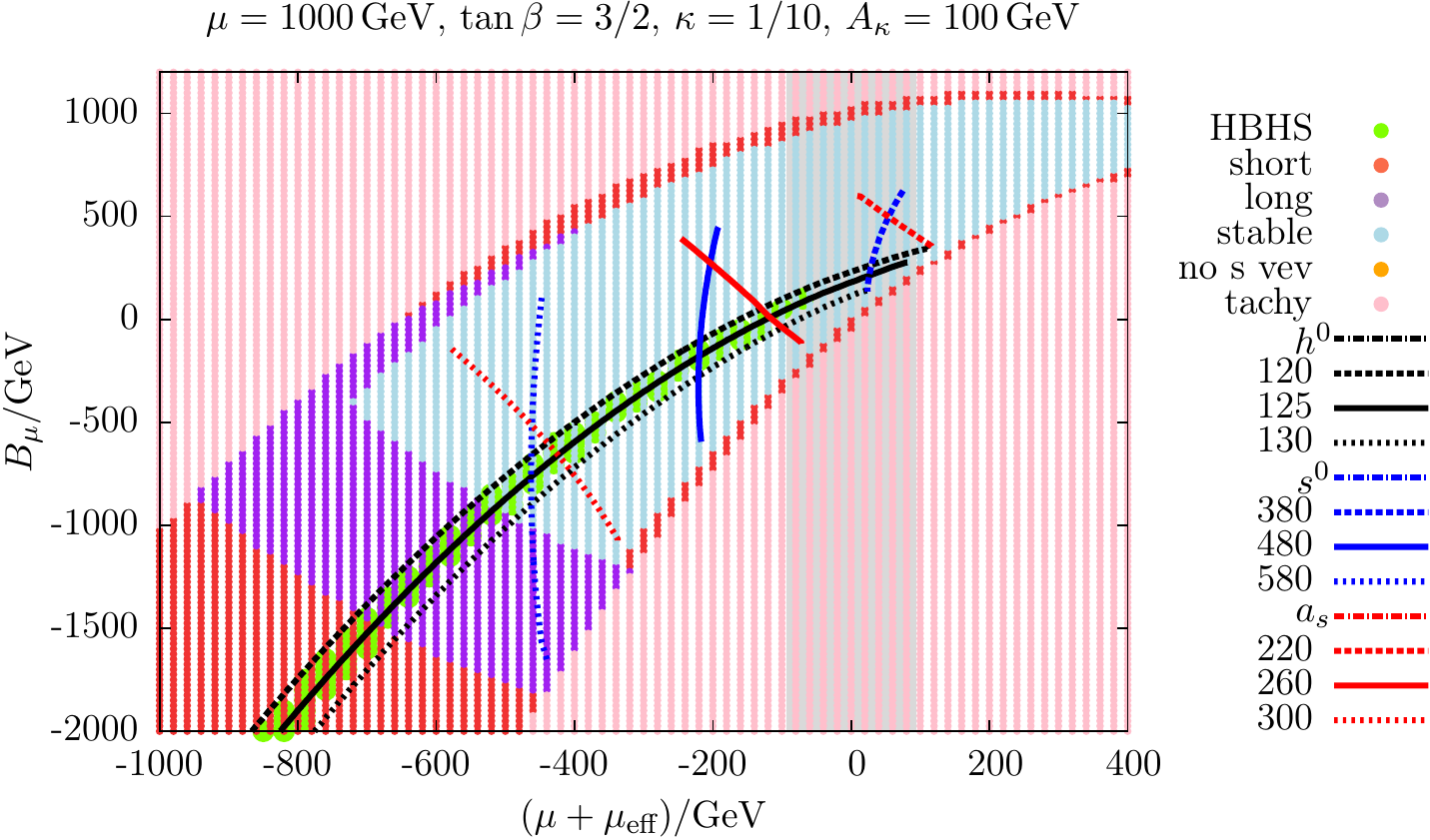}
\caption{\label{fig:example4} Dependence of mass contours and vacuum
  stability, see \fig{fig:example1} for an explanation of the color
  code, on the \(\mathbb{Z}_3\)-breaking
  soft \susy-breaking~\(B_\mu\)~term and~\mbox{$(\mu+\mue)$}
  for~\mbox{\(\lambda = 0.5\)}. On the left-hand side, the
  value~\mbox{\(A_\kappa = 0\,\GeV\)} was chosen, while on the
  right~\mbox{\(A_\kappa = 100\,\GeV\)}.}
\end{figure}

In \figs{fig:example1}--\ref{fig:example4}, we present a selection of
parameter regions. Before we discuss them individually, their common
features are explained. The colored dots in the background display
different states of the electroweak vacuum: we distinguish
stable~(blue), long-lived meta-stable~(purple), short-lived
meta-stable~(red), and tachyonic~(rose). As discussed above, we regard
not only tachyonic but also meta-stable regions as excluded in the
context of this inflationary scenario, but nevertheless display long-
and short-lived meta-stable regions for illustration. Furthermore, we
indicate those points that do not fulfill
Ineq.~\eqref{eq:Akappaconstr} and thus have no singlet vev~(orange),
although, as explained in \sct{sec:vacuum}, this constraint is not
relevant for the~\munmssm. We overlay mass contours for the~\sm-like
Higgs~$h^0$~(black), the~\cp-even singlet-like Higgs~$s^0$~(blue), and
the~\cp-odd singlet-like Higgs~$a_s$~(red). The spectrum is calculated
taking into account the full one-loop and the known~\mssm{}-like
two-loop contributions as described in \sct{sec:masscorrections}. The
assignment of the labels~$h^0$, $s^0$ and~$a_s$ to the loop-corrected
states is determined by the largest respective contribution in the
mixing matrix~$Z^{\text{\tiny mix}}_{ij}$. We emphasize again that the
parameters of the stop sector specified in Tab.~\ref{tab:param} for
the given scale of SUSY masses maximize the~\sm-like Higgs mass
for~\mbox{\(\mu+\mue = 0\,\GeV\)}; therefore, lower values for
the~\sm{}-like Higgs mass could easily be obtained by reducing the
mixing in the stop sector. Finally, we also indicate a na\"ive
exclusion bound from direct searches for charginos by the gray-shaded
band: Ref.~\cite{Patrignani:2016xqp} reports a lower bound on the
chargino mass of~$94\,\GeV$ which translates into the requirement
that~\mbox{$\lvert\mu+\mue\rvert$} must be above that value in
the~\munmssm. Lastly, all \figs{fig:example1}--\ref{fig:example4} show
the region of parameter points that successfully passed~\HB{}
and~\HS{} and thus, in particular, yield a~\sm{}-like Higgs boson
compatible with the observed state at~$125\,\GeV$. This region is
represented through the larger, light green dots in the background. We
refer to \sct{sec:parameters} for our statistical interpretation of
the results obtained from the two codes.

A large part of the parameter region that is consistent with the
measured~\sm-like Higgs mass is also in concordance with an absolutely
stable electroweak vacuum. Small intersections between stable regions
and regimes with tachyonic Higgs states exist, where there are
meta-stable non-standard vacua. The strongest constraints arise from
the existence of tachyonic masses for one of the physical Higgs states
at the tree level. In the remaining region only a small fraction of
points has a global minimum which does not coincide with the
electroweak vacuum whereas the majority has a true electroweak
vacuum. For the short-lived meta-stable regions, the vacuum lifetime
is longer than the age of the universe.

In \fig{fig:example1} we indicate the Higgs-mass contours and the
constraints from vacuum stability in the plane
of~\mbox{$\left(\mu+\mue\right)$} and~$\kappa/\lambda$ with
fixed~$\mu$ and~$\lambda$. Note that for this choice of variables the
tree-level doublet sector in Eqs.~\eqref{eq:MassS}, \eqref{eq:MassP}
and~\eqref{eq:a1prime} remains constant; any structure visible in the
prediction of the~\sm{}-like Higgs mass is thus induced by mixing with
the singlet state, or by loop corrections. The chosen parameter values
are indicated in the legends of the figures and in
Tab.~\ref{tab:param}; in the left plot~\mbox{$A_\kappa=0\,\GeV$} is
used, while in the right plot~\mbox{$A_\kappa=100\,\GeV$}. The value
of~$A_\kappa$ has an impact in particular on the mass scale of
the~\cp-odd singlet-like Higgs which is much lighter on the left-hand
side. In fact, for a light~\cp-odd singlet-like Higgs a parameter
region opens up where decays of the~\sm-like Higgs into a pair of them
become kinematically allowed. The~\cp-even singlet-like Higgs is also
somewhat lighter for~\mbox{$A_\kappa=0$\,\GeV}, while the~\sm-like
Higgs is scarcely affected. The contour lines of the Higgs masses stop
when one Higgs becomes tachyonic. The reason why this does not exactly
coincide with the border between the blue and pink dotted regions are
the loop corrections to the Higgs spectrum while the constraints from
vacuum stability were investigated at the tree level. It can be seen
that the boundaries at the left of the stable region are parallel to
one of the displayed Higgs-mass contours---the corresponding particle
becomes tachyonic at this boundary. The boundary to the right of the
stable region can be understood when comparing the right plots of
\fig{fig:example1} and \fig{fig:example2}, which differ from each
other by the value of~$\mu$: in the right plot of \fig{fig:example2} a
contour for the~\sm-like Higgs mass which is parallel to the tachyonic
border appears around~\mbox{$\mu+\mue=250\,\GeV$}
and~\mbox{$\kappa/\lambda=0.5$}. In \fig{fig:example1} such a contour
is not visible as this particular parameter region is excluded by a
tachyonic~\sm{}-like state at the tree level. Note that the~\nmssm-
and~\munmssm-specific one-loop contributions to the Higgs spectrum are
particularly large in that region~(about~$60\,\GeV$ additional shift
compared to the same scenario in the~\mssm{}-limit
with~\mbox{$\lambda\to 0$} and~$\kappa/\lambda$ constant), see also
\citere{Drechsel:2016jdg}; a dedicated analysis taking into account
two-loop effects beyond the~\mssm{}-limit might be necessary for a
robust prediction of the Higgs mass close to the right border of the
stable region, see \EG~\citere{Goodsell:2014pla}. It should be noted
that in \fig{fig:example1} the region where the Higgs mass is close to
the right border of the stable region is disfavored by the limits from
chargino searches at~\lep.

As expected, the region allowed by~\HB{} and~\HS{} is a subset of the
region where the~\sm{}-like Higgs has a mass in the vicinity
of~$125$\,GeV. In the green-marked region,~$\Delta\chi^2$~is at
maximum~$5.99$. The minimal value~$\chi_m^2$ from~\HS{} is~$74.6$ in
both figures. One can see on the left-hand side of \fig{fig:example1}
that this region is split into two: in between the two regions
the~\sm-like Higgs can decay into a pair of~\cp-odd singlet-like Higgs
bosons~\mbox{$h^0 \to a_sa_s$} with a branching ratio of up
to~$90\,\%$; this behavior is not compatible with the observed signal
strengths implying a limit on decays of the state at~$125\,\GeV$ into
non-\sm~particles. For a very light~\cp-odd singlet, the admixture
between the~\sm-like Higgs and the~\cp-even singlet component is
reduced, since the latter becomes heavier in this region. In the
scenario under consideration, the decay~\mbox{$h^0 \to a_sa_s$} is
dominated by the coupling among the two singlet
states,~$\lambda_{355}$ in \eqn{eq:lambda355}, such that a reduced
admixture between~$h^0$ and~$s^0$ also closes the decay~\mbox{$h^0 \to
a_sa_s$}. This is why---despite the very light~\cp-odd
Higgs~$a_s$---the region at~\mbox{$\mu+\mue \simeq -300$\,GeV}
and~\mbox{$\kappa/\lambda \simeq 0.4$} is allowed by the constraints
from both~\HS{} and~\HB.

In \fig{fig:example2} we present scenarios similar to the right-hand
side of \fig{fig:example1} with~\mbox{\(A_\kappa = 100 \, \GeV\)}, but
with different values of~$\mu$ (note the larger scale at
the~$x$-axis). Thus, the influence of this parameter that
distinguishes the~\munmssm{} from the~\nmssm{} can be seen
directly. Obviously, the parameter region with a stable vacuum is
enlarged: for a given value~\mbox{$(\mu+\mue)$} the tachyonic border
moves to smaller ratios of~$\kappa/\lambda$
as~$\mu$~increases. Concerning the Higgs spectrum, the most notable
difference is seen for the~\sm{}-like Higgs mass: for~\mbox{$\mu =
  1\,\TeV$} a turning point at about~\mbox{$\mu+\mue=-800\,\GeV$} is
visible, which moves to smaller values of~$\kappa/\lambda$
for~\mbox{$\mu = 1.5\,\TeV$}. For the larger value of~$\mu$ one can
see that the possibility emerges for scenarios with the
correct~\sm{}-like Higgs mass but positive~\mbox{$(\mu+\mue)$}. Again
all tested points which yield a~\sm{}-like Higgs boson close
to~$125$\,GeV successfully pass the constraints implemented in~\HB{}
and~\HS. The minimal values of~$\chi_m^2$ from~\HS{} are~$74.9$
and~$74.6$ on the left-hand and on the right-hand side of
\fig{fig:example2}, respectively.

\fig{fig:example3} shows scenarios with larger~$\tan\beta$ and
smaller~$\lambda$ compared to the previous figures. Like in
\fig{fig:example1} we set~\mbox{$A_\kappa=0$\,GeV} on the left,
and~\mbox{$A_\kappa=100$\,GeV} on the right-hand side, but~\mbox{$\mu
  = 1\,\TeV$} is used. We observe again that a larger value
of~$A_\kappa$ widens the allowed parameter region, because the mass of
the~\cp-odd singlet is lifted up, giving rise to a drastic effect in
this case. In fact, for~\mbox{$A_\kappa=0$\,GeV} only a rather small
area in the plane of~\mbox{$(\mu+\mue)$} and~$\kappa/\lambda$ is
allowed, while the allowed region is very significantly enhanced
for~\mbox{$A_\kappa=100$\,GeV}. In the plot on the right-hand side one
can see a (nearly) closed~$125\,\GeV$~contour for the mass of
the~\sm{}-like Higgs with even larger values in the enclosed
area. Adjusting the parameters of the stop sector in order to obtain a
smaller contribution to the~\sm{}-like Higgs mass can render
a~\sm{}-like Higgs with a mass of about~$125\,\GeV$ in the whole
enclosed region. Close to the tachyonic borders we find larger regions
with a long-lived meta-stable vacuum~(purple) than in
Figs.\,\ref{fig:example1} and \ref{fig:example2}. However, in this
part of the plot the prediction for the mass of the~\sm{}-like Higgs
is below the experimental value. On the right-hand side of
\fig{fig:example3} a large region is allowed by the constraints
from~\HB{} and~\HS. Only low values
of~\mbox{$\lvert\mu+\mue\rvert<m_h/2$} are excluded by~\HS{} due to
the decay of the~\sm{}-like Higgs boson into a pair of
higgsinos. However, this region is anyhow not compatible with
the~\lep~bound on light charginos. The minimal values of~$\chi_m^2$
from~\HS{} are~$74.7$ in both plots.

In \fig{fig:example4} we change the parameter on the~$y$-axis: $B_\mu$
is varied and~$\kappa$ is kept fixed. We set~\mbox{$A_\kappa=0\,\GeV$}
on the left-hand side, and~\mbox{$A_\kappa=100$\,GeV} on the
right-hand side. One can see that non-zero values for~$B_\mu$ can have
a significant impact on the predicted Higgs masses and might determine
whether or not a scenario is excluded. For larger negative values
of~\(B_\mu\), one can see an area where the electroweak vacuum is
meta-stable and long-lived, while the area in the lower left corner of
the plots indicates that the electroweak vacuum is unstable and
short-lived. The effect of a larger~\(A_\kappa\) mainly lifts the
tachyonic boundary at the top so that values of~\mbox{\(B_\mu =
1\,\TeV\)} are allowed for~\mbox{\(A_\kappa = 100\,\GeV\)} and leaves
the other regions invariant. However, towards the upper limit
of~\(B_\mu\), there is a small short-lived area. As a new feature, we
find large regions with a meta-stable vacuum but a~\sm{}-like Higgs
with a mass of~$125\,\GeV$ for both values
of~\(A_\kappa\). Accordingly, scenarios with too large negative values
of~$B_\mu$ are excluded due to a rapidly decaying vacuum despite
providing a~\sm{}-like Higgs boson close to the observed mass. The
constraints from~\HB{} and~\HS{} indicate that a large part of the
region with the correct Higgs mass is compatible with the experimental
data. For both plots~\HS{} yields a minimal value of~\mbox{$\chi_m^2 =
74.9$}. Only in those scenarios where the decay channel~\mbox{$h^0 \to
a_sa_s$} is kinematically allowed---which happens in the plot
for~\mbox{\(A_\kappa = 0\,\GeV\)} for~\mbox{\(\mu + \mue \gtrsim
-300\,\GeV\)} and~\mbox{\(\mu + \mue \lesssim -700\,\GeV\)}---the
parameter region is incompatible with the data on the detected Higgs
boson.

We briefly summarize the observed features and give an outlook for the
phenomenological studies in the following. The allowed parameter
region is mainly constrained by configurations where one Higgs field
is tachyonic at the tree level. It can be seen that the tachyonic
boundaries follow the Higgs mass contours in the
\figs{fig:example1}--\ref{fig:example4}; in addition, there are
effects from~\(\mue^{-1}\)~terms as discussed in \sct{sec:potential}
which enhance the doublet--singlet mixing and eventually cause
tachyons. This feature can be observed towards the right end of the
\figs{fig:example1}--\ref{fig:example3}. The experimental limits and
constraints confine the allowed regions further around the region
where the~\sm-like Higgs has a mass of about~\(125\,\GeV\) and exclude
parameter regions where for instance the decay of the~\sm-like Higgs
into a pair of light~\cp-odd singlets has a large branching ratio. In
this context, the singlet sector has a significant impact on the
features discussed in \figs{fig:example1}--\ref{fig:example4}.

In the~\nmssm, one usually expects to find the phenomenologically most
interesting regions (accommodating a~$125\,\GeV$~Higgs) for rather
large values of~\mbox{$\lambda \gtrsim 0.1$}, since
the~\nmssm{}~contribution to the~\sm-like Higgs mass at the tree level
is enhanced. In addition, large~$\lambda$ enhances the
doublet--singlet mixing. However, in the~\munmssm, there is another
way to obtain a large doublet--singlet mixing also for small values
of~\(\lambda\): this is the region of low~$\mue$ where terms
proportional to~$\mue^{-1}$ become large, as discussed in
\sct{sec:parameters}. We will investigate this class of scenarios,
which are not possible in the~\nmssm{} but generic to the~\munmssm{},
in \sct{sec:decays} in more detail.

Similar to the~\nmssm{}, the chosen value of~$A_\kappa$ has a strong
influence on the singlet-like Higgs masses, which is relevant for the
tachyonic regions. In a large part of the viable parameter space the
relation~\mbox{\(\sign{(A_\kappa)} = -\sign{(\mue)}\)} applies, where
for~\mbox{\(A_\kappa = 0\,\GeV\)} both signs of~\(\mue\) are allowed
in general. This dependence on the relative signs of~\(A_\kappa\)
and~\(\mue\) can be derived from the discussion in \sct{sec:potential}
about the Higgs singlets and especially the functional dependence
of~\(a_5\) in \eqn{eq:abbrev_a5} versus~\(a_4'\) in \eqn{eq:a4prime}:
large negative values of the sum~\mbox{\((a_4' + a_5)\)} drive
the~\cp-even singlet tachyonic. In the investigated scenarios above,
which have either~\mbox{$A_\kappa = 0\,\GeV$} or~\mbox{$A_\kappa =
100\,\GeV$}, the sign of~$\mue$ is negative in most of the viable
parameter space. Accordingly, there is a preference for
negative~\mbox{$(\mu + \mue)$}. The allowed region with small positive
values occurs where the negative value of~$\mue$ is overcompensated by
the positive value of~$\mu$. In \sct{sec:decays} we will investigate a
scenario where we keep~\mbox{$(\mu + \mue)$} fixed at a positive
value, while for~$A_\kappa$ small negative and small positive values
are used for~\mbox{$\mue > 0$\,\GeV} and~\mbox{$\mue < 0$\,\GeV},
respectively.  There we will also discuss the dependence of the
singlet masses on~$\mu$ and~$\mue$ in more detail.

\tocsubsection[\label{sec:production}]{Higgs-boson and electroweakino
  production}

In this and the next section we discuss phenomenological features of
Higgs-boson mixing and thus consequences on Higgs-boson production and
decays due to the~\(\mu\) parameter of the~\munmssm. For
vanishing~$\mu$ the phenomenology of the Higgs bosons equals the one
of the~\nmssm, for which typical benchmark scenarios can be found in
\citere{deFlorian:2016spz} (see also \citere{Domingo:2015eea}).
Naturally they differ from~\mssm-type benchmark scenarios through
singlet states modifying the phenomenology: since the singlet
states~$s^0$ and~$a_s$ neither directly couple to fermions nor to
gauge bosons, but only through their admixture with the doublet
states, their direct production---both at a hadron collider and a
lepton collider---is negligible in many scenarios. However, besides
their direct production light singlet states can also be potentially
observable via their production in cascade decays of heavier Higgs
bosons, as we will discuss in the following.

In most parts of our numerical study, we make use of the approximation
of \sm-normalized effective couplings of a Higgs boson to
gluons---calculated at leading order---which we insert into~\HB{} for
the evaluation of the Higgs-production cross-sections for the neutral
Higgs bosons at the~\lhc. This treatment should be sufficiently
accurate for determining the allowed regions in our scans over the
parameter space. In the following, however, we will investigate to
what extent the~\munmssm{} can accommodate the slight excesses in the
data over the background expectation at a mass
around~\mbox{$95$--$98\,\GeV$} that have been reported recently
by~\cms~\cite{CMS:2017yta} in
the~\(\gamma\gamma\)~channel\footnote{The results
  of~\atlas~\cite{ATLAS-CONF-2018-025} are presented in a fiducial
  region and are compatible with both the~\sm~expectation and the
  excess reported by~\cms.} and earlier at~\lep~\cite{Schael:2006cr}
in the~\(b\bar b\)~channel. For this purpose we use more sophisticated
predictions for the Higgs-production cross-sections in order to
compare with the experimental results. We obtain those predictions
from~\code{SusHi}~\cite{Harlander:2012pb, Harlander:2016hcx,Chetyrkin:2000yt,
Harlander:2002wh,Harlander:2003ai,Harlander:2005rq,
Anastasiou:2014lda,Anastasiou:2015yha,Anastasiou:2016cez}, for
which a dedicated version for the~\nmssm{}
exists~\cite{Liebler:2015bka}. The predictions
include~\nklo{3}~\qcd~corrections for the top-quark contribution of
the light~\cp-even Higgs bosons, while we have neglected contributions
from heavy squarks and gluinos beyond the resummed contributions in
the bottom-Yukawa coupling.

In the~\nmssm{}, the observed excesses in the data
around~\mbox{$95$--$98\,\GeV$} can be interpreted in terms of a
singlet-like state~\(s^0\), see \citere{Badziak:2016tzl} for a
discussion of the~\lep~result, and \citere{Tao:2018zkx} for a
discussion of the~\cms~data. At first sight it seems to be non-trivial
to describe both excesses simultaneously, since accommodating
the~\lep~excess would require a rather large rate~\mbox{$s^0\to b\bar
  b$}, which in turn would suppress the channel~\mbox{$s^0\to
  \gamma\gamma$} that is employed in the interpretation of
the~\cms~excess. As it was pointed out in \citere{Domingo:2018uim}
based on a detailed analysis of the Higgs mixing properties, this is
nevertheless possible---albeit in a relatively narrow region of the
parameter space, which is somewhat enlarged if the possibility of
non-vanishing phases giving rise to~\cp-violating effects is taken
into account. We investigate in the following to which extent the
additional freedom that is present in the~\munmssm{} with respect to
the possible values of the masses in combination with the mixing
properties has an impact regarding a possible interpretation of the
observed excesses. In Tab.~\ref{tab:97scenarios} we present four
scenarios with~$s^0$~masses in the range~\mbox{$95$--$98$\,GeV} that
have a phenomenology addressing the excesses observed both at~\lep{}
and~\cms. Scenarios~1~and~3 have a small value of~$\mu$ and
are~\nmssm-like (inspired by the scenarios investigated in
\citere{Domingo:2018uim}), while Scenarios~2 and~4 both
have~$\mu$~values that significantly differ from~zero, and Scenario~4
furthermore has a non-zero value of~$B_\mu$. These
two~\munmssm~scenarios are intrinsically different from the~\nmssm.
Similar scenarios could also be obtained by changing the signs
of~\mbox{$(\mu+\mue)$} and~$A_\kappa$ simultaneously.

\begin{table}[t!]
\centering
\caption{\label{tab:97scenarios} Scenarios that yield a light~\cp-even
  singlet-like Higgs boson. The Higgs boson at about~$125$\,GeV
  is~\sm-like. All other parameters are chosen in accordance to
  Tab.~\ref{tab:param}.  }
\begin{tabular}{c|cccc}
\hline
Scenario               & 1      & 2      & 3      & 4       \\\hline
 $\lambda$             & $0.08$ & $0.08$ & $0.28$ & $0.08$  \\
 $\kappa$              & $0.04$ & $0.023$ & $0.08$ & $0.0085$\\
 $\tan\beta$           & $12$   & $12$   & $2.5$  & $2$     \\
 $(\mu+\mue)$\,[GeV]   & $-140$ & $-140$ & $-300$ & $-400$  \\
 $\mu$\,[GeV]          & $5$    & $195$  & $5$    & $150$   \\
 $B_\mu$\,[GeV]        & $0$    & $0$    & $0$    & $-300$  \\
 $m_{H^\pm}$\,[GeV]    & $800$  & $800$  & $800$  & $1000$  \\
 $A_\kappa$\,[GeV]     & $130$  & $265$  & $250$  & $32$    \\
 $A_{f}$\,[GeV]        & $400$  & $450$  & $3200$ & $4000$  \\\hline
 $m_{s^0}$\,[GeV]      & $97.6$ & $95.7$ & $97.2$ & $97.1$  \\
 $m_{h^0}$\,[GeV]      & $124.7$& $126.8$& $124.6$& $125.0$ \\
 $m_{a^s}$\,[GeV]      & $168.2$& $277.0$& $257.2$& $75.6$  \\\hline
 $\frac{\sigma{\left(e^+e^-\to Zs^0\right)}\cdot\text{BR}{\left(s^0\to b\bar
b\right)}}{\sigma^{\text{\sm{}}}{\left(e^+e^-\to ZH\right)}\cdot\text{BR}^{\text{\sm{}}}{\left(H\to b\bar b\right)}}$
                           & $0.28$  & $0.31$  & $0.14$  & $0.35$ \\
 $\sigma{\left(gg\to s^0\right)}$\,[pb] & $25.3$  & $28.1$  & $14.4$  & $31.5$  \\
 BR${\left(s^0\to \gamma\gamma\right)}$ & $0.0020$& $0.0016$& $0.0024$&  $0.0005$\\\hline
\(\chi^2(\text{\HS})\) & $97$ & $96$ & $82$ & $101$ \\ \hline
\end{tabular}

\vspace{2ex}
\capstart

\caption{\label{tab:97ewinos} Cross-sections for electroweakinos at an
  electron--positron collider for Scenario~1 defined in
  Tab.~\ref{tab:97scenarios}.}
\begin{tabular}{c|ccccc}
\hline
 Scenario 1       & $\tilde\chi^0_1$& $\tilde \chi^0_2$ & $\tilde\chi^0_3$ & $\tilde \chi_1^\pm$ & \\
 Masses [GeV]           & $127.3$ & $138.3$ & $155.9$ & $138.4$ & \\\hline
 $\sigma(e^+e^-\to \tilde\chi_i\tilde\chi_j)$ [fb] for $\sqrt{s}=350$\,GeV &
    $\tilde\chi^0_1\tilde\chi^0_2$ & $\tilde\chi^0_1\tilde\chi^0_3$ & $\tilde\chi^0_2\tilde\chi^0_3$ & $\tilde\chi^0_2\tilde\chi^0_2$ & $\tilde\chi^+_1\tilde\chi^-_1$ \\
 Unpolarized                  & $141$ & $195$ & $0.08$ & $0.19$ & $795$ \\
 Pol($e^+,e^-)=(+30\%,-80\%)$ & $208$ & $287$ & $0.12$ & $0.28$ & $1620$ \\
 Pol($e^+,e^-)=(-30\%,+80\%)$ & $142$ & $196$ & $0.08$ & $0.19$ & $352$ \\\hline
 $\sigma(e^+e^-\to \tilde\chi_i\tilde\chi_j)$ [fb] for $\sqrt{s}=500$\,GeV &
    $\tilde\chi^0_1\tilde\chi^0_2$ & $\tilde\chi^0_1\tilde\chi^0_3$ & $\tilde\chi^0_2\tilde\chi^0_3$ & $\tilde\chi^0_2\tilde\chi^0_2$ & $\tilde\chi^+_1\tilde\chi^-_1$ \\
 Unpolarized                  & $74$  & $109$ & $0.12$ & $0.22$ & $459$ \\
 Pol($e^+,e^-)=(+30\%,-80\%)$ & $110$ & $161$ & $0.19$ & $0.32$ & $926$ \\
 Pol($e^+,e^-)=(-30\%,+80\%)$ & $75$  & $110$ & $0.13$ & $0.22$ & $212$ \\\hline
\end{tabular}
\end{table}

Interpreting the~\lep~excess as the contribution of a singlet-like
state~$s^0$ in the considered mass range yields a ``signal strength''
of
\begin{align}
 \frac{
 \sigma{\left(e^+e^-\to Zs^0\right)}\cdot\text{BR}{\left(s^0\to b\bar b\right)}}
 {\sigma^{\text{\sm{}}}{\left(e^+e^-\to ZH\right)}\cdot
 \text{BR}^{\text{\sm{}}}{\left(H\to b\bar b\right)}}
 &\simeq \text{$0.2$--$0.3$}\,,
\end{align}
while a ``signal rate'' of~\mbox{$\sigma(pp\to
s^0\to \gamma\gamma)\simeq 0.1$\,pb} would be compatible with
the~\cms~observation. As mentioned above, the
cross-section~\mbox{$gg\to s^0$} in our analysis is obtained
from~\code{SusHi}~\cite{Harlander:2012pb, Harlander:2016hcx} for
the~$13$\,TeV~\lhc{} at~\nklo{3}{}~\qcd. The renormalization- and
factorization-scale uncertainties amount to about~$\pm 5\%$. Sizable
values for the cross-sections~\mbox{$gg\to s^0$} and~\mbox{$e^+e^-\to
Zs^0$} as well as the branching ratio~\mbox{BR$(s^0\to b\bar b)$}
arise if the admixture of~$s^0$ with the~\sm-like Higgs boson is
sufficiently large. A sizable~\mbox{BR$(s^0\to \gamma\gamma)$} can
occur as a consequence of a significant~$H_u$~component of the singlet
state~$s^0$, whereas a small~$H_d$~component suppresses the decay
into~$b\bar b$. In all the listed scenarios the~\cp-odd singlet-like
Higgs boson~$a_s$ has a mass below~$300\,\GeV$. It should be noted
that the occurrence of the state~$s^0$ at low masses in combination
with a very heavy~$a_s$ state through a large value of~$A_\kappa$
would usually yield a meta-stable (long-lived) vacuum. The listed
scenarios involve a certain amount of tuning in the choice
of~$A_\kappa$ since an increase in~$A_\kappa$ by a few~GeV yields a
tachyonic~$s^0$~state. It is well-known from the~\nmssm{} that a too
large~$A_\kappa$ yields a tachyonic~\cp-even singlet-like Higgs
boson~$s^0$, see
Eq.~(\href{https://arxiv.org/pdf/hep-ph/0304049.pdf#equation.37}{37})
in \citere{Miller:2003ay} or
Eq.~(\href{https://arxiv.org/pdf/0903.3596.pdf#page=7}{26}) in
\citere{Bartl:2009an} for lower and upper bounds on~$A_\kappa$.
Similarly, we have noted a very pronounced dependence of the masses of
both states,~$s^0$ and~$a_s$, on~$A_\kappa$ for the~\munmssm~scenarios
investigated here.

Of course, a large admixture of~$s^0$ with the~\sm-like Higgs boson in
turn has an impact on the~\sm-like Higgs properties, visible through
the increase in~$\chi^2$ returned by~\HS. In fact, from the listed
scenarios only Scenario~3 with~\mbox{$\chi^2=82$} is compatible with
the~\sm-like Higgs boson at the~$95$\%~C.L. The other scenarios
have~$\chi^2$~values outside of the~$95$\%~C.L.~region, as they have a
slightly larger mixing of the singlet state with the~\sm-like Higgs
boson. The enhanced mixing increases the~$s^0$~cross-sections, but on
the other hand yields reduced relative couplings to fermions and gauge
bosons for the~\sm-like Higgs boson~$h^0$. It is thus apparent that
explaining the excesses through a singlet state that only couples
to~\sm~particles through its admixture with the~\sm-like Higgs boson
is under a certain tension from the measured~\sm-like Higgs-boson
properties for both the~\munmssm{} and the~\nmssm, if one requires
signal rates that fully saturate the amount of deviation from
the~\sm{} indicated by the excesses observed by~\lep{} and~\cms.

Scenarios with light singlet-like Higgs bosons tend to have a light
singlino. For Scenario~1 we provide the light electroweakino spectrum,
\IE~the masses of~$\tilde\chi^0_{1,2,3}$ and~$\tilde\chi^\pm_1$, in
Tab.~\ref{tab:97ewinos}. Due to~\mbox{$\mu+\mue=-140$\,GeV} the
scenario has light higgsino-like states, whereas the gauginos are
close in mass to~\mbox{$M_1=239$\,GeV} and~\mbox{$M_2=500$\,GeV}. The
higgsino-like states are strongly admixed with the singlino, \EG~the
singlino-fraction of~$\tilde\chi^0_2$ is~$59$\%, the singlino-fraction
of~$\tilde \chi^0_3$ is~$40$\%. It is apparent that the three lightest
neutralinos and the light chargino are very close to each other in
mass. At the~\lhc,~\atlas{} and~\cms{} have only recently started to
probe such compressed mass spectra by dedicated analyses, see
\EG~\citeres{Aaboud:2017leg,Sirunyan:2018iwl}. In fact, an
electron--positron collider may be required to ultimately probe
scenarios of this kind, see for instance \citere{Berggren:2013vfa}
tackling such compressed higgsino-like scenarios at the International
Linear Collider~(ILC). For Scenario~1 we provide the cross-sections
for the two center-of-mass energies~\mbox{$\sqrt{s}=350$\,GeV}
and~\mbox{$\sqrt{s}=500$\,GeV}, which are considered for Higgs-boson
and top-quark precision studies at
the~ILC~\cite{Moortgat-Picka:2015yla}, in
Tab.~\ref{tab:97ewinos}. Although in this scenario the~LSP is the
gravitino, the lightest neutralino~$\tilde\chi_1^0$ has a lifetime of
a few milliseconds such that it only gives rise to a missing-energy
signature. Besides the possibility to
tag~\mbox{$e^+e^-\to\tilde\chi_1^0\tilde\chi_1^0$} through
initial-state radiation~(ISR), the production of one or more heavier
neutralinos or charginos results in detectable~\sm~particles. The
possibility to polarize the initial state is an important tool to
enhance the signal-to-background ratio, and allows one to minimize
systematic uncertainties. This capability is mandatory for performing
precision measurements. In Tab.~\ref{tab:97ewinos} we provide results
for three different polarizations: an unpolarized initial state (as
reference only), and polarizations of~$\pm 80$\% and~$\mp 30$\% for
the initial-state electron and positron, respectively. Such
polarizations are foreseen in the current baseline design of the~ILC.
As one can see from Tab.~\ref{tab:97ewinos}, polarized beams
with~\mbox{Pol$(e^+,e^-)=(+30\%,-80\%)$}, corresponding to the
so-called effective
polarization~\cite{MoortgatPick:2005cw}~\mbox{Pol$_{\text{eff}}=89\,\%$},
enhance the production cross-sections
of~$\tilde{\chi}^0_1\tilde{\chi}^0_2$
and~$\tilde{\chi}^0_1\tilde{\chi}^0_3$ by about a factor~$1.5$ as well
as the one of~$\tilde{\chi}^+_1\tilde{\chi}^-_1$ by about a
factor~$2$. The fact that the production cross-sections
of~$\tilde\chi^0_2\tilde\chi^0_2$ and~$\tilde\chi^0_2\tilde\chi^0_3$
are significantly smaller than the other quoted cross-sections is due
to a cancellation between the higgsino components~$\tilde h_d^0$
and~$\tilde h_u^0$.

As discussed above, the electroweakino spectrum of the~\munmssm{} is
a~priori indistinguishable from the~\nmssm{} if one restricts the
analysis to information from the electroweakino sector and employs
tree-level predictions, see \sct{sec:higgspheno}. Previous studies of
the electroweakino sector, see
\EG~\citere{Berggren:2013vfa} and
\citeres{MoortgatPick:2005vs,Moortgat-Pick:2014uwa}, discussed
the~ILC~capabilities for distinguishing the~\mssm{} from the~\nmssm{}
electroweakino sector. From such studies one can infer that a
determination of the parameters of the electroweakino sector with an
accuracy at the percent~level is possible using the measurements of
the light electroweakino masses and the corresponding production
cross-sections for different polarizations,
see~\EG~\mbox{\citeres{Choi:2001ww,Desch:2003vw} and references
  therein}. This holds even if only the lightest electroweakinos are
accessible. Based on earlier comprehensive studies where similar rates
as in the scenarios of Tab.~\ref{tab:97ewinos} were considered, the
input parameters of the corresponding sector can be extracted: as an
example, the values of~$M_1$ and~$M_2$ can be determined from the
measurement of light gauginos, or the value of~\mbox{$(\mu+\mue)$}
from the measurement of light higgsinos. In this regard beam
polarization plays a crucial role: it allows one to even resolve
scenarios where only a few light particles are kinematically
accessible. Furthermore, the clean environment at an
electron--positron collider allows the application of
an~ISR~method~\cite{Berggren:2013vfa} to detect and precisely measure
scenarios where the light spectrum is close together in mass, as it is
the case for instance for the compressed electroweakino spectrum in
Scenario~1 leading to very soft decay characteristics. Complementing
the particle spectrum via measuring additional heavier electroweakino
masses and parts of the scalar and colored sector at the~\lhc{} would
allow global fits of the model parameters, so that a model distinction
between the~\munmssm, \nmssm{} and the~\mssm{} might be feasible.

\tocsubsection[\label{sec:decays}]{Higgs-boson mixing and decays}

We now extend our previous discussion on Higgs-boson mixing and
consider Higgs-boson decays. In this context we investigate in
particular the influence of~$\mu$ and~$\mue$ on the masses of the two
light singlets. For our discussion of the possibilities for
distinguishing the~\munmssm{} from the~\nmssm{} we assume that the
sum~\mbox{$\left(\mu+\mue\right)$} is identified with the~$\mue$~term
of the~\nmssm, and~$\kappa$ is rescaled according to
Eq.~\eqref{eq:Lieblerrescaling}. As discussed in \sct{sec:higgspheno},
the rescaling of~$\kappa$ significantly reduces the dependence of all
Higgs masses on~$\mu$ and~$\mue$ over a large region of the parameter
space. Light~\cp-even singlets and decays of the~\sm-like Higgs boson
into them were already part of the discussion
in \figs{fig:example1}--\ref{fig:example4}, but therein we focused on
constraints from vacuum stability and general features in the
parameter space. Now we investigate Higgs-boson decays in more detail.

\begin{figure}[t!]
\centering
\includegraphics[width=.6\textwidth]{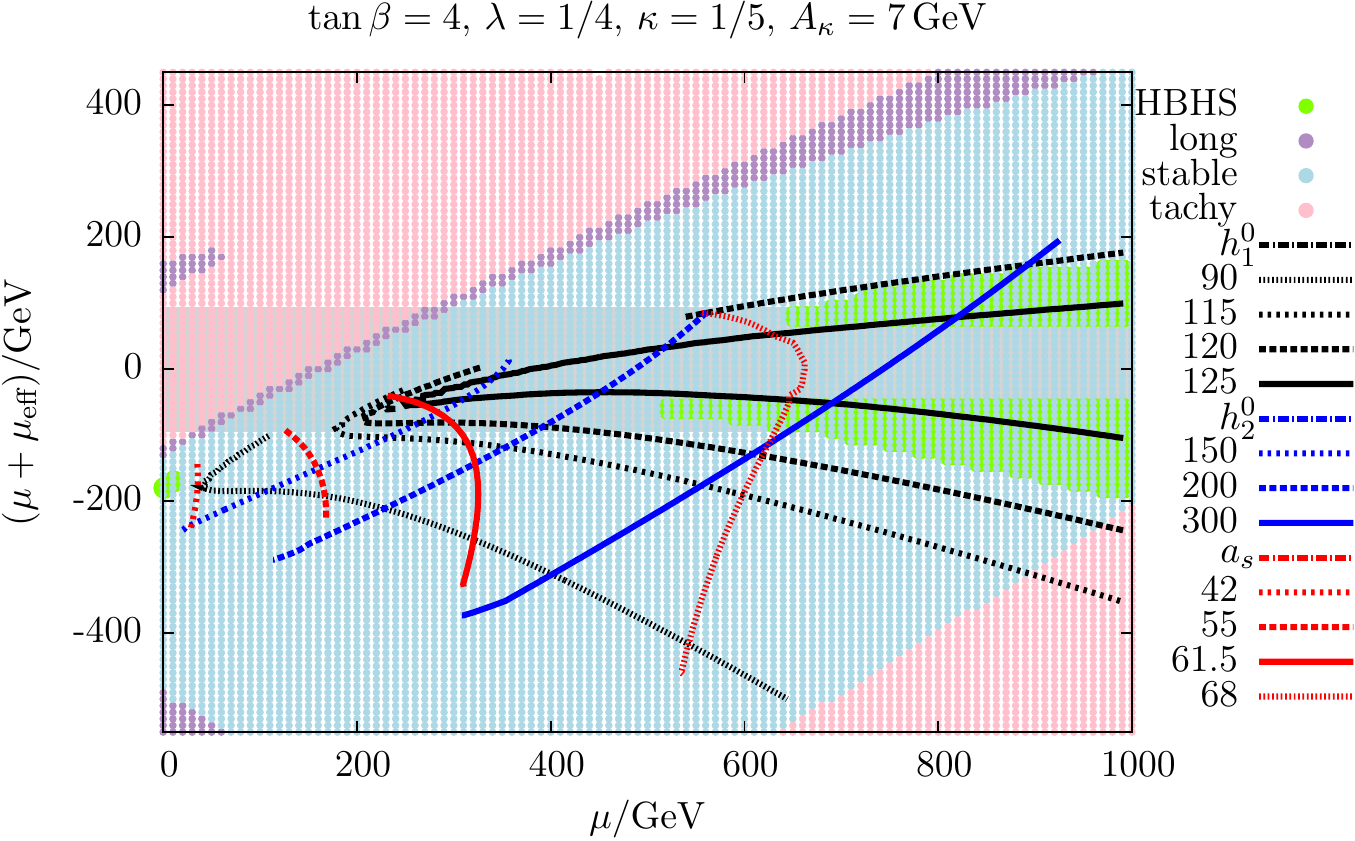}
\caption{\label{fig:examplemumu} The behavior of the masses
of the three lightest neutral Higgs states in a scenario
with~\mbox{$\lambda=1/4$}, \mbox{$\kappa=1/5$}, \mbox{$\tan\beta=4$}
and~\mbox{$A_\kappa=7$\,GeV} is shown. In contrast
to \figs{fig:example1}--\ref{fig:example4}, the black and blue lines
show the mass contours of the lightest and second lightest~\cp-even
Higgs mass eigenstate, respectively. The remaining parameters are
chosen according to Tab.~\ref{tab:param}.}
\end{figure}

For the discussion of the dependence of the masses of the light
singlet states on~$\mu$ and~$\mue$ we choose a scenario based on
Tab.~\ref{tab:param} and fix in addition~\mbox{$\lambda=1/4$},
\mbox{$\kappa=1/5$}, \mbox{$\tan\beta=4$}, \mbox{$A_\kappa=7\,\GeV$}.
We vary~\mbox{$(\mu+\mue)$} between~$-600$ and~$450$\,GeV and~$\mu$
between~$1$ and~$1000$\,GeV. The lower end of the range of~$\mu$
values corresponds to the~\nmssm-limit,~\mbox{$\mu\to 0$\,\GeV}. The
results are depicted in \fig{fig:examplemumu}, where we show the
masses of the three lightest neutral Higgs bosons as a function
of~$\mu$ and~\mbox{$(\mu+\mue)$}. The background colors indicate the
constraints from vacuum stability and from the experimental results on
the Higgs sector using the same color coding as in
Figs.~\ref{fig:example1}--\ref{fig:example4}. It is apparent that with
increasing~$\mu$ the range in~\mbox{$(\mu+\mue)$} that is allowed by
the constraints from vacuum stability is also increasing, which is in
accordance to our observations in the previous section. Since the
mixing between the light~\cp-even doublet state and the~\cp-even
singlet-like state is large in the parts of the displayed parameter
plane where the masses of the two lightest~\cp-even states are close
to each other, we label the~\cp-even states in \fig{fig:examplemumu}
as the mass eigenstates~$h_1^0$ and~$h_2^0$ rather than as~$h^0$
and~$s^0$. However, for large values of~$\mu$ the two
lightest~\cp-even Higgs bosons are sufficiently separated in mass so
that the light~\cp-even doublet state~$h_1^0$ can be identified
with~$h^0$ and exhibits only a mild dependence on~$\mu$, since such a
dependence is only induced through the mixing with the singlet at
tree-level. For smaller values~\mbox{$200\,\GeV\lesssim\mu \lesssim
  300$\,GeV} the contour lines of~\(m_{h_1^0}\) become very dense, and
the mass of the singlet-like state~\(h_2^0\) approaches values
below~\(150\,\GeV\), implying a large mixing between the~\cp-even
light doublet state and the singlet state. With further
decreasing~$\mu$ the mass eigenstates~$h_1^0$ and~$h_2^0$ flip their
role, \IE~$h_1^0$ is singlet-like, and~$h_2^0$ corresponds to
the~\sm-like doublet state in this region.

We now focus on the region of large~$\mu$ where both states can be
clearly separated: the region with a~\sm-like Higgs mass
of~\mbox{$m_{h_1^0}\sim 125\,\GeV$} is strongly affected by the values
of~$\lambda$ and/or~$\tan\beta$ through their impact on
the~\nmssm-like tree-level contribution to the doublet states, see the
quantity~$a'_1$ in \eqn{eq:abbrevprime}. In addition, it is well-known
that this state receives large radiative corrections that depend on
the mass splitting in the stop sector, which is proportional
to~\mbox{\(X_t = A_t - (\mu + \mue) / \tan\beta\)}. As discussed
above, we have chosen~\(A_t\) in such a way that the contribution is
maximized at~\mbox{\(\mu + \mue = 0\)\,GeV} and thus decreases to both
directions. This behavior is visible in \fig{fig:examplemumu} for the
contours displaying the mass of~$h_1^0$ at values of~\mbox{$\mu\gtrsim
  300$\,GeV}. The behavior of the singlet-like states is important for
the phenomenology: as explained in \sct{sec:potential}, for a scenario
with~\mbox{$\kappa\sim\lambda$} the mass of the~\cp-even singlet~$s^0$
is mainly controlled by~$a_5$, see Eq.~\eqref{eq:abbrev_a5}, and
therefore proportional to~$\mue$. As in our scenario~$A_\kappa$ is
small, the mass of the~\cp-odd singlet~$a_s$ is dominated
by~$a_4^\prime$, see Eq.~\eqref{eq:a4prime}, and therefore
proportional to~$\sqrt{\lvert\mu/\mue\rvert}$. Those mass dependences
of the singlet-like~\mbox{$m_{h_2^0} \gtrsim 150\, \GeV$}~(blue)
and~$m_{a_s}$~(red) can be clearly identified in
\fig{fig:examplemumu}: for~\makebox(20,-0.5){$m_{h_2^0}$}, the lines are roughly
diagonal, and thus the mass contours follow lines with
constant~$\mue$. For the mass of~$a_s$, the dependence on the square
root~$\sqrt{\lvert\mu/\mue\rvert}$ leads to the shape of the
contours. We emphasize that the behavior displayed in
\fig{fig:examplemumu} is specific to a small value
of~\mbox{$A_\kappa\ll \mu$} or~$\mue$ for~\mbox{$\kappa\sim\lambda$}.

As above, the parameter range allowed by constraints from~\HB{}
and~\HS{} is indicated by light-green dots in the background. For
values of~\mbox{$\mu\gtrsim 300$\,GeV}, the light~\cp-even
Higgs~$h_1^0$ corresponds to the~\sm-like state. In this region, the
decay~$h_1^0$ into higgsinos forbids low values
of~\mbox{$\lvert\mu+\mue\rvert$}, while the decay~\mbox{$h_1^0\to
  a_sa_s$} is kinematically closed. For~\mbox{$\mu\lesssim500$\,\GeV}
the mixing between the~\cp-even doublet state and the singlet state
becomes larger which is not compatible with the observation of the
properties of the~\sm-like Higgs. The minimal value of~\(\chi^2\) in
this figure is~\mbox{$\chi_m^2 = 76.4$} and thus slightly worse than
the scenarios studied in Section~\ref{sec:numerics}. There is also a
small allowed region with~\mbox{$\Delta \chi^2<5.99$} at low values
of~$\mu$, where a~\sm-like Higgs boson is present. In this region the
mass of the singlet state~$s^0$ has crossed the mass of the doublet
state~$h^0$, and the states~\(h_1^0\) and~\(h_2^0\) have changed their
character as discussed previously. The doublet--singlet mixing in this
case yields a positive contribution to the mass of the~\sm-like
state~$h_2^0$, lifting the tree-level value towards the experimentally
allowed mass window (in the allowed region at low values of~$\mu$ the
mass of~$h_2^0$ is about~$126\,\GeV$). In this region the
decay~\mbox{$h_2^0\to a_sa_s$} is kinematically open, but sufficiently
suppressed to be in accordance with experimental observations.

We conclude that the additional~$\mu$~term of the~\munmssm{} lifts up
the~\cp-odd Higgs mass and enlarges the allowed parameter space
compared to the~\nmssm. Still, in particular due to the large
admixture of the singlet and doublet states, such a scenario is
difficult to distinguish from the standard~\nmssm, if not all Higgs
states are fully determined. As a consequence of the strong admixture
of the Higgs bosons and the influence of their masses on the
kinematics, all decay modes show a non-trivial dependence on the
coupling structure. The decay rates of the heavy Higgs bosons~$H^0$
and~$A^0$ into any combination of the three light Higgs bosons remain
small throughout the parameter plane, \IE~the branching ratios are
below~$3\%$. The maximal branching ratios for~\mbox{$A^0\to h^0 a_s$}
and~\mbox{$A^0\to s^0 a_s$} are reached at large~$\mu$
and~$\lvert\mue\rvert$, \IE~in the lower right corner of
\fig{fig:examplemumu}. The two decays show a different dependence
on~$\mu$ and~$\mue$, which is in accordance with our discussion in
\sct{sec:selfcoupling}. Whereas~\mbox{$h_2^0\to a_s a_s$} is
kinematically only allowed for very low~$\mu$ in this scenario, the
decay~\mbox{$h_2^0\to h_1^0h_1^0$} is---when kinematically
open---strongly dependent on~$\mue$. We will demonstrate below the
dependence of the different decay modes on~$\mu$ and~$\mue$ in a
scenario with essentially fixed Higgs-boson masses.

\begin{figure}[tp!]
\centering
\includegraphics[width=.49\textwidth]{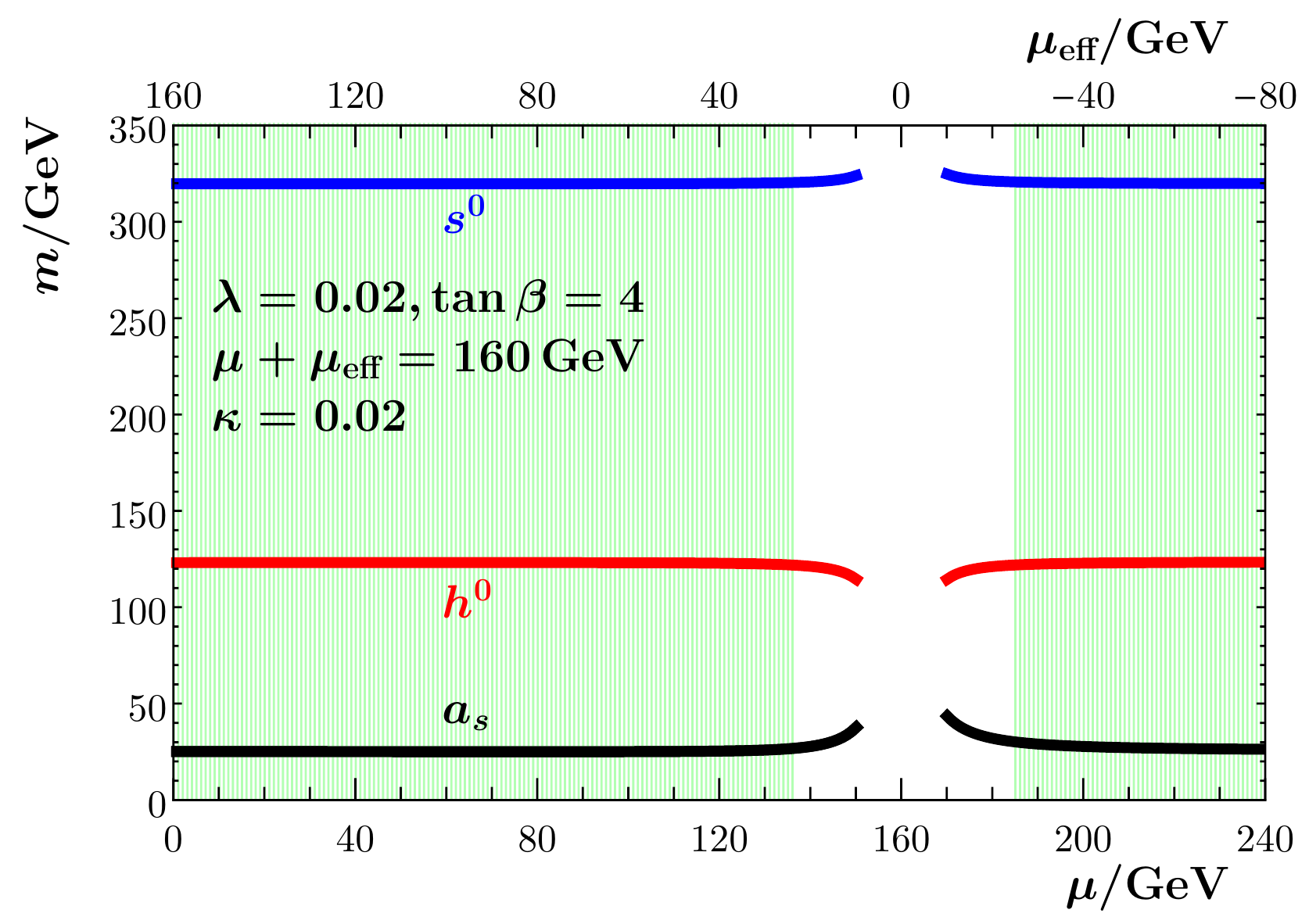}
\hfill
\includegraphics[width=.49\textwidth]{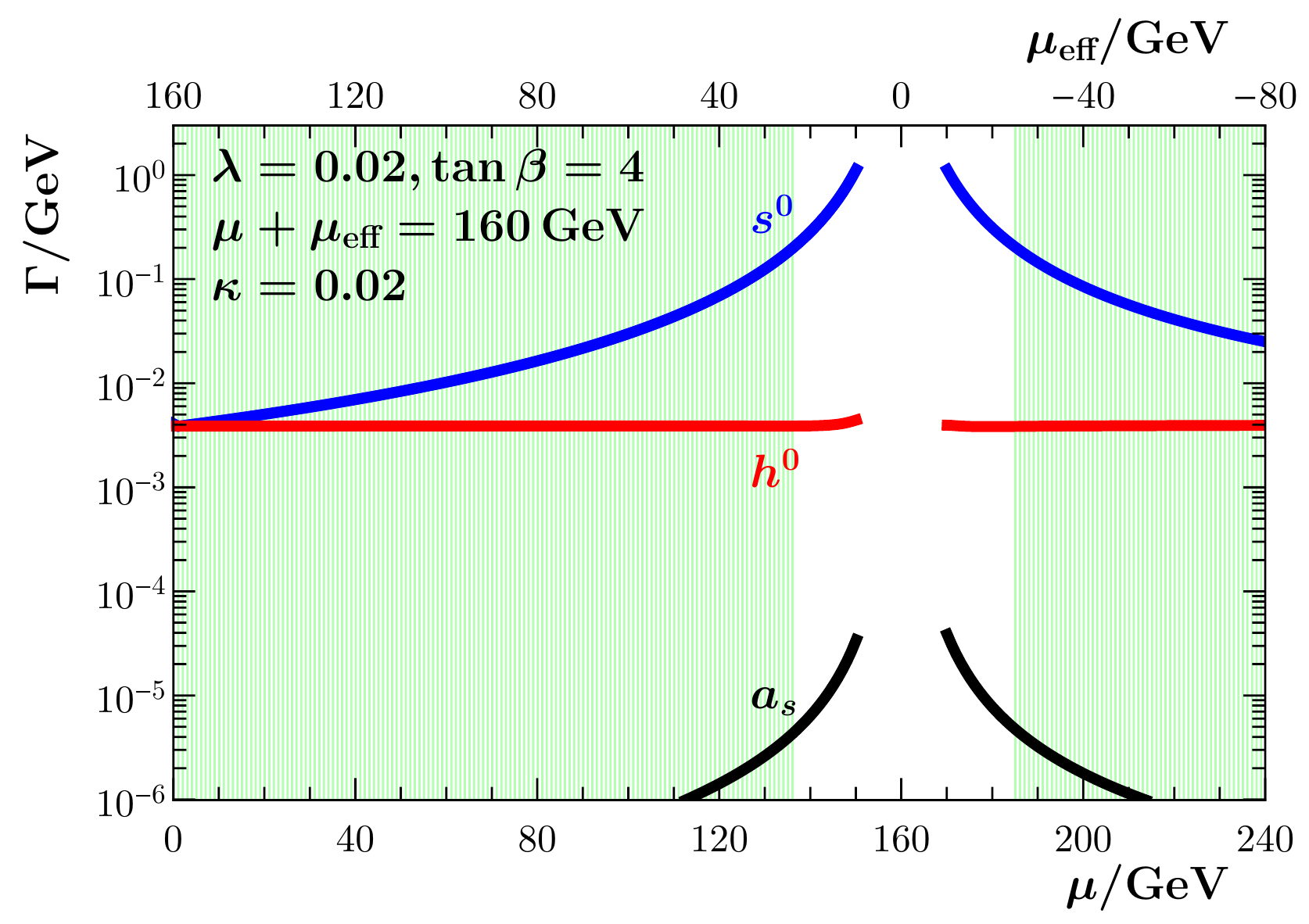}
\caption{\label{fig:example5a}The mass spectrum~(left) and total decay
  widths~(right) for the three lightest Higgs bosons in a scenario
  with small~\mbox{$\lambda=0.02$} and
  fixed~\mbox{$\mu+\mue=160$\,GeV} are shown. The value of~\(\kappa\)
  is rescaled as~\mbox{\(\kappa \to \tilde \kappa\)} according
  to \eqn{eq:Lieblerrescaling} with~\mbox{\(\kappa =
  0.02\)}. Furthermore,~\mbox{\(\lvert A_\kappa\rvert = 1.3\,\GeV\)}
  with the opposite sign of~\((\mue\,\tilde{\kappa})\). The other
  parameters are given in Tab.~\ref{tab:param}. The green area
  indicates compatibility with the constraints from~\HS.}

\vspace{1.5ex} \hrule \vspace{1ex} \capstart

\includegraphics[width=.49\textwidth]{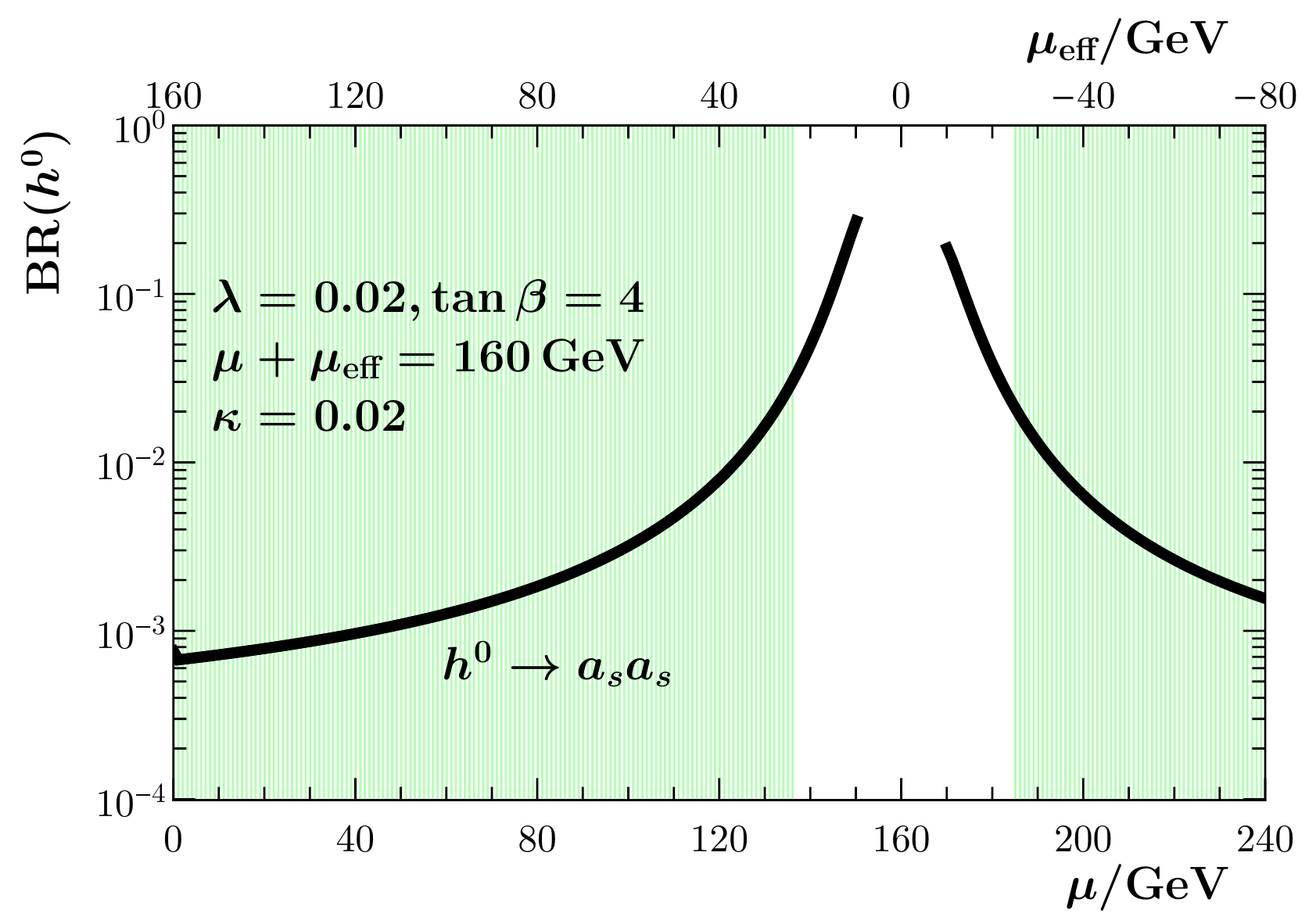}
\hfill
\includegraphics[width=.49\textwidth]{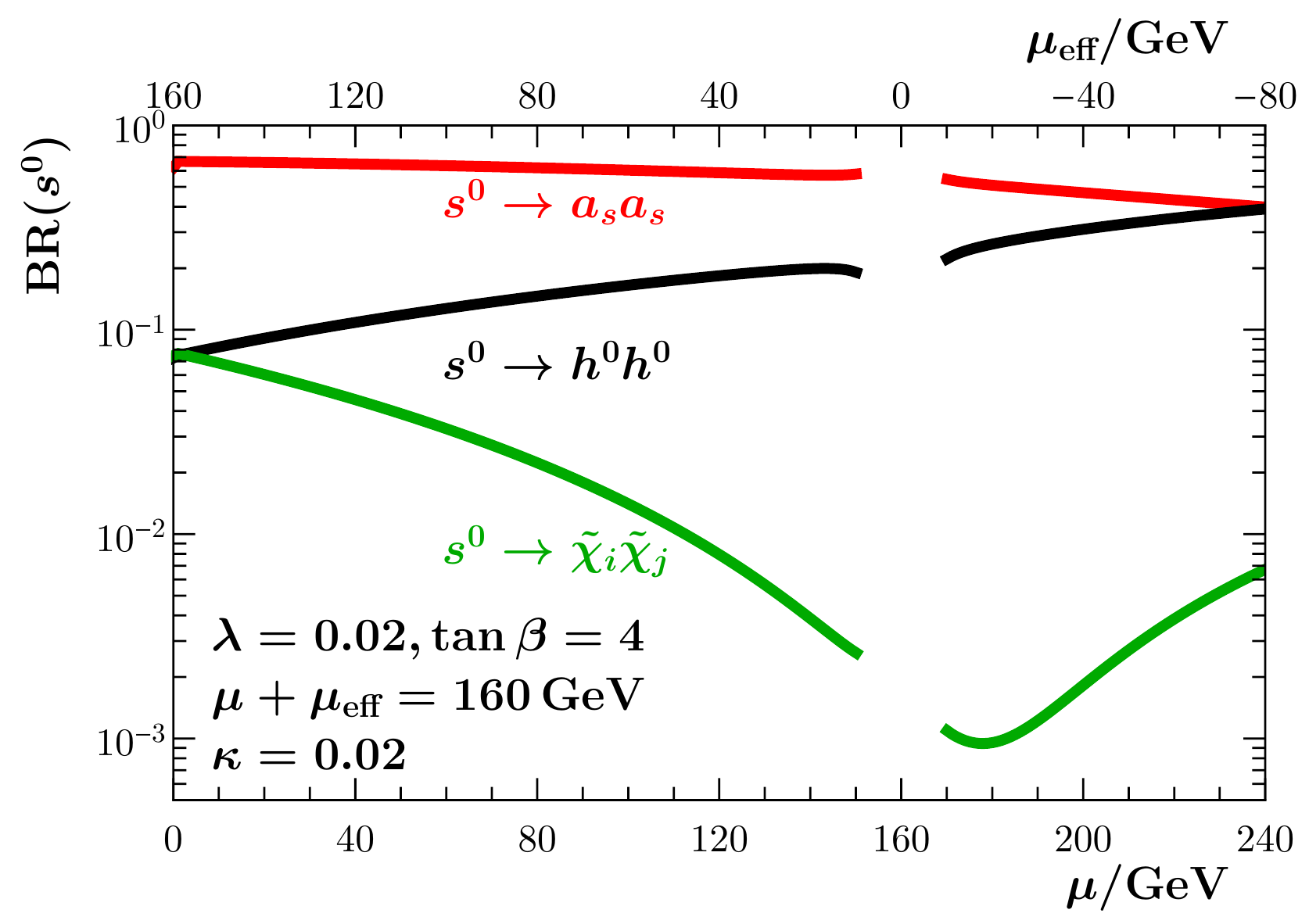}
\caption{\label{fig:example5b}(left) Branching ratio of~$h^0$ into a
  pair of light~\cp-odd singlets; (right) Branching ratios of~$s^0$
  into non-\sm~particles and Higgs bosons; both for the same scenario
  as in \fig{fig:example5a}. The green area indicates compatibility
  with the constraints from~\HS. The branching ratio~\mbox{BR$(s^0\to
    \tilde\chi_i\tilde\chi_j)$} includes all branching ratios of~$s^0$
  into pairs of neutralinos and charginos.}

\vspace{1.5ex} \hrule \vspace{1ex} \capstart

\includegraphics[width=.49\textwidth]{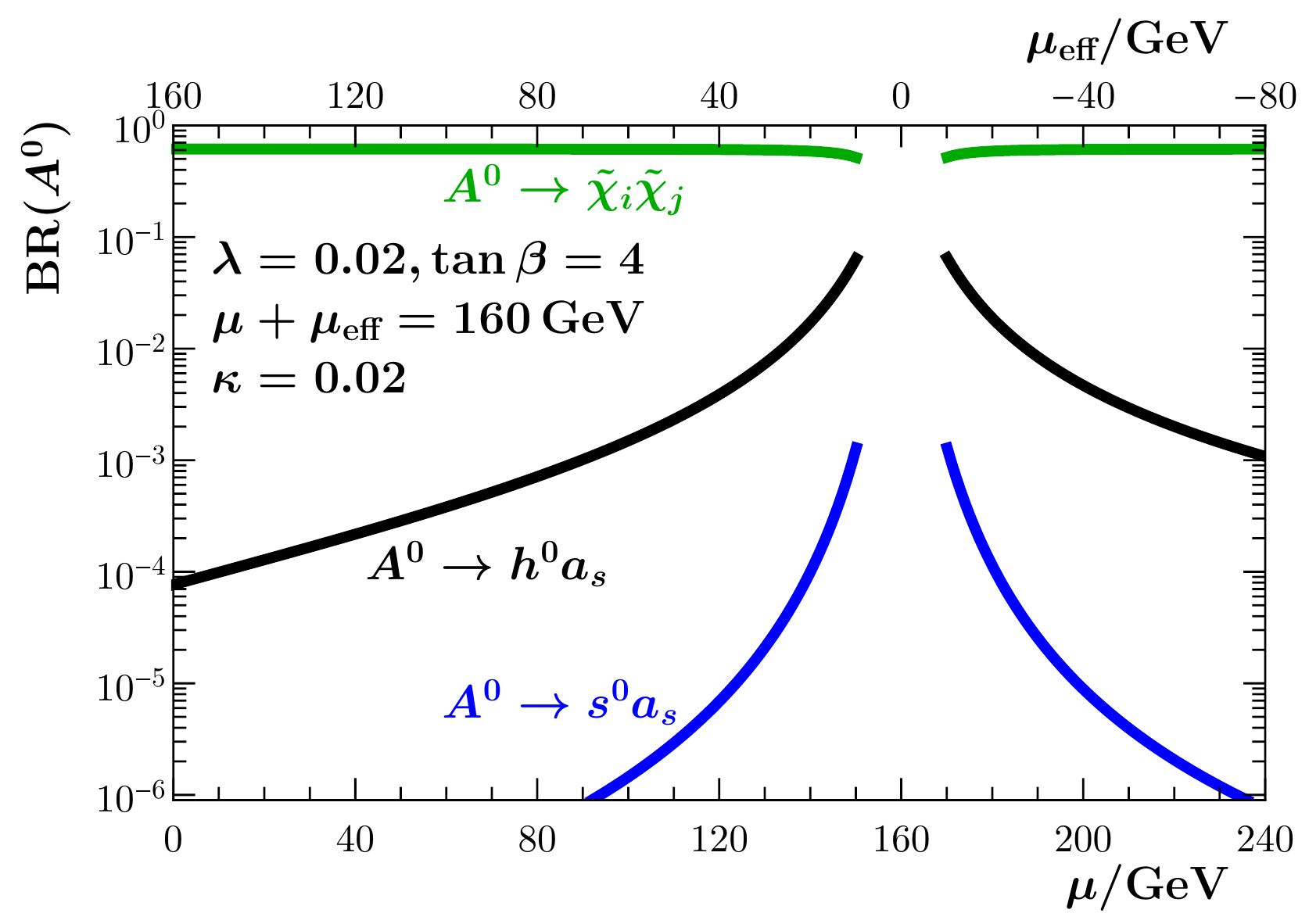}
\hfill
\includegraphics[width=.49\textwidth]{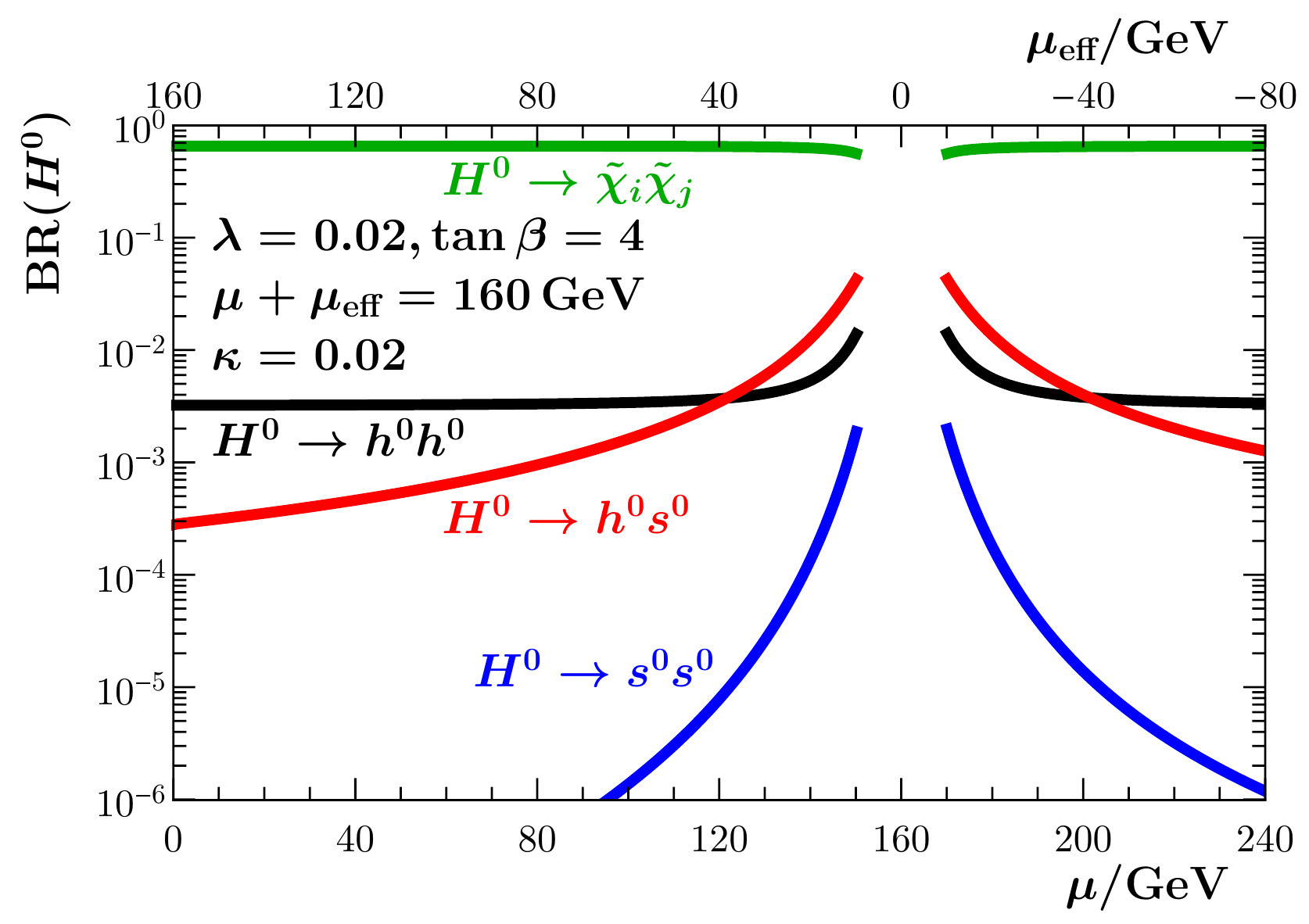}
\caption{\label{fig:example5c} Branching ratios of~$A^0$~(left)
  and~$H^0$~(right) into pairs of lighter Higgs bosons; both for the
  same scenario as in \fig{fig:example5a}. The branching
  ratios~\mbox{BR$(A^0\to \tilde\chi_i\tilde\chi_j)$}
  and~\mbox{BR$(H^0\to \tilde\chi_i\tilde\chi_j)$} include all
  kinematically allowed channels into pairs of neutralinos and
  charginos.}
\end{figure}

We now discuss a scenario that is intrinsically different from
the~\nmssm{} and shows a peculiar dependence of Higgs mixing and thus
Higgs-boson decays on~$\mu$ and~$\mue$. As indicated in the third item
of \sct{sec:parameters} the~\munmssm{} allows large values of~$\kappa$
in combination with low values of~$\mue$ and~$\lambda$ without being
constrained by higgsino-like states. In the following we vary~$\mu$
from~$0$~to~$240$\,GeV and fix~\mbox{$\mu+\mue=160$\,GeV} and thus
simultaneously reduce~$\mue$ from~$160$~to~$-80$\,GeV. We choose a
very small value of~\mbox{$\lambda=0.02$} and a value
of~\mbox{$\kappa=0.02$}, which we rescale as~\mbox{\(\kappa \to \tilde
  \kappa\)} according to \eqn{eq:Lieblerrescaling}. The Higgs bosons
therefore stay almost constant in mass, such that differences in
Higgs-boson decays are solely induced by differences in the mixing
among the Higgs bosons and not by kinematics. Note that in the
limit~\mbox{$\mue\to 0$\,\GeV} the rescaled parameter~$\tilde\kappa$
gets pushed beyond the perturbativity limit. This and the fact that at
the tree-level one scalar mass becomes tachyonic are the reasons why
the region around~\mbox{\(\mue = 0\,\GeV\)} is omitted for the lines
in \figs{fig:example5a}--\ref{fig:example5c}. Besides the parameters
in Tab.~\ref{tab:param} we set~\mbox{$\tan\beta=4$},
\mbox{$A_\kappa=-\sign{(\mue\,\tilde{\kappa})}\,1.3$\,GeV},
which is like in \fig{fig:examplemumu} small compared to~$\mu$
and~$\lvert\mue\rvert$, and~\mbox{$B_\mu=0$\,GeV}. Note that the
rescaling procedure for~\(\kappa\) according
to \eqn{eq:Lieblerrescaling} turns~\(\tilde\kappa\) negative in this
scenario, when~\(\mu+\mue > 0\) is fixed and~\(\mu\) takes on values
larger than~\mbox{$(\mu+\mue)$}. This case might be unsuitable for
inflation, see \citere{Ferrara:2010yw}. The opposite
case~\mbox{$\mue>0$} and~\mbox{$\mu+\mue<0$} cannot appear in our
model, since~$\mu$ is always positive. The green-shaded area, which we
show in the figures depicting \sm-like Higgs properties, indicates the
region that is compatible with the constraints from \HS, where as
before we demand~$\Delta\chi^2<5.99$ with a minimum
of~$\chi_m^2=77.5$. For~$\mue\in[-22,22]\,\GeV$ the decay~$h^0\to
a_sa_s$ is enhanced such that this region is not compatible with the
allowed fraction of non-\sm{}~decays of the~\sm{}-like Higgs boson.
It should be noted that all of the shown area is allowed by the
constraints from \HB.  We emphasize that the effects that will be
discussed in the following are related to a small value of~$\mue$:
\IE~in the scenario discussed in \fig{fig:Higgsspecrescale}, despite
the same rescaling procedure for~$\kappa$ with respect to \(\mue\),
the mixing among the Higgs states is much less influenced by the
choice of~$\mu$, since a large value of~$\mu$ results in an even
larger negative value of~$\mue$.

In the standard~\nmssm{} a measurement of the masses of the whole
neutralino and neutral Higgs spectrum would fix all free parameters,
in particular~$\mue$, $\lambda$, $\kappa$ and~$A_\kappa$. With these
parameters also the Higgs mixing is completely determined (at the tree
level). A small value of~$\lambda$ in any case implies a small mixing
between the singlet and doublet states of the Higgs sector. This is
not the case in the~\munmssm: we show our results in
\figs{fig:example5a}--\ref{fig:example5c}. As explained above, in the
considered parameter region the Higgs-boson masses are almost
constant, see \fig{fig:example5a} on the left-hand side. The two heavy
Higgs bosons~$H^0$ and~$A^0$ both have a mass very close to~$800$\,GeV
within a range of~$3$\,GeV. The neutralino masses are constant, in
detail~\mbox{$m_{\tilde\chi_i^0}=\{134.7,163.9,252.1,320.0,516.1\}\,\GeV$},
where the particle with mass~\mbox{$m_{\tilde\chi_3^0}=m_{\tilde
    s}=320$\,GeV} corresponds to the singlino-like state with a purity
of~$99.9$\%. The two lightest neutralinos are higgsino-like states.
Though the mixing in the neutralino sector remains constant, the
mixing between the light~\cp-even Higgs boson~$h^0$ and the singlet
component~$s^0$ is strongly enhanced for~\mbox{$\mue\to 0$\,GeV}. We
depict the total widths for the three lightest Higgs bosons on the
right-hand side of \fig{fig:example5a}. The enhancement of the total
width of~$h^0$ for~\mbox{$\mue\to 0$\,GeV} is due to the
decay~\mbox{$h^0\to a_sa_s$}. This is also apparent in the left plot
of \fig{fig:example5b}, where the branching ratio for the decay of
the~\sm-like state~$h^0$ into a pair of light~\cp-odd singlets is
displayed. For~$s^0$ both the decays into~$h^0h^0$ and~$a_sa_s$ are of
relevance, whereas other non-standard decay modes---\EG~into a pair of
higgsinos---have a small rate, see the right-hand sides of
\figs{fig:example5a} and \ref{fig:example5b}. Apart from decays into
Higgs bosons,~$s^0$~decays into massive~\sm~gauge~bosons. As mentioned
above, all of the shown area in \fig{fig:example5b} is allowed by the
constraints from~\HB, \IE~both~$a_s$ and~$s^0$ are compatible with
searches for additional Higgs bosons. However, for the state~$s^0$ the
region around~\mbox{$\mue=\pm 10$\,GeV} is close to the boundary of
the region that is excluded by the limits from Higgs searches, see
below. As a result, we conclude that in this scenario with
small~$\mue$ the singlet~$s^0$ can again be directly produced at a
hadron collider through its admixture with the two~\cp-even doublets,
see the discussion in
\sct{sec:production}. For~\mbox{$\mu=\{150,170\}$\,GeV} the mass of
the singlet is~\mbox{$m_{s^0}=\{323.9,324.8\}$\,GeV}, and the
gluon-fusion production cross-section is~\mbox{$\sigma(gg\to
s^0)=\{270,274\}$\,fb}. The production rates through bottom-quark
annihilation is negligible. Given the large branching
ratios~\mbox{BR($s^0\to a_sa_s)\sim 57\%$}, \mbox{BR($s^0\to
h^0h^0)\sim 19\%$}, \mbox{BR($s^0\to W^+W^-)\sim 15\%$}
and~\mbox{BR($s^0\to ZZ)\sim 7\%$}, the most sensitive searches are
those with a decay into a pair of~\sm-like Higgs or gauge bosons. As
an example, for~\mbox{$m_X\sim 320$\,GeV} the upper
limits~\mbox{$\sigma(pp\to X\to h^0h^0)\lesssim
500$\,fb}~\cite{CMS:2018obr} and~\mbox{$\sigma(pp\to X\to ZZ)\lesssim
200$\,fb}~\cite{Sirunyan:2018qlb} are already within a factor~$10$ of
the signal rates that can be obtained
at~\mbox{$\mu=\{150,170\}$\,GeV}. Lastly, also the decays of the heavy
Higgs bosons---whose total decay widths only vary within~$10$\% for
the considered scenario---show potentially observable branching ratios
into pairs of lighter Higgs bosons in the limit~\mbox{$\mue\to
0$\,GeV}, see
\fig{fig:example5c}. At such low values of~$\tan\beta$ both heavy
Higgs bosons~$H^0$ and~$A^0$ are not predominantly decaying into
bottom~quarks or tau~leptons, but decay into a pair of top~quarks with
a branching ratio of about~$30\%$. Thus, decay modes into Higgs bosons
could actually serve as discovery modes. However, note that our
scenario includes light electroweakinos, into which heavy Higgs bosons
tend to decay with large branching fractions. The branching
ratios~\mbox{BR$(A^0\to \tilde\chi_i\tilde\chi_j)$}
and~\mbox{BR$(H^0\to \tilde\chi_i\tilde\chi_j)$}, shown in
\fig{fig:example5c}, both exceed~$60$\% except for small values
of~$\mue$. Both branching ratios include all kinematically allowed
decays into pairs of neutralinos and charginos. This adds to the
motivation for dedicated searches for heavy Higgs bosons decaying
either into a pair of lighter Higgs bosons or into supersymmetric
particles, see also the discussion in
\citeres{Carena:2013ytb,Bahl:2018zmf}.

\enlargethispage{2ex}
We conclude that a small value of~$\mue$ in the discussed scenario
strongly enhances the mixing among the Higgs bosons despite a low
value of~$\lambda$, which makes both singlet states potentially
accessible at colliders. We have demonstrated that the Higgs-boson
decays are not only controlled through the self-coupling dependences
given in \sct{sec:selfcoupling} for gauge eigenstates, but are also
strongly dependent on the mixing of the Higgs bosons. In the
standard~\nmssm, light singlet states are usually associated
with~\mbox{$\kappa<\lambda$}, since the limits from chargino searches
at~\lep{} imply~\mbox{$\lvert\mue\rvert\gtrsim 120$\,GeV}, and
therefore~\mbox{$v_s\gg 120$\,GeV}. Accordingly, only
small~\mbox{$\kappa \ll \lambda<1$} results in two light singlet
states. However, in the~\munmssm~scenario that we have
considered~$\kappa/\lambda$ and~$\mu$ can be large in combination with
a small~$\mue$. Whereas the~\cp-odd singlet~$a_s$ can be as light as
a~few~GeV, the~\cp{}-even singlet~$s^0$ is usually in the ballpark of
a~few~hundred~GeV in such scenarios. This scenario is intrinsically
different from the behavior of Higgs masses and mixing known in
the~\nmssm.

\begin{figure}[t!]
\centering
\includegraphics[width=.6\textwidth]{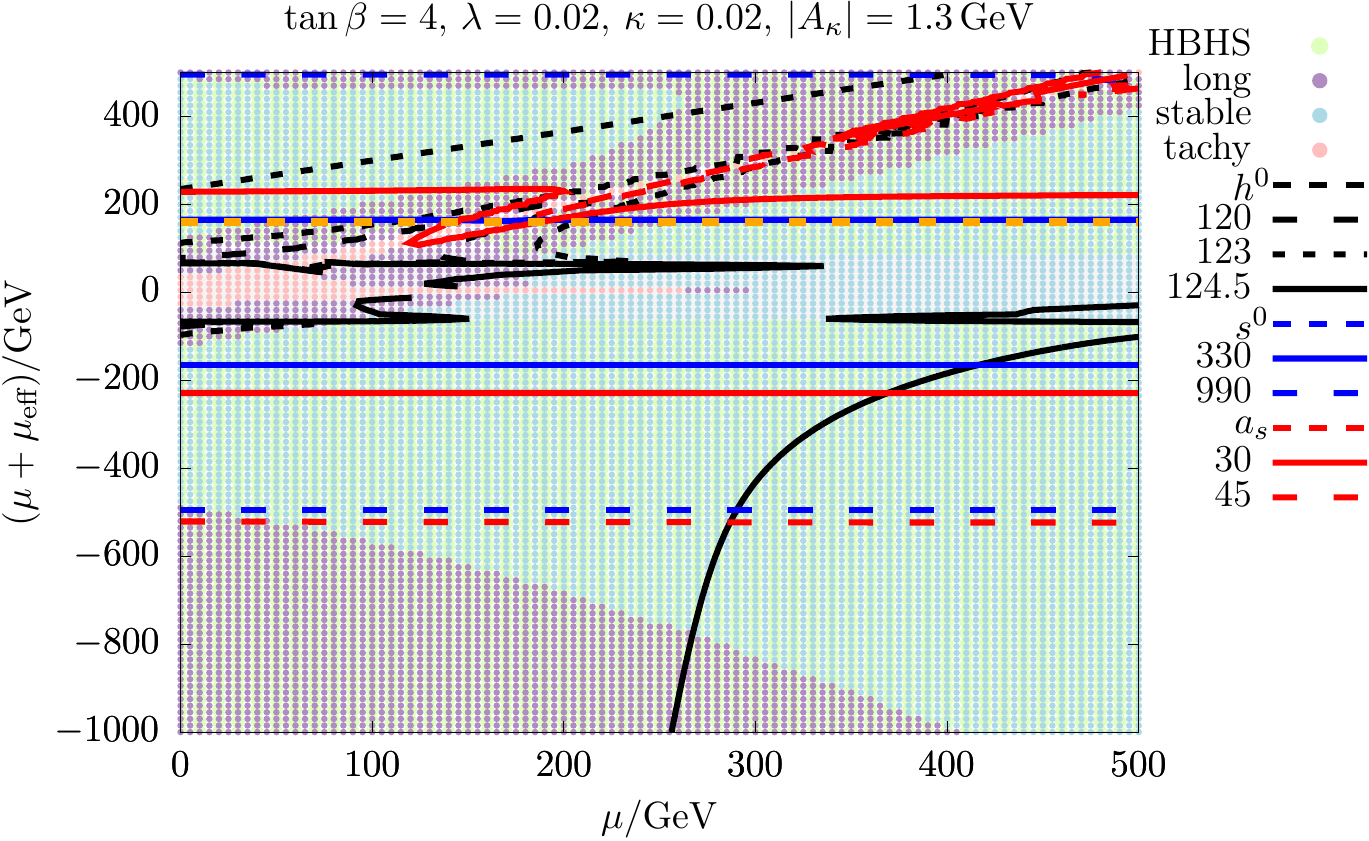}
\caption{\label{fig:smallla} 
Vacuum stability analysis for a scenario with small value
of~$\lambda$, but potentially large values of~$\tilde\kappa$ along the
diagonal line of~\mbox{$\mue\approx 0$}\,GeV. The masses of
the \sm-like Higgs boson and the singlet-like \cp-even and -odd states
are indicated in black, blue and red, respectively. The green area is
allowed by \HB{} and \HS{}; with the gray band, the direct exclusion
bound on light charginos is shown. Superimposed with orange dashes is
the line of constant~\mbox{\(\mu+\mue = 160\,\GeV\)}, along
which \figs{fig:example5a}--\ref{fig:example5c} are defined.}
\end{figure}

We are left with a discussion of vacuum stability in this scenario of
large values of~$\kappa$ together with small values of~$\lambda$
and~$\mue$, for which we consider a wider range of parameters, \IE~we
allow for different values of~\mbox{$\left(\mu+\mue\right)$}. Our
results are shown in \fig{fig:smallla}, where we varied both~$\mu$
and~\mbox{$\left(\mu+\mue\right)$} similar to \fig{fig:examplemumu}
but for small~\mbox{\(\lambda = 0.02\)} and initial~\mbox{\(\kappa
= \lambda\)}. In contrast to \fig{fig:examplemumu} we
rescale~\(\kappa\) according to \eqn{eq:Lieblerrescaling} in order to
achieve a flat neutralino spectrum. Except at vanishing values
of~$\mue$, which correspond to a diagonal line
from~\mbox{$(\mu,\mu+\mue)=(0,0)\,\GeV$}
to~\mbox{$(\mu,\mu+\mue)=(500,500)\,\GeV$}, also the Higgs spectrum is
almost constant throughout the plane due to the rescaling
of~$\kappa$. The behaviour discussed
in \figs{fig:example5a}--\ref{fig:example5c} appears along the
horizontal (orange, dashed) line at~\mbox{$\mu+\mue=160\,\GeV$}
indicated in \fig{fig:smallla}. Very similar results to the ones
described in \figs{fig:example5a}--\ref{fig:example5c} are obtained
for smaller or larger values of~\mbox{$\mu+\mue>0\,\GeV$} close
to~\mbox{$\mue\approx 0\,\GeV$}. \fig{fig:smallla} demonstrates that a
large fraction of the~\mbox{$(\mu,\mu+\mue)$}-plane yields a stable
vacuum. The vacuum appears to be unstable, but long-lived, in a valley
around the diagonal line of~\mbox{$\mue\approx 0\,\GeV$} discussed
above as well as the lower left corner of the plot and
along~\mbox{\(\mu+\mue \simeq 500\,\GeV\)}. Along the same line and
for small values of~\mbox{$\mu+\mue$} tachyonic states rule out the
parameter space. Those two bands, \IE~\mbox{$\mue\approx 0\,\GeV$}
and~\mbox{$\mu+\mue\approx 0\,\GeV$}, are also the only regions in
parameter space, which do not include a \sm-like Higgs boson
compatible with experimental measurements, indicated by the green area
obtained with \HB{} and \HS. Two more comments are on order: In the
upper right triangle
of \fig{fig:smallla}, \IE~\mbox{$\mu+\mue>0\,\GeV$}
and~\mbox{$\mue<0\,\GeV$},~$\tilde\kappa$ is negative, which might not
be compatible with inflation.  Moreover we want to point out that a
very similar scenario can be found for even smaller~\(\lambda\), as
long as the initial value~\mbox{$\kappa\approx\lambda$} is kept.

\tocsection[\label{sec:conclusions}]{Conclusions}

We have analyzed the phenomenology at the electroweak scale of an
inflation-inspired extension of the Next-to-Minimal Supersymmetric
Standard Model~(\nmssm{}). We have put special emphasis on the spectra
of additional, non-\sm-like Higgs bosons and the branching ratios of
their decays. This model has the same field content as the~\nmssm, but
at early times in the universe the~\(D\)-flat direction of the Higgs
doublet plays the role of the inflaton. Such a model can successfully
describe inflation without the need of introducing a dedicated
inflaton field. The singlet superfield~\(\hat S\) of the~\nmssm{} is
needed to stabilize the inflationary direction at the origin
of~\mbox{\(\hat S = 0\)}. Inflation occurs due to a non-minimal
coupling of the doublet Higgs fields to
gravity~\mbox{\(\simord \chi\,H_u \cdot H_d\)}, where the
proportionality factor involves the gravitino mass~\(m_{3/2}\) at low
energies. Thus, this model is characterized by
an~\mbox{\mssm-like~\(\mu\)~term}, which is generated from the
coupling~\(\chi\) and involves~\(m_{3/2}\), in addition to the usual
effective~$\mue$~term of the~\nmssm. The latter arises since the
scalar component of the singlet superfield acquires a vacuum
expectation value as in the~\nmssm. At low energies, \IE~the
electroweak scale, this model differs from the~\nmssm{} by the
additional~\(\mu\)~term which breaks the
accidental~\(\mathbb{Z}_3\)~symmetry of the~\nmssm; we denote this
model as the~\munmssm. The higgsino-mass term in the~\munmssm{} is
composed of the sum~\mbox{\((\mu + \mue)\)}. We have classified and
discussed various scenarios regarding the prospects to distinguish
the~\munmssm{} from the~\nmssm{}, where the latter corresponds to the
limit~\mbox{\(\mu = 0\)\,GeV} of the~\munmssm. We have derived
constraints on the model parameters from theoretical and
phenomenological considerations. For that purpose, we have computed
the~\sm-like Higgs mass at the one-loop order in the~\munmssm{} and
added as approximation at the two-loop level the known two-loop
results from the~\mssm{} which are implemented in~\FH. We have probed
our scenarios against the rate measurements of the~\sm-like Higgs
boson and the limits from searches for additional Higgs bosons at
colliders with the codes~\HB{} and~\HS. Furthermore, we have checked
whether the electroweak ground state of the Higgs potential
corresponds to the absolute minimum of the theory, \IE~the true
vacuum, or whether the Higgs potential has a deeper non-standard
minimum such that the electroweak vacuum eventually decays. In the
inflationary scenario considered here, configurations with a
meta-stable electroweak vacuum in general do not yield a viable
phenomenology. In fact, the most stringent constraints arise from the
possible appearance of tachyonic Higgs states at the tree level.

The additional freedom of varying~$\mu$ and~$B_\mu$ in the~\munmssm{}
allows one to choose values for the parameters of the~\nmssm{} which
would otherwise be excluded. In this extended parameter space, we have
focused on relatively small values of~$\tan\beta$, since in this
case---like in the~\nmssm---the light doublet-like Higgs mass squared
is increased by a shift~${\propto}\,\lambda^2\,v^2$; in this way the
loop corrections which are required in order to acquire a~\sm-like
Higgs at~$125$\,GeV can be smaller.  As expected, in particular the
requirement of a~\sm-like Higgs boson at about $125$\,GeV yields
important constraints on the parameter space.  Concerning the
constraints from vacuum stability, we find that the region with a
phenomenologically viable Higgs spectrum is strongly correlated with
the region of a stable electroweak vacuum, where the electroweak
ground state corresponds to the true vacuum at the electroweak
scale. An exception is the case where the soft
\susy{}-breaking~\(B_\mu\,\mu\)~term is large. We have demonstrated
that large negative values of~\(B_\mu\,\mu\) destabilize the vacuum.

For most of the numerical analyses in this paper we have fixed the
sum~\mbox{$(\mu+\mue)$}, since~$\mu$ enters at the tree level only in
this combination in the mass matrices for the charginos and sfermions
as well as in the~\mssm-like part of the neutralino mass matrix.
Accordingly, the particle spectrum of the~\munmssm{} in those sectors
resembles the one of the~\nmssm{} if the sum~\mbox{$(\mu+\mue)$} in
the~\munmssm{} is identified with the~$\mue$~term of the~\nmssm.
Moreover, we have pointed out the possibility to further reduce the
influence of the non-minimal coupling to supergravity~\(\simord \mu\)
on the neutralino sector by a rescaling of the
parameter~\(\kappa\). This rescaling compensates the dependence of the
singlino component of the neutralino mass matrix on~$\mue$, so that
the neutralino, chargino and sfermion sectors of the~\munmssm{} and
the~\nmssm{} become indistinguishable from each other at tree
level. We have demonstrated that the dependence of the Higgs masses
on~$\mu$ is significantly weakened after this transformation, but the
individual dependences on~$\mu$ and~$\mue$ still have a large impact
on the Higgs mixing and thus the branching ratios of Higgs decays. The
modified value of~\(\kappa\) resulting from the rescaling can also
have an important influence on Higgs phenomenology.

Since with the above parameter settings the neutralino sector of
the~\munmssm{} is~\nmssm-like, we have not performed a detailed
numerical analysis of the neutralino sector---besides our discussion
of Higgs decays into electroweakinos.  In general, the gravitino is
found to be the~LSP since it is tightly connected to the size
of~\(\mu\). For phenomenological reasons in our scenarios it typically
has a mass of~\(\mathcal{O}(10\,\MeV)\). The~NLSP, which is either
singlino- or bino-like, tends to be sufficiently long-lived such that
it only gives rise to missing-energy signatures in collider searches.
Accordingly, typical constraints from SUSY~searches including missing
energy apply without large modifications.  The character of the~NLSP
is influenced by a variation of the corresponding parameters,
\IE~\((\mu+\mue)\) for the higgsino mass and~\(M_1\) or~\(M_2\) for
the bino or wino mass, respectively. Our choices for~\(M_{1,2}\) are
rather arbitrary in this context. Their impact could be scrutinized
with a dedicated study of the neutralino phenomenology in
the~\munmssm.

In some of our analyses we have kept~$\lambda$ large in order to lift
up the mass of the~\sm-like Higgs boson at the tree level through
genuine~\nmssm~effects, and in order to allow for sizable
doublet--singlet mixing. However, we emphasize that large mixing
between the doublet and singlet fields can also be achieved through
small~$\lambda$ in combination with nearly vanishing~$\mue$.  Such a
scenario is viable in the~\munmssm{} and gives rise to a phenomenology
that significantly differs from the~\nmssm.

A phenomenologically very interesting set of scenarios includes light
singlet states. The direct production of these states at colliders
suffers from their nature as gauge singlets: couplings
to~\sm~particles only emerge through the admixture with doublet-like
Higgs states. Similarly, Higgs-to-Higgs decays involving doublet and
singlet fields are strongly correlated with Higgs mixing. We have
shown this effect exemplarily for decays of the~\sm-like Higgs boson
into a pair of light~\cp-odd singlets, which depends on the fraction
of the~\cp-even singlet component in the~\sm-like Higgs
boson~$h^0$. In the~\munmssm, this mixing is not only controlled
through~$\lambda$, but also depends sensitively on the values of~$\mu$
and~$\mue$. We conclude that in order to distinguish the Higgs sectors
of the~\munmssm{} and the~\nmssm, the detection of singlet states in
the Higgs spectrum and their couplings to other Higgs bosons and
the~\sm~particles will be crucial. We have discussed four scenarios
that yield a light~\cp-even singlet-like Higgs around~$97$\,GeV,
motivated by slight excesses in experimental searches performed
with~\cms{} and at~\lep. These scenarios are associated with a
compressed spectrum of light electroweakinos. We have pointed out that
searches at a future electron-positron collider would provide
complementary information to the results achievable at the~\lhc{} in
scenarios of this kind.

\section*{\tocref{Acknowledgments}}
The authors thank S.~Abel, P.~Basler, F.~Domingo, K.~Schmidt-Hoberg,
T.~Stefaniak, A.~Westphal and J.~Wittbrodt for helpful discussions, and
I.~Ben-Dayan, A.~Ringwald and A.~Salvio for insights on Higgs
inflation. This project has been supported by the Deutsche
Forschungsgemeinschaft through a lump sum fund of the SFB~676
``Particles, Strings and the Early Universe''. S.~P.\ acknowledges
support by the~ANR grant ``HiggsAutomator''~(ANR-15-CE31-0002).

\appendix

\tocsection[\label{sec:betaf}]{Beta functions}

The beta~functions for the parameters of the Higgs sector in the
superpotential of the~\gnmssm{} in \eqn{eq:GNMSSM} and their
corresponding soft SUSY-breaking parameters in \eqn{eq:break} can be
found in~\citeres{King:1995vk, Masip:1998jc, Ellwanger:2009dp};
however, since we employ different conventions we list them in the
following. At the one-loop order, we define~\mbox{$\betaf{x}=16\,
  \pi^2\, \frac{\operatorname{d}{x}}{\operatorname{d}{\ln{\mu_r^2}}}$}
as the beta function of parameter~$x$ in the~\drbar~scheme with the
renormalization scale~$\mu_r$ of mass dimension~one.

The symbols~\mbox{$y_t=m_t/v_u$}, \mbox{$y_b=m_b/v_d$}
and~\mbox{$y_\tau=m_\tau/v_d$} denote the top, bottom and tau Yukawa
couplings, respectively. The trilinear soft-breaking parameters are
denoted as~$A_t$ for the stops, $A_b$ for the sbottoms and~$A_\tau$
for the staus. Analog contributions from the first and second
generation of quarks and squarks are not depicted. The
parameters~$M_1$ and~$M_2$ denote the soft-breaking bino and wino
masses.\newline
\begin{subequations}
\begin{align}
  \betaf{\xi} &= \xi\left(\lambda^2 + \kappa^2\right),\\
  \betaf{C_\xi} &= 2\left(\lambda^2\,A_\lambda + \kappa^2\,A_\kappa\right) + 2\,\frac{\lambda\,\mu}{\xi}\left[m_{H_d}^2 + m_{H_u}^2 + B_\mu\left(\nu + A_\lambda\right)\right] + \frac{\kappa\,\nu}{\xi}\left[2\,m_S^2 + B_\nu\left(\nu + A_\kappa\right)\right],\\
  \betaf{\mu} &= 2\,\mu\left(-g_1^2 - 3\,g_2^2 + 2\,\lambda^2 + 3\,y_t^2 + 3\,y_b^2 + y_\tau^2\right),\\
  \betaf{B_\mu} &= g_1^2\,M_1 + 3\,g_2^2\,M_2 + 2\,\lambda^2\,A_\lambda + 3\,y_t^2\,A_t + 3\,y_b^2\,A_b + y_\tau^2\,A_\tau + \frac{\lambda}{\mu}\left(2\,\lambda\,B_\mu\,\mu + \kappa\,B_\nu\,\nu\right),\\
  \betaf{\nu} &= 2\,\nu\left(\lambda^2 + \kappa^2\right),\\
  \betaf{B_\nu} &= 4\left(\lambda^2\,A_\lambda + \kappa^2\,A_\kappa\right) + 2\,\frac{\kappa}{\nu}\left(2\,\lambda\,B_\mu\,\mu + \kappa\,B_\nu\,\nu\right),\\
  \betaf{\lambda} &= \tfrac{1}{2}\,\lambda\,\left(-g_1^2 - 3\,g_2^2 + 4\,\lambda^2 + 2\,\kappa^2 + 3\,y_t^2 + 3\,y_b^2 + y_\tau^2\right),\\
  \betaf{A_\lambda} &= g_1^2\,M_1 + 3\,g_2^2\,M_2 + 4\,\lambda^2\,A_\lambda + 2\,\kappa^2\,A_\kappa + 3\,y_t^2\,A_t + 3\,y_b^2\,A_b + y_\tau^2\,A_\tau\,,\\
  \betaf{\kappa} &= 3\,\kappa\left(\lambda^2 + \kappa^2\right),\\
  \betaf{A_\kappa} &= 6\left(\lambda^2\,A_\lambda + \kappa^2\,A_\kappa\right)\,.
\end{align}
\end{subequations}
In addition we list the one-loop beta functions for the electroweak
vevs\footnote{The results are given in 't Hooft--Feynman gauge. For
details we refer to~\citere{Sperling:2013eva}.} and the soft-breaking
Higgs masses:%
\begin{subequations}
\begin{align}
  \betaf{v_d} &= -v_d\left(-\tfrac{1}{2}\,g_1^2 - \tfrac{3}{2}\,g_2^2 + \lambda^2 + 3\,y_b^2 + y_\tau^2\right),\\
  \betaf{m_{H_d}^2} &= 2\left(-g_1^2\,M_1^2 - 3\,g_2^2\,M_2^2 + \lambda^2\,M_\lambda^2 + 3\,y_b^2\,M_b^2 + y_\tau^2\,M_\tau^2 - \tfrac{1}{2}\,g_1^2\,M_\xi^2\right),\\
  \betaf{v_u} &= -v_u\left(-\tfrac{1}{2}\,g_1^2 - \tfrac{3}{2}\,g_2^2 + \lambda^2 + 3\,y_t^2\right),\\
  \betaf{m_{H_u}^2} &= 2\left(-g_1^2\,M_1^2 - 3\,g_2^2\,M_2^2 + \lambda^2\,M_\lambda^2 + 3\,y_t^2\,M_t^2 + \tfrac{1}{2}\,g_1^2\,M_\xi^2\right),\\
  \betaf{v_s} &= -2\,v_s\left(\lambda^2 + \kappa^2\right),\\
  \betaf{m_S^2} &= 4\left(\lambda^2\,M_\lambda^2 + \kappa^2\,M_\kappa^2\right)\,,
\end{align}
\end{subequations}
where we use the following abbreviations containing the bilinear
soft-breaking parameters~$m_{\tilde{Q}}$, $m_{\tilde{t}}$
and~$m_{\tilde{b}}$ for the squarks, and~$m_{\tilde{L}}$
and~$m_{\tilde{\tau}}$ for the sleptons of the third generation (while
suppressing the first and second generation; the notation is
introduced in Section~\ref{sec:sfermions})
\begin{subequations}
\begin{align}
  M_\lambda^2 &= m_{H_d}^2 + m_{H_u}^2 + m_S^2 + A_\lambda^2\,,\\
  M_\kappa^2 &= 3\,m_S^2 + A_\kappa^2\,,\\
  M_\xi^2 &= m_{H_u}^2 - m_{H_d}^2 + m_{\tilde{Q}}^2 - 2\,m_{\tilde{t}}^2 + m_{\tilde{b}}^2 - m_{\tilde{L}}^2 + m_{\tilde{\tau}}^2\,,\\
  M_t^2 &= m_{H_u}^2 + A_t^2 + m_{\tilde{Q}}^2 + m_{\tilde{t}}^2\,,\\
  M_b^2 &= m_{H_d}^2 + A_b^2 + m_{\tilde{Q}}^2 + m_{\tilde{b}}^2\,,\\
  M_\tau^2 &= m_{H_d}^2 + A_\tau^2 + m_{\tilde{L}}^2 + m_{\tilde{\tau}}^2\,\,.
\end{align}
\end{subequations}

\newpage

\begingroup
\let\secfnt\undefined
\newfont{\secfnt}{ptmb8t at 10pt}
\setstretch{.5}
\bibliographystyle{utcaps}
\bibliography{NMSSM_inflation}
\endgroup

\end{document}